\pdfoutput=1
\documentclass[nonamebreak]{aa}
\usepackage{graphicx}
\usepackage{txfonts}
\usepackage[flushleft]{threeparttable}
\usepackage{multirow}
\usepackage{multicol}
\usepackage{orcidlink}
\usepackage{amsmath}
\usepackage{natbib}
\bibpunct{(}{)}{;}{a}{}{,}
\usepackage{rotating}
\usepackage[T1]{fontenc}
\usepackage{lmodern}
\usepackage{enumitem}
\usepackage{subcaption}
\usepackage{caption}
\usepackage{dcolumn}
\usepackage{xcolor}
\usepackage{url}
\usepackage{array}
\usepackage{tabularx}
\usepackage{makecell}
\usepackage{ragged2e}
\usepackage{booktabs}
\usepackage{standalone}
\usepackage{tikz}
\usepackage{adjustbox}
\usepackage{tocloft}
\usepackage[normalem]{ulem}
\usetikzlibrary{arrows.meta, positioning, shapes.geometric, fit, backgrounds, calc}
\usepackage{bm}
\usepackage{hyperref}
\usepackage{float}       
\usepackage{placeins}    
\providecommand{\nolinenumbers}{}
\usepackage{fancyhdr}
\fancypagestyle{plain}{%
	\fancyhf{}
	\fancyfoot[C]{\thepage}

}

\definecolor{inputcol}{HTML}{534AB7}
\definecolor{proccol}{HTML}{0F6E56}
\definecolor{compcol}{HTML}{185FA5}
\definecolor{deccol}{HTML}{BA7517}
\definecolor{corrcol}{HTML}{D85A30}
\definecolor{addcol}{HTML}{1D9E75}
\definecolor{outcol}{HTML}{993556}
\definecolor{notecol}{HTML}{5F5E5A}
\definecolor{bgloop}{HTML}{F5F4FE}
\definecolor{bgIE}{HTML}{EFF3FF}
\definecolor{bgspc}{HTML}{FFF7EE}

\newcolumntype{Y}{>{\RaggedRight\arraybackslash}X}
\newcolumntype{C}[1]{>{\Centering\arraybackslash}p{#1}}

\newenvironment{ruledtabular}{\par}{\par}

\hypersetup{
	colorlinks=true,
	citecolor=blue,
	linkcolor=blue,
	urlcolor=blue
}

\providecommand{\aap}{A\&A}

\providecommand{\apjl}{ApJ}
\providecommand{\apjs}{ApJS}

\makeatletter

\renewcommand\@biblabel[1]{[#1]} 

\makeatother

\makeatletter

\renewcommand*\@fnsymbol[1]{
	\ensuremath{
		\ifcase#1
		\or \ast
		\or \dagger
		\or \ddagger
		\or \mathsection
		\or \mathparagraph
		\or \|
		\or \ast\ast
		\or \dagger\dagger
		\or \ddagger\ddagger
		\else
		\@ctrerr
		\fi
	}
}

\def\aainstitutename{
	\par
	\begingroup
	\setcounter{aa@institutecnt}{0}
	\renewcommand*\and{
		\par\vspace{2pt}
		\refstepcounter{aa@institutecnt}
		\ifnum\aa@nbinstitutes=1\relax
		\else
		\@textsuperscript{\normalfont\theaa@institutecnt}\enspace
		\fi
		\ignorespaces
	}
	{
		\aa@institutefont
		\centering
		\refstepcounter{aa@institutecnt}%
		\ifnum\aa@nbinstitutes=1\relax
		\else
		\@textsuperscript{\normalfont\theaa@institutecnt}\enspace
		\fi
		\ignorespaces
		\aa@institute\par
	}
	\endgroup
}

\makeatother

\begin{document}
	\raggedbottom
	
	\title{
		Mapping the Milky Way in Six Dimensions: A contiguous, homogenised, phase-space catalogue of \textit{Gaia} DR3 tracers up to 250 kpc.
	}
	\titlerunning{Mapping the Milky Way in Six Dimensions}
	\authorrunning{Mukherjee et al.}
		
		\author{
			Sutirtha Mukherjee \orcidlink{0009-0005-0872-6716}\inst{1,3}\fnmsep
			\thanks{E-mail: \href{mailto:sutirtham@tifrh.res.in}{sutirtham@tifrh.res.in}}
			\fnmsep
			\and
			Amogh Srivastav \orcidlink{0009-0006-1163-1600}\inst{2}\fnmsep
			\thanks{E-mail: \href{mailto:23b1826@iitb.ac.in}{23b1826@iitb.ac.in}}
			\and
			Subhabrata Majumdar\inst{1}\fnmsep
			\thanks{E-mail: \href{mailto:subha@tifr.res.in}{subha@tifr.res.in}}
		}
		
		\institute{
			Department of Theoretical Physics, Tata Institute of Fundamental Research, Mumbai
			\and
			Department of Physics, Indian Institute of Technology Bombay
			\and
			Tata Institute of Fundamental Research, Hyderabad
		}
		
		\date{}
		
		\abstract{
			We present a comprehensive, quality-assured, homogenised catalogue of $32{,}552{,}876$ stellar sources with full 6D phase-space information, spanning a contiguous range of Galactocentric distances from the inner galaxy ($\gtrsim 5$\,kpc) to the outer halo ($\sim\!250$\,kpc). The catalogue cross-matches \textit{Gaia} DR3 astrometry with spectrophotometric distances and line-of-sight velocities from 14 large-scale surveys (including \textsc{desi}, \textsc{sdss-boss}, \textsc{apogee}, \textsc{lamost}, \textsc{galah}, and \textsc{ges}). We achieve a median fractional distance uncertainty of $\approx2.7\%$ for tracers with heliocentric distances, $d_{\rm helio}\le15$\,kpc, increasing to $\approx29\%$ for $15<d_{\rm helio}\le250$\,kpc. The catalogue-wide median fractional radial velocity uncertainty is $\approx8.31\%$. \textit{Gaia}-anchored, survey-specific zero-point corrections and statistical normalisation suppress both random and systematic errors before aggregation. We further derive spectrophotometric parallaxes by jointly exploiting multi-band photometry and spectroscopic observations. For sources with discrepant or missing measurements, superior data from overlapping surveys supersede unreliable values and supplement absent radial velocities, improving both completeness and internal consistency. This multi-survey synthesis yields precise parameters and kinematics for confirmed members of globular clusters, dwarf galaxies, and stellar streams when validated with external catalogues. Of the final sample, $517{,}123$ halo stars, predominantly faint ($G > 17$) and metal-poor ($[\mathrm{Fe/H}] < -1$) tracers including RR~Lyrae, blue horizontal-branch, and K~giant stars, extend well beyond the \textit{Gaia} parallax limit ($\gtrsim\!10$\,kpc), making this the largest, well-calibrated, statistical sample of halo tracers. This unique dataset is ideally suited for accurate multi-component mass modelling, cluster kinematics, chemical abundance mapping, galactic archaeology, dark matter searches, etc. over almost the entire Milky Way volume.
		}
		
		\maketitle
		\nolinenumbers
		{\footnotesize
			\setlength{\parskip}{0pt}
		}
		
		\section{Introduction}
		\label{sec:intro}
		
		Over the past two decades, there has been substantial advancement in stellar
		astrophysics, galactic kinematics and dynamics, and galactic archaeology,
		facilitated largely by the arrival of extensive large-scale digital surveys
		spanning photometric, astrometric, and spectroscopic domains.
		Early photometric surveys such as 2MASS~\citep{Skrutskie2006},
		SDSS~\citep{York2000}, and Pan-STARRS~\citep{Chambers2016} established the
		first wide-area maps of stellar density and colour across the Galaxy, while
		the \textit{Gaia} space mission~\citep{GaiaCollaboration2016} has since
		delivered high-precision proper motions and parallaxes for $\sim$1.5~billion
		sources~\citep{GaiaDR3Summary, GaiaEDR3}, radial velocities for
		$\sim$33~million stars from its Radial Velocity
		Spectrometer~\citep{2019A&A...622A.205K, Katz2023}, and BP/RP spectrophotometry for
		$\sim$220~million objects~\citep{DeAngeli2023}. Complementing
		\textit{Gaia}'s astrometry, ground-based spectroscopic campaigns, including
		SEGUE~\citep{Yanny2009, Rockosi2022}, LAMOST~\citep{Zhao2012, Luo2016,
			Xiang2019}, GALAH~\citep{DeSilva2015}, APOGEE~\citep{Majewski2017}, the
		Gaia-ESO Survey~\citep{Gilmore2012}, RAVE~\citep{Steinmetz2006,
			Kordopatis2013}, and the Dark Energy Spectroscopic Instrument (DESI) Milky
		Way Survey~\citep{Cooper:2023mws, DESI:2022instrument, Koposov2024, Li2025SpecDis}
		now provide radial velocities, stellar atmospheric parameters, and chemical
		abundances for millions of stars across diverse Galactic environments.
		
		The integration of radial velocities with astrometric proper motions and
		distances yields complete six-dimensional (6D) phase-space coordinates
		essential for identifying ancient accretion debris in integral-of-motion
		space~\citep{Myeong2018, Koppelman2019, Naidu2020}, determining the Milky
		Way's mass distribution to its virial radius~\citep{Xue2008, Deason2021_Mass100kpc,
			Bird2022_MWmass, Koposov2023_Potential}, characterising the stellar halo density profile and
		shape~\citep{Deason2011, Deason2014, Sesar2010_HaloShape, Pila-Diez,
			Amarante2024_AnisotropicHalo}, and mapping tidal streams and satellite
		kinematics~\citep{Vasiliev2021, Belokurov2016Streams, Li2019S5}. The principal
		halo tracers employed for these studies, namely blue horizontal branch (BHB)
		stars~\citep{Xue2008, Deason2011, Das2016, Amarante2024_AnisotropicHalo},
		K~giants~\citep{Xue2014, Xu2018}, near-main-sequence turnoff (nMSTO)
		stars~\citep{Sesar2010, Juric2008, Ivezic2008}, and RR~Lyrae as well as
		other variable stars~\citep{Watkins2009, Liu2020, Cohen2017, Feng2024,
			Hernitschek2017, Clementini2019}, each offer complementary advantages but
		also characteristic limitations: BHB stars are intrinsically rare and
		challenging to separate from blue stragglers; K~giant luminosities span two
		orders of magnitude depending on age and metallicity, complicating distance
		inference; and nMSTO stars are too faint to probe $d \gtrsim 20$\,kpc.

		\begin{table*}[htbp]
			\centering
			\caption{Frequently used abbreviations and acronyms in this paper.}
			\label{tab:abbreviations}
			\scriptsize
			\setlength{\tabcolsep}{3.0pt}
			\renewcommand{\arraystretch}{1.2} 
			
			\begin{tabularx}{\textwidth}{@{} l X @{\hskip 6pt} l X @{\hskip 6pt} l X @{}}
				\hline\hline
				\textbf{Abbrev.} & \textbf{Meaning} & \textbf{Abbrev.} & \textbf{Meaning} & \textbf{Abbrev.} & \textbf{Meaning} \\
				\hline
				2MASS & Two Micron All Sky Survey &
				6D & Six-dim.\ phase space (3 pos.\ + 3 vel.) &
				AGB & Asymptotic Giant Branch \\
				
				AGN & Active Galactic Nucleus &
				APOGEE & Apache Point Obs.\ Gal.\ Evol.\ Exp. &
				BHB & Blue Horizontal Branch \\
				
				BOSS & Baryon Osc.\ Spectroscopic Survey &
				BP/RP & Blue/Red Photometer (\textit{Gaia}) &
				CatWISE & \textit{WISE}-based IR source catalogue \\
				
				CMD & Colour--Magnitude Diagram &
				DESI & Dark Energy Spectroscopic Instrument &
				DR & Data Release \\
				
				EM & Expectation--Maximisation &
				EMP & Extremely Metal-Poor &
				GALAH & Gal.\ Archaeology with HERMES \\
				
				GCVS & Gen.\ Cat.\ of Variable Stars &
				GES & Gaia-ESO Survey &
				HB & Horizontal Branch \\
				
				HEALPix & Hier.\ Equal Area isoLat.\ Pixelization &
				KG & K giant &
				LAMOST & Large Sky Area MOS Telescope \\
				
				LEGUE & LAMOST Exp.\ for Gal.\ Und.\ \& Expl. &
				LMC & Large Magellanic Cloud &
				LONEOS & Lowell Obs.\ Near-Earth-Object Search \\
				
				MIST & MESA Isochrones \& Stellar Tracks &
				MWS & Milky Way Survey &
				NGVS & Next Generation Virgo Survey \\
				
				nMSTO & Near-Main-Sequence Turnoff &
				NSIDE & HEALPix resolution parameter &
				NSVS & Northern Sky Variability Survey \\
				
				PARSEC & Padova--Trieste Stellar Evol.\ Code &
				PDR1 & Pristine Data Release 1 &
				PGS & Pristine-Gaia Survey \\
				
				PL/PLZ & Period--Luminosity (--Metallicity) &
				PS1 & Pan-STARRS1 &
				PSF & Point Spread Function \\
				
				PTR & Palomar Transient Factory RRL sample &
				QSO & Quasi-Stellar Object &
				RAVE & Radial Velocity Experiment \\
				
				RC & Red Clump &
				RGB & Red Giant Branch &
				RRL & RR~Lyrae \\
				
				RRab/RRc & Fundamental/first-overtone RRL &
				RUWE & Renorm.\ Unit Weight Error &
				RVS & Radial Velocity Spectrometer \\
				
				SDSS & Sloan Digital Sky Survey &
				SEGUE & Sloan Ext.\ for Gal.\ Und.\ \& Expl. &
				Sgr & Sagittarius \\
				
				SMC & Small Magellanic Cloud &
				SoS / SoS-I & Survey of Surveys (I) &
				SX~Phe & SX Phoenicis stars \\
				
				TOPCAT & Tool for Ops.\ on Cat.\ \& Tables &
				UMP & Ultra Metal-Poor &
				VAC & Value-Added Catalogue \\
				
				VMP & Very Metal-Poor &
				WISE & Wide-field IR Survey Explorer &
				XP & \textit{Gaia} low-res.\ XP spectra \\
				
				BIC  & Bayesian Information Criterion & 
				DUP & duplicate/repeated-measurement calibration route &
				TCH  & three-cornered-hat method \\
				
				GSR  & Galactic Standard of Rest &
				LSR  & Local Standard of Rest &
				KDE  & kernel-density estimate \\
				\hline\hline
			\end{tabularx}
		\end{table*}
		
		\begin{table*}[htbp]
			\centering
			\caption{Frequently used symbols and variables in this paper.}
			\label{tab:variables_compact}
			\scriptsize
			\setlength{\tabcolsep}{2.5pt}
			\renewcommand{\arraystretch}{1.0}
			\begin{tabular}{@{}l l @{\hskip 6pt} l l @{\hskip 6pt} l l@{}}
				\hline\hline
				\textbf{Symbol} & \textbf{Meaning} & \textbf{Symbol} & \textbf{Meaning} & \textbf{Symbol} & \textbf{Meaning} \\
				\hline

				$\varpi$ & Measured parallax &
				$\varpi_c$ & Zero-point-corrected parallax &
				$Z_{\mathrm{EDR3}}$ & \textit{Gaia} EDR3 plx.\ zero-point corr. \\
				$\sigma_\varpi$ & Parallax uncertainty &
				$\sigma_{\mathrm{sys}}$ & Systematic parallax floor &
				$\varpi_{\mathrm{est}}$ & Model-predicted parallax \\
				$\varpi_{\mathrm{phot}}$ & Spectrophotometric parallax &
				& & & \\[5pt]

				$d$ & Generic distance &
				$d_{\rm helio}$ & Heliocentric distance &
				$d_{\rm plx}$ & Parallax-based distance \\
				$d_{\rm spec}$ & Spectroscopic distance &
				$d_{\rm phot}$ & Photometric distance &
				$d_{\rm phot}$ & Spectrophotometric distance \\
				$d_{\rm ref}$ & Reference distance &
				$\sigma_{d_{\rm ref}}$ & Uncertainty in $d_{\rm ref}$ &
				& \\[5pt]

				$r_{\rm gc}$ & Galactocentric radius &
				$R$ & Galactocentric / cylindrical radius &
				$R_{xy}$ & Cylindrical Galactocentric radius \\
				$R_\odot$ & Solar Galactocentric distance &
				$z$ & Height above Galactic plane &
				$l,b$ & Galactic longitude, latitude \\[5pt]

				$\mu_{\alpha^*}$ & PM in RA ($\times\cos\delta$) &
				$\mu_\delta$ & PM in declination &
				$\boldsymbol{\mu}$ & Proper-motion vector \\
				$\mathbf{C}_\mu$ & PM covariance matrix &
				& & & \\[5pt]

				$G$ & \textit{Gaia} broad-band mag. &
				$G_{\rm BP},G_{\rm RP}$ & \textit{Gaia} BP/RP magnitudes &
				$C^*$ & Corrected BP/RP excess \\
				$A_G$ & Extinction in \textit{Gaia} $G$ &
				$E(B\!-\!V)$ & Reddening / colour excess &
				$E$ & \texttt{phot\_bp\_rp\_excess\_factor} \\
				$M_G$ & Abs.\ mag.\ in \textit{Gaia} $G$ &
				$M_V$ & Abs.\ mag.\ in Johnson $V$ &
				$M_R$ & Abs.\ mag.\ in $R$ band \\
				$M_{W1}$ & Abs.\ mag.\ in WISE $W1$ &
				$M_{i'}$ & Abs.\ mag.\ in $i'$ band &
				$M_g^{\rm BHB}$ & BHB abs.\ mag.\ in SDSS $g$ \\[5pt]

				$T_{\rm eff}$ & Effective temperature &
				$T_{\rm eff}^{\rm Ref}$ & Reference $T_{\rm eff}$ (RC cut) &
				$\log g$ & Log surface gravity \\
				$[\mathrm{Fe/H}]$ & Iron metallicity (Solar ref.) &
				$[\alpha/\mathrm{Fe}]$ & Alpha-element abundance &
				$Z$ & Metal mass fraction \\[5pt]
				
				$P$ & Pulsation period &
				$P_F$ & Fundamentalised period &
				$\Phi$ & Pulsation phase \\[5pt]

				$v_r$ & Radial velocity &
				$\sigma_{v_r}$ & RV uncertainty &
				$\Delta v_r$ & Calibrated$-$reference RV diff. \\
				$V_{\rm helio}$ & Heliocentric RV &
				$V_{\rm los}$ & Line-of-sight velocity (GSR) &
				$V_{\rm LSR}$ & LSR circular speed \\
				$(U,V,W)$ & Solar motion w.r.t.\ LSR &
				& & & \\[5pt]

				${\bf A}$ & Design matrix (sp.phot.\ model) &
				${\boldsymbol\theta}$ & Coefficient vector &
				$\lambda_j$ & Regularisation hyperparameter \\
				$y(\lambda)$ & S\'ersic-profile line model &
				$a,b,c$ & S\'ersic fit parameters &
				$\lambda_0$ & Central wavelength \\[5pt]

				$N$ & Total catalogue entries &
				$N_\star$ & Unique stars after dedup. &
				$n$ & Sigma-threshold parameter \\
				$\eta$ & Core / member fraction &
				$\langle P\rangle$ & Mean purity &
				$x_{\rm cut}$ & Generic quality threshold \\
				$\Pi_{\rm RC}(x)$ & Polynomial / purity function &
				$\mathcal{C}(x_{\rm cut})$ & Completeness vs.\ threshold &
				$\mathcal{P}(x_{\rm cut})$ & Purity vs.\ threshold \\
				$\mathcal{Q}(x_{\rm cut})$ & Simple quality function &
				$\mathcal{Q}_{\beta}(x_{\rm cut})$ & Compl.--purity utility ($\beta$-weighted) &
				$\beta$ & Compl./purity weight \\[5pt]

				$u$ & Astrometric quality stat. &
				$\chi^2$ & Chi-squared statistic &
				$\nu$ & Degrees of freedom / $N_{\rm obs}$ \\
				$q_d$ & Relative distance error &
				$q_v$ & Relative RV error &
				$\mathrm{MAD}$ & Median absolute deviation \\
				$\tilde{d}$ & Median recovered distance &
				$\tilde{d}_\varpi$ & Median plx.-inferred distance &
				$\tilde{v}_r$ & Median recovered RV \\[5pt]

				$P_{\mathrm{mem},i}^{(t)}$ & Posterior membership prob. &
				$p_{\mathrm{mem}}^{(t)}$ & Member spatial term &
				$p_{\mathrm{field}}^{(t)}$ & Field spatial term \\
				$\mathcal{L}_{\mathrm{mem},i}^{(t)}$ & Member likelihood &
				$\mathcal{L}_{\mathrm{field},i}^{(t)}$ & Field likelihood &
				$n_{\mathrm{ok}}$ & Valid stars in EM update \\
				$\mathcal{S}_{\mathrm{ok}}$ & Valid-star set (EM) &
				$\bm{\mu}_{\mathrm{mem}}$ & Member PM centroid &
				$w_i^{(t)}$ & EM weight for star $i$ \\
				$v_{\mathrm{sys,EM}}$ & Systemic velocity (EM) &
				& & & \\[5pt]
				
				$f_{\rm DUP}$ & DUP RV scale factor &
				$f_{\rm tch}$ & TCH RV scale factor &
				$f_\star$ & Adopted RV norm.\ factor \\
				$C$ & Colour var.\ (Pristine dwarf--giant) &
				$grz$ & Combined colour index (BHB sel.) &
				& \\
				\hline\hline
			\end{tabular}
		\end{table*}
		
		However, assembling a complete 6D sample from the solar neighbourhood to the
		outer halo ($d \gtrsim 50$\,kpc) remains a formidable challenge owing to two
		fundamental barriers. First, \textit{Gaia}'s geometric parallaxes become
		unreliable beyond $\sim$5--10\,kpc as fractional uncertainties exceed
		unity~\citep{Lindegren2021, BailerJones2021, BailerJones2023}, a regime we
		term the ``\textit{Gaia} barrier,'' where simple parallax inversion is
		dominated by noise and the Lutz--Kelker bias~\citep{BailerJones2021}.
		Second, the \textit{Gaia} RVS is magnitude-limited to $G \lesssim
		14$~\citep{Katz2023}, leaving the vast majority of faint halo tracers without
		radial velocities. Existing outer-halo studies have consequently been forced
		to rely on small, heterogeneous samples of a few hundred to a few thousand
		tracers with full 6D information~\citep{Cohen2017, Feng2024, Xue2011KinematicSubstructure}, or to
		work with incomplete phase-space coordinates.
		
		Significant progress has been made on both fronts individually. For
		distances, spectrophotometric methods now achieve $\sim$10--25\% precision
		through Bayesian isochrone fitting~\citep{Burnett2010, Burnett2011,
			Binney2014, Xue2014, Carlin2015, DasSanders2019, Queiroz2018, Queiroz2020,
			Anders2022, Yang2025, Li2025SpecDis}, data-driven models applied to
		\textit{Gaia} XP spectra~\citep{Zhang2023XP, Andrae2023, Ting2019,
			Xiang2021, Xiang2022}, and the SpecDis neural-network catalogue for
		$>$4~million DESI stars~\citep{Li2025SpecDis}, while the BOSS-HALO
		MINESweeper survey~\citep{2025Mainesweeper} extends Bayesian distances to
		$\sim$100\,kpc for metal-poor halo giants. Period--luminosity--metallicity
		relations deliver $\sim$5\% precision for RR~Lyrae~\citep{Muraveva2015,
			Muraveva2018, Sollima2006, Braga2015} and Cepheids~\citep{Madore1991,
			Matsunaga2006}, with Mira variables~\citep{Whitelock2008, Feast2000,
			Soszynski2007} and $\delta$~Scuti stars~\citep{McNamara2011, McNamara1997}
		extending the ladder to complementary populations. For radial velocities, the
		Survey of Surveys~\citep[SoS;][]{Tsantaki2022SurveyI} established the standard
		homogenisation framework by merging $\sim$11~million RV measurements from five
		surveys onto a common \textit{Gaia} reference frame. More recently,
		\citet{Verberne2024} derived radial velocities from \textit{Gaia} BP/RP
		spectra for $\sim$125~million sources ($G < 17.65$), and DESI provides
		$\sim$1\,km\,s$^{-1}$ precision RVs for millions of stars through its MWS
		value-added catalogue~\citep{Koposov2024}. For each contributing survey, stars in common with \textit{Gaia} define the radial-velocity calibration sample. Survey-dependent RV offsets and smooth trends with the available stellar/observational parameters are fitted on this overlap and then applied to place the survey velocities on the common \textit{Gaia} RV scale.
		
		Existing catalogues address complementary but distinct parts of this problem. SoS-I~\citep{Tsantaki2022SurveyI} homogenised radial velocities from \textit{Gaia}, APOGEE, GALAH, Gaia-ESO, RAVE, and LAMOST onto a common scale for nearly 11 million stars, with reported zero-point accuracies of $\sim0.16$--$0.31\,\mathrm{km\,s^{-1}}$. SoS-II~\citep{Turchi2025SurveyII} extends the survey-combination philosophy to $T_{\rm eff}$, $\log g$, and $[\mathrm{Fe/H}]$, using a PASTEL-calibrated spectroscopic reference and machine-learning estimates to provide atmospheric parameters for $\sim23$ million stars. Our catalogue is complementary: its principal aim is a homogenised 6D phase-space sample that combines \textit{Gaia} DR3 astrometry with DESI, other spectroscopic surveys, photometric information, and specialised halo tracers over $\sim3$--$250\,\mathrm{kpc}$. Atmospheric parameters from SoS-II can be incorporated in future releases where available without changing the broader phase-space and outer-halo scope of the present catalogue. SpecDis delivers accurate spectrophotometric distances but only for DESI targets~\citep{Li2025SpecDis}; \textit{Gaia} XP catalogues offer large, parallax-calibrated distance samples yet lack homogenised radial velocities, since \textit{Gaia} RVS covers only $\sim 33$~million bright stars~\citep{Zhang2023XP, Andrae2023, Katz2023}; and dedicated outer-halo tracer studies remain limited to at most a few thousand stars with non-uniform error properties~\citep{Cohen2017, Feng2024, Liu2020, Xue2014, Amarante2024_AnisotropicHalo, Xue2011KinematicSubstructure}. The most complete existing combination, the DESI DR1~$\times$~\textit{Gaia} DR3 cross-match~\citep{Koposov2024}\footnote{The official \textit{Gaia} cross-match identifies the most probable \textit{Gaia} counterpart of a source in an external catalogue through a local source-to-source procedure, in which nearby candidates are ranked using angular separation and the local source density. Our catalogue construction instead solves the distinct many-to-many survey problem of grouping repeated detections across heterogeneous survey products into one physical star. For catalogues already tied to dense reference catalogues such as \textit{Gaia}, we adopt direct \texttt{source\_id} matches, while specialised tracer catalogues and external systems are validated using the physical-parameter likelihood framework of Section~\ref{sec:membership}.}, provides internally consistent radial velocities and spectrophotometric distances for $\sim 4$~million stars, but remains confined to the DESI footprint and target classes, does not extend beyond $\sim 100$\,kpc, and offers no homogenisation framework for non-DESI surveys. Consequently, applications requiring large-scale, internally consistent 6D kinematics, including mass modelling, stream dynamics~\citep{Vasiliev2021}, cluster kinematics, and substructure identification, still rely on \emph{ad hoc} combinations of heterogeneous surveys that target distinct Galactic regions using different instrumentation and pipelines. Unifying these demands reconciling distance estimates that operate in fundamentally different precision regimes with distinct systematic sources: astrometric, spectrophotometric, photometric, and period-luminosity distances cannot be directly homogenised. It also requires propagating astrometric quantities to a common epoch, enforcing a consistent coordinate framework, carefully propagating uncertainties, rejecting outliers, and externally validating the catalogue before robust scientific inferences can be drawn.

		In this paper, we address this gap by presenting a unified catalogue of
		$\sim$32.5~million unique Milky Way stars with complete 6D phase-space
		information, constructed by cross-matching, calibrating, deduplicating, and
		statistically aggregating 52 million input entries from 14 large-scale surveys spanning
		heliocentric distances from $\sim$3\,kpc to $\sim$250\,kpc, including
		$\sim$517,000 halo stars ($R_{xy} > 25$\,kpc or $|z| > 5\,\mathrm{kpc}$). Three
		principal methodological contributions distinguish this work. The first is a
		scalable cross-survey deduplication pipeline combining HEALPix spatial
		partitioning (NSIDE$\,=\,$32), $k$-d tree pair-finding on unit-sphere
		Cartesian coordinates with parallax-consistency validation, and Union-Find
		connected-component extraction, achieving a 37.5\% deduplication rate
		($N_\star = 32{,}552{,}876$ unique stars) with inverse-variance weighted
		aggregation and outlier rejection. The second is an extended radial-velocity
		homogenisation framework that builds upon the SoS methodology of
		\citet{Tsantaki2022SurveyI} to encompass DESI and 15~additional catalogues not in
		the original SoS, deriving per-survey error normalisation and removing
		parameter-dependent zero-point offsets through weighted least-squares
		polynomial fitting against \textit{Gaia}, reducing inter-survey systematics
		to $\lesssim$0.1--0.5\,km\,s$^{-1}$. The third is a seven-term Bayesian
		expectation-maximisation membership framework applied to 1,722~discrete
		stellar systems: 189~globular clusters~\citep{baumgardt2021},
		1,481~open clusters~\citep{CantatGaudin2018, CantatGaudin2019,
			CantatGaudinAnders2020, CastroGinard2020, Soubiran2018},
		and 52~dwarf galaxies~\citep{pace2022, Koposov2011, Mateo2008,
			Geha2026}, together with the Sagittarius stream~\citep{Vasiliev2021}, jointly fitting spatial position, proper motion,
		radial velocity, spectrophotometric distance, CMD location, [Fe/H], and
		$\log g$ so that the most informative observable at each heliocentric distance
		automatically dominates the membership decision.
		
		The construction is guided by a simple observational fact: no single distance or radial-velocity channel is uniformly reliable across the full Galactic volume. At small heliocentric distance, high-significance \textit{Gaia} parallaxes provide the most direct distance information. As parallax precision deteriorates, we instead use spectrophotometric distances, inferred by combining spectroscopy with broad-band photometry and stellar models or data-driven calibrations, and specialised photometric distance relations for tracers such as BHB stars and RR~Lyrae. Radial velocities are taken from \textit{Gaia} and ground-based spectroscopic surveys, but their quoted uncertainties and zero points are first placed on a common scale. The catalogue therefore does not force one estimator to work everywhere; it selects and combines the most informative measurements available for each star while retaining provenance and uncertainty information.
		
		The paper is organised as follows. Section~\ref{sec:data} describes the input surveys, common quality cuts, and the parallax-based, spectrophotometric, photometric, and radial-velocity channels. The construction of the spectrophotometric distance estimator and its validation are described in the corresponding distance-method section. Section~\ref{sec:pipeline} then explains error normalisation, duplicate identification, RV zero-point calibration, and the final aggregation into one record per physical star. Section~\ref{sec:validation} tests the resulting catalogue against literature-based globular clusters, open clusters, dwarf galaxies, and the Sagittarius stream. Finally, Section~\ref{sec:completeness_purity} quantifies the retained-fraction--purity trade-off and Section~\ref{subsec:cp_guidance} gives practical recommendations for science use.

		\begin{figure*}[htbp]
			\centering
			\includegraphics[width=0.9\textwidth]{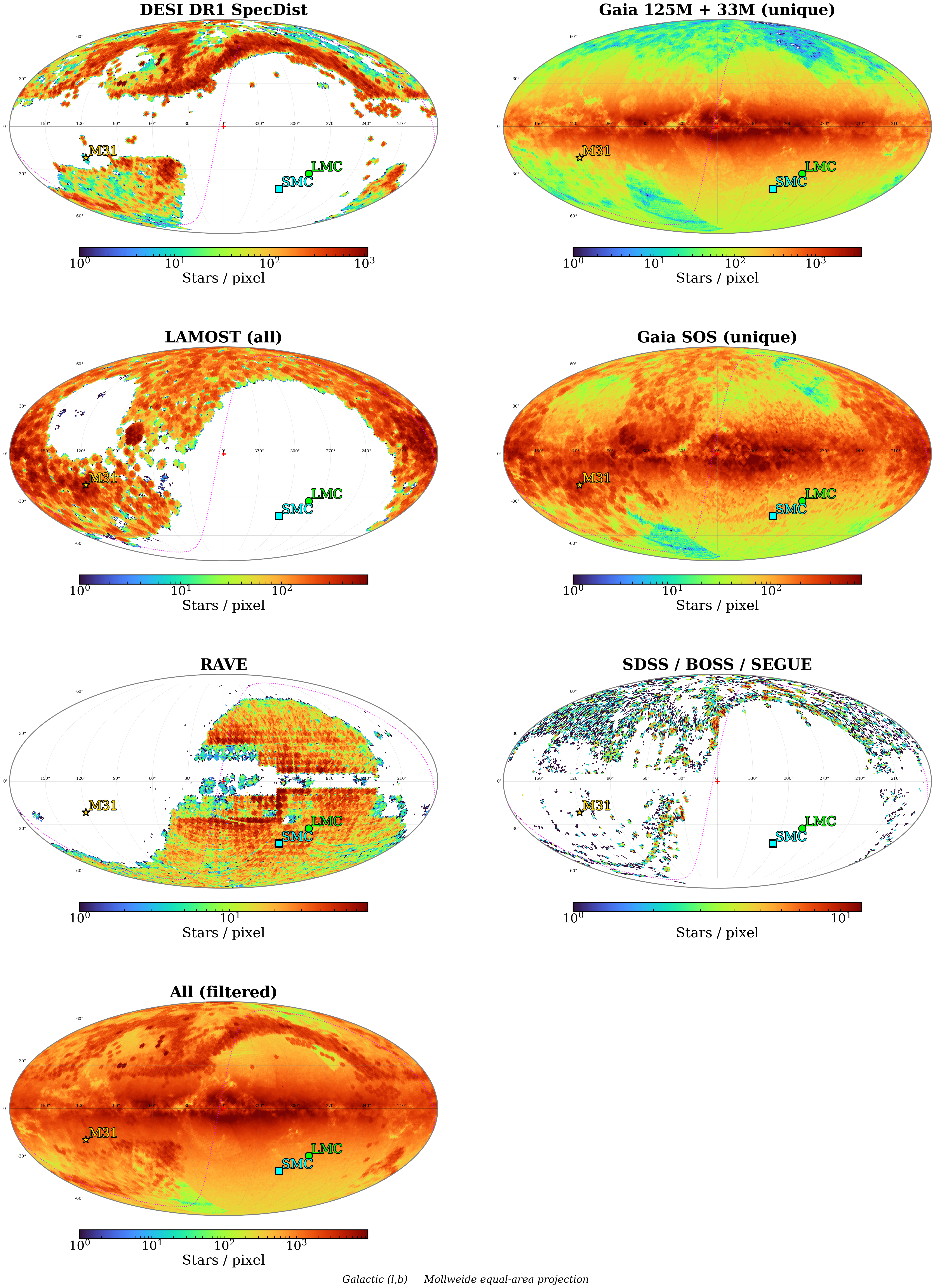}
			\caption{\small Sky coverage of the main components of the unified stellar
				catalogue in Galactic coordinates, shown in a Mollweide equal-area
				projection. Each panel shows a HEALPix density map with logarithmic colour
				scaling in stars per pixel. Shown are DESI DR1 SpecDist, Gaia 125M+33M
				(unique), LAMOST, Gaia SoS-I (unique), RAVE, SDSS/BOSS/SEGUE, and the final
				filtered catalogue. The Gaia-only unique panels isolate sources without
				external-survey counterparts. M31, the LMC, and the SMC are marked in every
				panel.}
			\label{fig:skymap_1}
		\end{figure*}
		
		\begin{figure*}[htbp]
			\centering
			\includegraphics[width=0.9\textwidth]{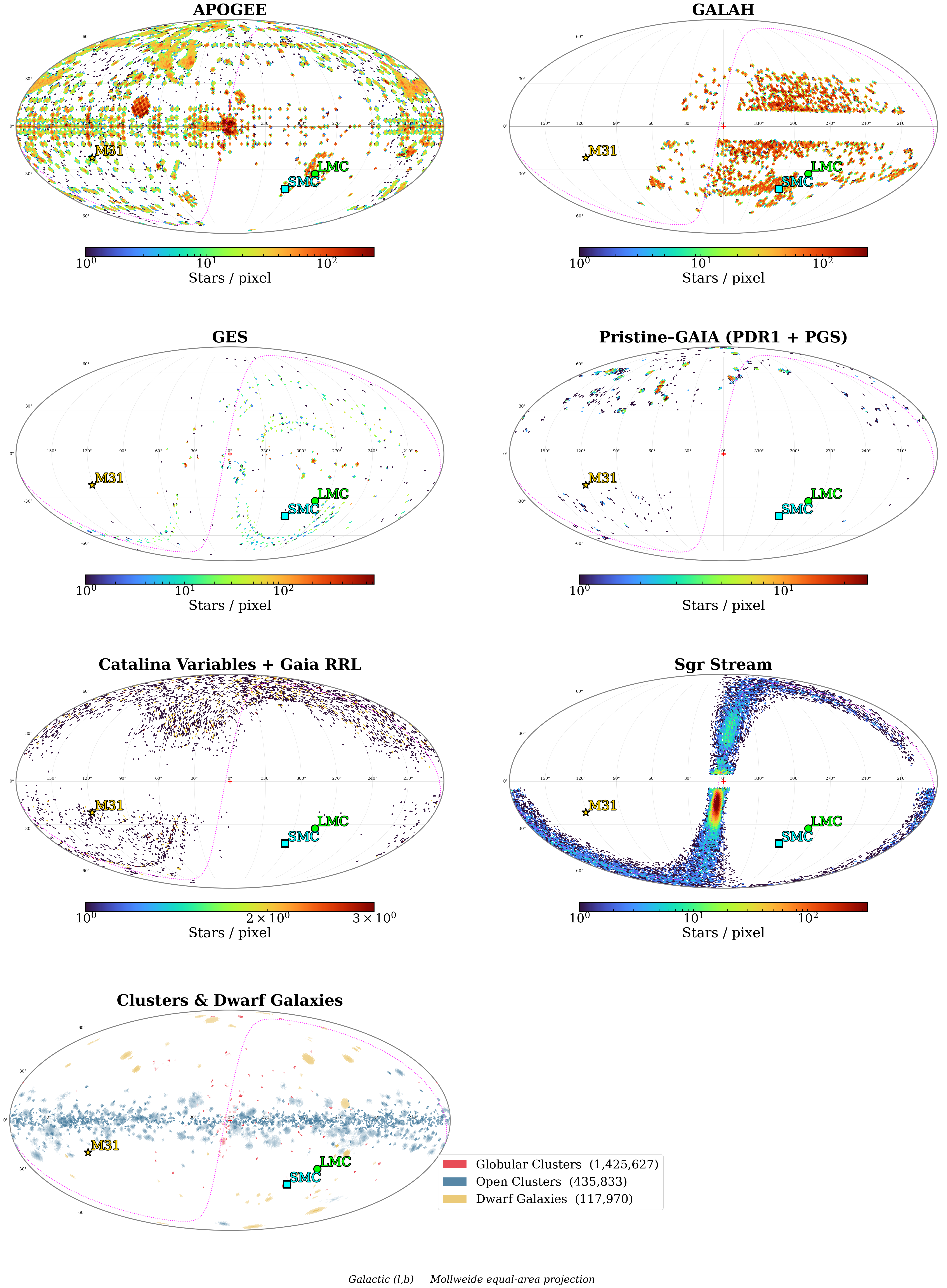}
			\caption{\small Sky coverage of the specialised sub-samples of the unified
				stellar catalogue in Galactic coordinates, shown in a Mollweide equal-area
				projection. Each panel presents a HEALPix density map (NSIDE$=64$) with
				logarithmic colour scaling in stars per pixel. Shown are APOGEE, GALAH,
				GES, Pristine-Gaia, Catalina Variables + Gaia RRL, the Sgr Stream, and
				Clusters \& Dwarf Galaxies. The Pristine-Gaia panel includes sources
				associated with \texttt{PRISTINE+DESI\_RGB}, \texttt{PGS+DESI\_RGB}, and
				\texttt{PTR}, while Catalina Variables + Gaia RRL combines tracers tagged
				as \texttt{RRL}, \texttt{Lamost\_Var}, \texttt{AGB}, \texttt{CAT}, and
				\texttt{CATALINA}, where \texttt{Lamost\_Var} denotes LAMOST variable-star
				candidates and \texttt{CAT}/\texttt{CATALINA} denote Catalina-derived
				variable-star flags. The Sagittarius orbital great-circle path is shown as a magenta dotted line.}
			\label{fig:skymap_2}
		\end{figure*}
		
		\begin{figure*}[htbp]
			\centering
			\begin{subfigure}[t]{0.49\textwidth}
				\centering
				\includegraphics[width=\linewidth]{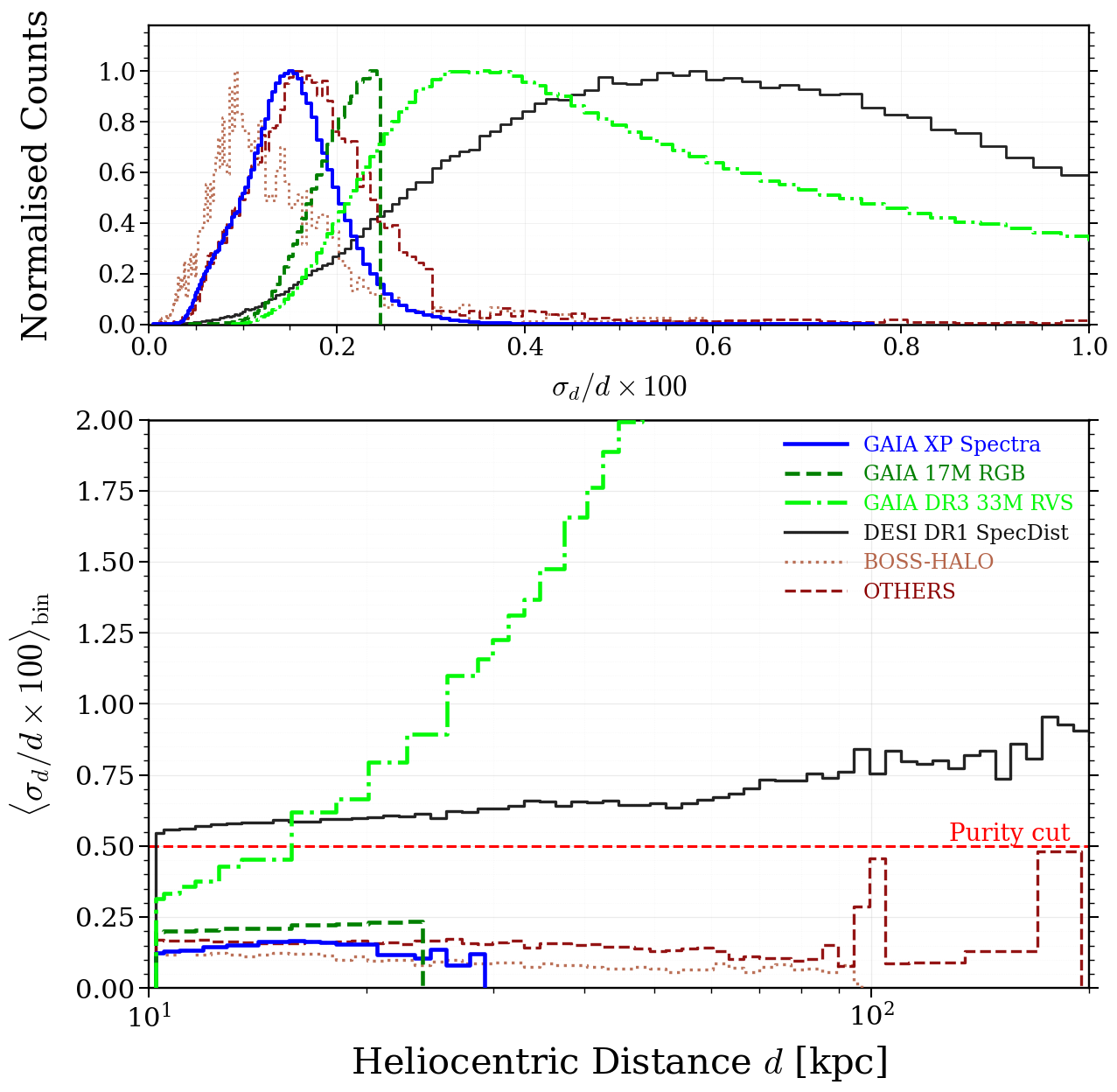}
				\caption{Beyond $10\,\mathrm{kpc}$.}
				\label{fig:dist_far}
			\end{subfigure}
			\hfill
			\begin{subfigure}[t]{0.49\textwidth}
				\centering
				\includegraphics[width=\linewidth]{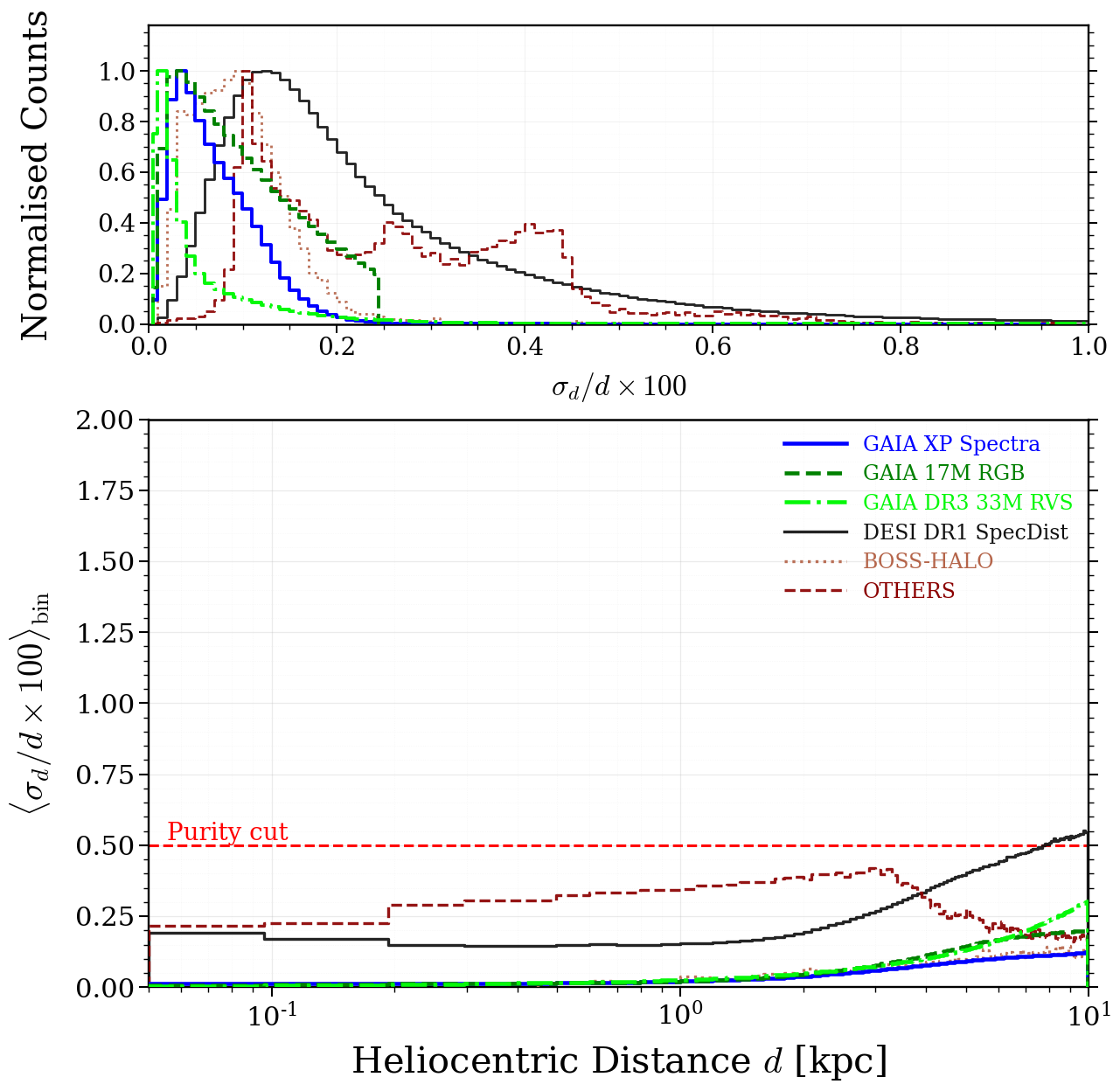}
				\caption{Within $10\,\mathrm{kpc}$.}
				\label{fig:dist_near}
			\end{subfigure}
			\caption{\small Mean relative distance error as a function of heliocentric distance for different tracer samples, shown separately for stars beyond $10\,\mathrm{kpc}$ (left) and within $10\,\mathrm{kpc}$ (right). The horizontal red dashed line marks the reference threshold $q_d=0.5$ ($50\%$ relative distance error). The upper panels show the corresponding normalised distributions of relative distance errors.
			}
			\label{fig:dist_combined}
		\end{figure*}

		\begin{figure*}[htbp]
			\centering
			\begin{subfigure}[t]{0.49\textwidth}
				\centering
				\includegraphics[width=\linewidth]{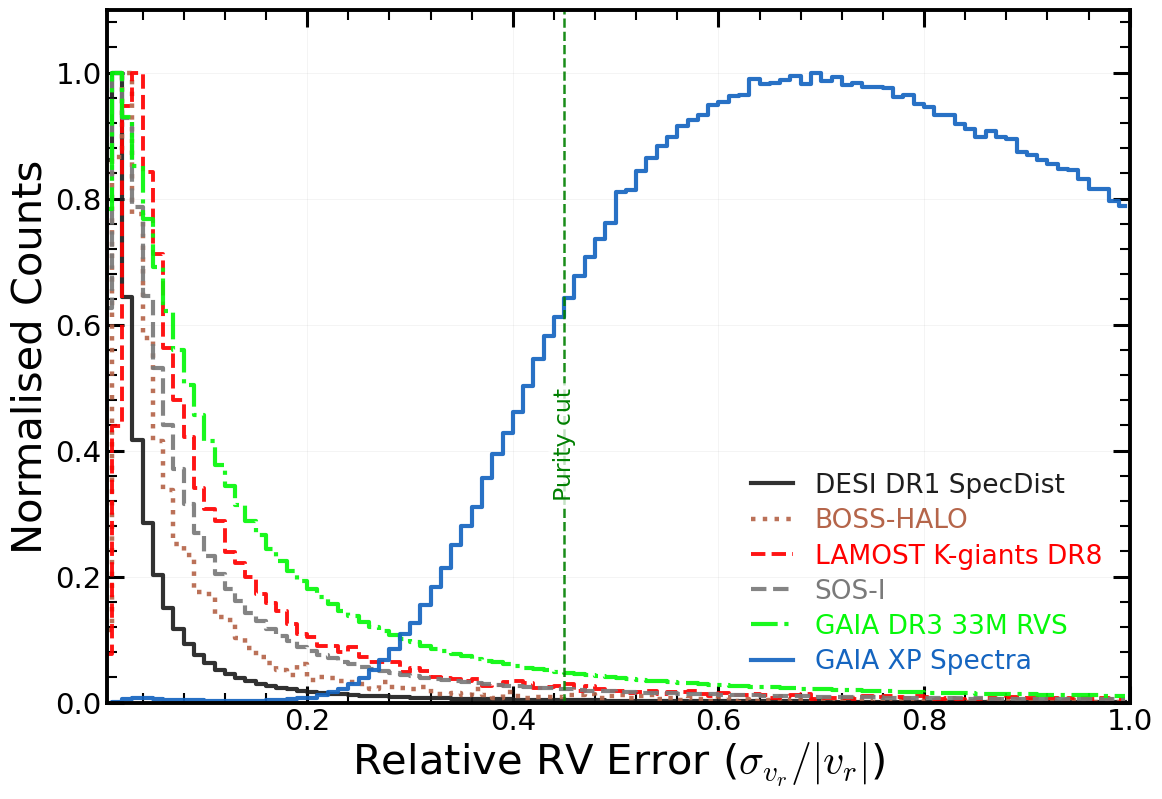}
				\caption{Distribution of relative RV errors.}
				\label{fig:rv_error_dist}
			\end{subfigure}
			\hfill
			\begin{subfigure}[t]{0.49\textwidth}
				\centering
				\includegraphics[width=\linewidth]{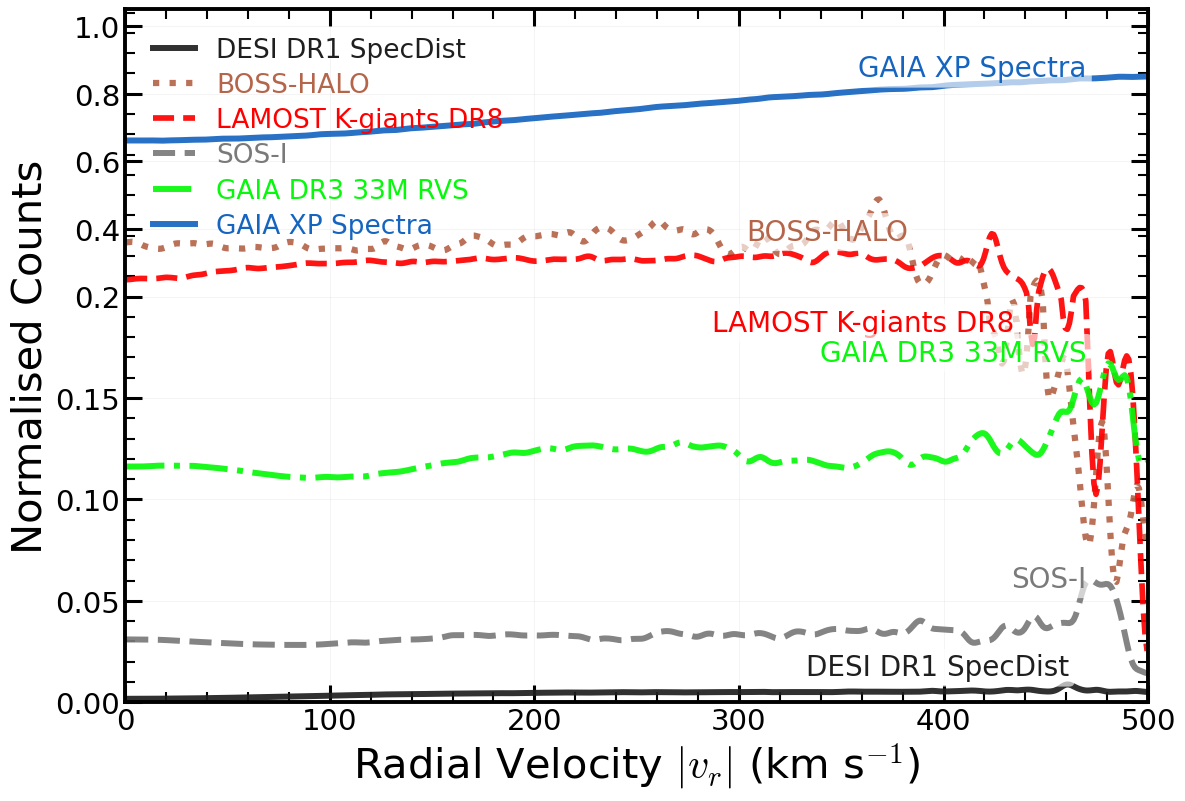}
				\caption{Mean relative RV error versus $|v_r|$.}
				\label{fig:mean_rv_error_vs_vr}
			\end{subfigure}
			\caption{\small Relative radial-velocity error properties for the survey groups contributing RV measurements. The left panel shows the normalised distribution of $\sigma_{v_r}/|v_r|$, with the vertical red line marking the reference threshold $q_v=0.5$ and dotted lines indicating the peak of each survey distribution. The right panel shows the mean relative RV error as a function of absolute radial velocity, with the same survey colours and line styles. 
			}
			\label{fig:rv_combined}
		\end{figure*}

		\section{Data Selection Criteria}\label{sec:data}

		The unified catalogue is assembled from 14 large-scale surveys, comprising
		20 parent catalogues in total (Table~\ref{tab:survey_summary_20}). Each
		derived catalogue is named using the convention
		\texttt{[Survey/Tracer]\_[Observable]\_[Method]\_[Cut]}, where
		\texttt{[Survey/Tracer]} identifies the parent survey or tracer sample,
		\texttt{[Observable]} specifies the measured or inferred quantity,
		\texttt{[Method]} denotes the procedure used to derive it, and
		\texttt{[Cut]} records the selection criteria applied, as described in
		Table~\ref{tab:survey_summary_20}. The allowed observables are
		\texttt{PlxDist}, \texttt{SpecDist}, \texttt{PhotDist}, and \texttt{rv},
		corresponding respectively to parallax-based distances, spectroscopic
		distances, photometric distances, and radial velocities. Thus, each derived
		column name encodes both the source catalogue and the method by which the
		corresponding observable was obtained. The selection cuts are either adopted
		from the relevant literature, with the corresponding references provided
		wherever those cuts are introduced, or are defined in this work to ensure
		the reliability and quality of the final catalogue. The subsections below
		describe the common \textit{Gaia} quality filters, distance catalogues,
		radial velocity catalogues, specialised tracer samples, and the assembly
		procedure.

		\subsection{Common \textit{Gaia} Astrometric Quality Cuts}
		\label{sec:gaia_cuts}

		Following \citet{Lindegren2021} and \citet{Maiz2022}, \textit{Gaia}
		parallaxes are corrected via
		\begin{equation}
			\varpi_c = \varpi - Z_{\mathrm{EDR3}},
			\label{eq:zp_correction}
		\end{equation}
		where $Z_{\mathrm{EDR3}}$ is the instrument bias dependent on magnitude,
		colour, and ecliptic latitude. We impose a 20 per cent relative
		parallax-uncertainty cut to select reliable parallax-based distances and
		reduce the contribution of poorly constrained sources:
		\begin{equation}
			\varpi_c > 0,
			\qquad
			\frac{\sqrt{\sigma^2_{\varpi} + \sigma^2_{\mathrm{sys}}}}{\varpi_c} \le 0.2,
			\label{eq:rel_error}
		\end{equation}
		where $\sigma_{\varpi}$ is the reported parallax uncertainty and $\varpi_c$
		is the corrected parallax. We adopt $\sigma_{\mathrm{sys}} = 0.015$\,mas as
		a conservative systematic uncertainty floor, motivated by residual
		\textit{Gaia} EDR3/DR3 parallax systematics at the level of several to a
		few tens of $\mu$as \citep{Lindegren2021}, and by angular-covariance
		analyses indicating a minimum parallax uncertainty of order $10\,\mu$as
		\citep{Maiz2022}. This floor prevents unrealistically small 
		uncertainties from producing overconfident distance estimates.
		
		To reduce contamination from nearby foreground stars, we remove sources
		whose \textit{Gaia} astrometry is inconsistent with lying beyond
		3\,kpc. Since $3\,\mathrm{kpc}$ corresponds to
		$\varpi_{3\,\mathrm{kpc}} = 1/3\,\mathrm{mas} \simeq 0.333\,\mathrm{mas}$,
		we reject sources satisfying
		\begin{equation}
			\frac{\varpi_c - 0.333\,\mathrm{mas}}{\sigma_{\varpi_c}} > 3,
			\label{eq:fg_plx}
		\end{equation}
		where $\sigma_{\varpi_c}$ is the uncertainty on the corrected parallax.
		This corresponds to a conservative $3\sigma$ foreground rejection, removing
		only sources whose parallaxes are significantly larger than the value
		expected at 3\,kpc.
		
		We also remove sources with highly significant proper motions using the
		two-dimensional \textit{Gaia} proper-motion covariance matrix
		$\mathbf{C}_{\mu} \equiv \mathrm{Cov}(\mu_{\alpha^*}, \mu_{\delta})$,
		rejecting sources for which
		\begin{equation}
			\boldsymbol{\mu}^{T}\mathbf{C}_{\mu}^{-1}\boldsymbol{\mu} > 9,
			\label{eq:fg_pm}
		\end{equation}
		where $\boldsymbol{\mu} = (\mu_{\alpha^*}, \mu_{\delta})^T$. This
		Mahalanobis-distance cut removes objects lying outside a Mahalanobis radius of 3
		from zero proper motion, using the full proper-motion covariance rather than
		treating the two components independently. Such parallax- and proper-motion-based foreground rejection
		is commonly applied to distant stellar samples constructed with
		\textit{Gaia} astrometry \citep[e.g.][]{pace2022, Vasiliev2021}.
		
		We further apply a common set of \textit{Gaia} astrometric and photometric
		quality cuts, following the standard quality-control recommendations of
		\citet{Lindegren2021, GaiaEDR3, GaiaDR3Summary} and the corrected BP/RP
		flux-excess prescription of \citet{Riello2021}:
		\begin{equation}
			\begin{aligned}
				\texttt{RUWE} &< 1.4, \\
				\texttt{visibility\_periods\_used} &> 8, \\
				\texttt{ipd\_frac\_multi\_peak} &\le 2, \\
				\texttt{ipd\_gof\_harmonic\_amplitude} &< 0.1, \\
				\texttt{duplicated\_source} &= \texttt{False}, \\
				\texttt{astrometric\_excess\_noise\_sig} &\le 2, \\
				|C^*| &\le 3\sigma_{C^*}, \\
				\texttt{astrometric\_sigma5d\_max} &< 1.5,
			\end{aligned}
			\label{eq:gaia_quality_cuts}
		\end{equation}
		where \texttt{RUWE} (Renormalised Unit Weight Error) is a normalised
		goodness-of-fit statistic for the astrometric solution ($\texttt{RUWE} \gg
		1$ indicates binaries or problematic fits), and $C^*$ is the corrected
		BP/RP flux-excess factor required to lie within the colour-dependent bounds
		of \citet{Riello2021}. The meaning of each diagnostic and the motivation
		for the adopted thresholds are summarised in
		Appendix~\ref{app:gaia_quality_diagnostics}.

		\subsection{Distance Catalogues}\label{sec:dist_cats}

		\noindent Throughout Sections~\ref{sec:dist_cats}--\ref{sec:variable},
		each sub-catalogue entry is labelled with a number in parentheses, e.g.\ (No.~1a), which corresponds to the survey number assigned in
		Table~\ref{tab:survey_summary_20} and may be used to cross-reference the quality cuts, initial counts, and final source numbers listed therein.
		
		\medskip
		
		\noindent\textbf{\texttt{GAIA\_PlxDist\_cut}} (No.~1a): We use the
		\textit{Gaia} XP-based spectrophotometric distance catalogue of
		\citet{10.1093/mnras/stad1941}, derived from a feed-forward neural-network
		model trained on spectroscopic parameters and multi-band photometry from
		LAMOST, 2MASS, and WISE, applied to approximately $220$ million
		\textit{Gaia} XP sources. We first retain sources satisfying the
		recommended model-quality cuts,
		\begin{equation}
			\chi^2_{\mathrm{opt}}/61 < 2, \qquad \ln(\mathrm{prior}) > -7.43,
		\end{equation}
		where $\chi^2_{\mathrm{opt}}$ is the neural-network optimisation
		chi-squared normalised by its 61 free parameters, and $\ln(\mathrm{prior})$
		is the log prior probability under the model's spatial density prior. These
		reduce the parent sample to $180{,}344{,}401$ sources. We then remove
		objects whose model-inferred parallax $\varpi_{\mathrm{est}}$ is strongly
		inconsistent with the zero-point-corrected \textit{Gaia} parallax
		\citep{Lindegren2021} by requiring
		\begin{equation}
			\frac{|\varpi_{c,\mathrm{Gaia}}-\varpi_{\mathrm{est}}|}
			{\sigma_{\varpi_{c,\mathrm{Gaia}}}} < 10,
			\label{eq:tension}
		\end{equation}
		where $\varpi_{c,\mathrm{Gaia}}$ and $\sigma_{\varpi_{c,\mathrm{Gaia}}}$
		are the corrected parallax and its uncertainty. This removes severe
		astrometric--spectrophotometric outliers while accommodating genuine
		systematics between the two estimators, yielding $90{,}390{,}241$ sources. We further impose
		\begin{equation}
			12 < G < 17, \qquad -0.5 < (G_{\rm BP}-G_{\rm RP}) < 3.0,
		\end{equation}
		together with $\texttt{RUWE} < 1.4$, proper-motion uncertainties below
		$1.0\,\mathrm{mas\,yr^{-1}}$, and $\varpi/\sigma_{\varpi} > 3$, where $G$
		is the \textit{Gaia} broad-band magnitude, $G_{\rm BP}$ and $G_{\rm RP}$
		are the blue- and red-photometer magnitudes. The signal-to-noise cut
		$\varpi/\sigma_{\varpi} > 3$ keeps the fractional parallax-inversion
		distance error below $\sim$33\%. We supplement the catalogue with radial
		velocities from \texttt{DESI\_vac\_rv\_cut}~\citep{Koposov2024},
		\texttt{SOS\_I\_rv\_cut}, \texttt{GAIA\_rv\_RVS\_cut}, and
		\texttt{GAIA\_rv\_BPRP\_cut} (Table~\ref{tab:survey_summary_20}).
		
		\medskip
		
		\noindent\textbf{\texttt{GAIA\_PlxDist\_RGBSubset\_cut}} (No.~1d): We
		include the 17~million quality-screened red giant branch (RGB) sample of
		\citet{Andrae2023} as a subset of \texttt{GAIA\_PlxDist\_cut}. It is
		derived via XGBoost, a gradient-boosted decision-tree algorithm mapping
		photometric features to stellar parameters from $\approx$175~million
		stars with complete \textit{Gaia}, CatWISE2020, and synthesised photometry.
		
		\emph{Giant selection.} We retain sources with surface gravity $\log g <
		4.0$ and effective temperature $T_{\rm eff} < 5500$\,K, placing them in
		the cool giant region of the Hertzsprung--Russell diagram. We additionally
		impose two luminosity-floor cuts in the absolute $W1$-band magnitude,
		\begin{align}
			M_{W1} &> -0.3 - 0.006\,(5500 - T_{\rm eff}), \label{eq:MW1_cut1}\\
			M_{W1} &> -0.01\,(5300 - T_{\rm eff}), \label{eq:MW1_cut2}
		\end{align}
		where $M_{W1} \equiv W1 + 5\log_{10}(\varpi/100)$ is the parallax-based
		absolute magnitude in the CatWISE2020 $W1$ band ($3.4\,\mu$m), with
		$\varpi$ in mas. These cuts remove reddened hot stars and early-type dwarfs
		whose dust-reddened colours mimic the giant locus but whose near-infrared
		luminosity falls well short of a genuine RGB star at the same $T_{\rm eff}$.
		
		We assign each source a deduplication priority
		tier based on $[\mathrm{Fe/H}]_{\rm XP}$: tier~1 for
		$[\mathrm{Fe/H}] < -2.0$, tier~2 for $< -1.5$, and tier~3 for $< -1.0$.
		When the same star appears in multiple sub-catalogues
		(Section~\ref{sec:assembly}), we retain the most metal-poor classification,
		preventing rare very-metal-poor halo stars from being displaced by more
		numerous metal-rich disc giants.
		
		We remove blended, reddened, or astrometrically
		unreliable sources by requiring
		\begin{align}
			\texttt{phot\_bp\_rp\_excess\_factor}
			&< 1.3 + 0.06\,(G_{\rm BP}-G_{\rm RP})^2, \label{eq:rgb_clean}\\
			\varpi/\sigma_\varpi &> 5, \\
			\texttt{RUWE} &< 1.4, \\
			A_G &< 0.5\,\text{mag}, \\
			\epsilon_a &< 0.3\,\text{mas},
		\end{align}
		where \texttt{phot\_bp\_rp\_excess\_factor} is the $BP{+}RP$-to-$G$ flux
		ratio (elevated values indicate blending or background contamination),
		$A_G$ is the $G$-band line-of-sight extinction, and
		$\epsilon_a \equiv \texttt{astrometric\_excess\_noise}$ is the residual
		scatter in the astrometric fit beyond the  model (large values signal
		unresolved binaries or non-stellar morphology). We adopt the stricter
		threshold $\varpi/\sigma_\varpi > 5$ compared to $> 3$ in
		\texttt{GAIA\_PlxDist\_cut} because the RGB absolute-magnitude calibration
		is more sensitive to distance errors.

		\begin{table*}[htbp]
			\centering
			\caption{Summary of 20 parent catalogues, grouped by survey.}
			\label{tab:survey_summary_20}
			\scriptsize
			\setlength{\tabcolsep}{4pt}
			\renewcommand{\arraystretch}{1.4}
			\setlength{\extrarowheight}{2pt}
			\begin{adjustbox}{max width=\textwidth,max totalheight=1.0\textheight,center}
				\begin{tabular}{@{}p{1.3cm}p{0.5cm}p{2.45cm}p{1.35cm}p{6.55cm}p{2.9cm}p{1.55cm}@{}}
					\hline\hline
					\textbf{Survey} & \textbf{No.} & \textbf{Parent Catalogue} &
					\textbf{Initial Count} & \textbf{Quality Cuts \& Selection Criteria} &
					\textbf{Derived catalogue\footnote{%
							\texttt{[Survey/Tracer]-[Observable]-[Method]-[Cut]}}} &
					\textbf{Final Count} \\
					\hline

					\multirow{4}{*}{\textit{Gaia}}
					& 1a & \texttt{GAIA XP Spectra} \newline \citep{10.1093/mnras/stad1941}
					& 220M
					& $\chi^2_{\rm opt}/61 < 2$; $\ln(\text{prior}) > -7.43$;
					$|\varpi_{\rm Gaia} - \varpi_{\rm est}|/\sigma_{\varpi,\rm Gaia} < 10$;
					$12 < G < 17$; RUWE $< 1.4$;
					proper-motion errors $< 1.0$\,mas\,yr$^{-1}$;
					$-0.5 < (G_{\rm BP}-G_{\rm RP}) < 3.0$; parallax S/N $> 3$
					& \texttt{GAIA\_PlxDist\_cut}
					& 90,390,241  \\
					\noalign{\vskip 2pt}\cline{2-7}\noalign{\vskip 2pt}
					& 1b & \texttt{GAIA RV BP-RP} \newline \citep{Verberne2024}
					& 125M
					& $E(B-V) < 0.5$; \texttt{rv\_err} $< 300$\,km\,s$^{-1}$;
					\texttt{CMD\_outlier\_fraction} $< 0.1$;
					\texttt{bad\_measurement} $< 0.1$;
					Random Forest classifier quality parameters; $G < 17.65$
					& \texttt{GAIA\_rv\_BPRP\_cut}
					& 6,367,355 \\
					\noalign{\vskip 2pt}\cline{2-7}\noalign{\vskip 2pt}
					& 1c & \texttt{GAIA DR3 33M RVS} \newline \citep{Katz2023}
					& 33M
					& Parallax S/N $> 2.3$; RUWE $< 1.4$;
					proper-motion errors $< 1.0$\,mas\,yr$^{-1}$
					& \texttt{GAIA\_rv\_RVS\_cut}
					& 2,314,443 \\
					\noalign{\vskip 2pt}\cline{2-7}\noalign{\vskip 2pt}
					& 1d & \texttt{GAIA 17M RGB} \newline \citep{Andrae2023}
					& 17M
					& See Section~\ref{sec:dist_cats}
					& \texttt{GAIA\_PlxDist\_RGBSubset\_cut}
					& 17M \\
					\midrule
					
					\textit{Pristine}
					& 2 & \texttt{Pristine-GAIA (PDR1 + PGS)}
					\newline \citep{Viswanathan2025Pristine, Viswanathan2024PristineXXVII}
					& 180,314 (PDR1); 2,420,898 (PGS)
					& $-4.0 < [\text{Fe/H}] < +0.1$; $\log g < 4.2$;
					$0.5 < (G_{\rm BP}-G_{\rm RP})_0 < 1.5$;
					parallax uncertainty $< 10\%$;
					piecewise dwarf--giant separation; RV not null
					& \texttt{PDR1\_PGS\_PhotDist\_cut}
					& 1,406 \\
					\midrule
					
					\textit{PTR}
					& 3 & \texttt{PTR RRLyrae ab} \newline \citep{Cohen2017}
					& 446
					& RRab at $d_{\rm helio} \geq 50$\,kpc;
					classification probability $> 0.70$;
					$|v_r| < 200$\,km\,s$^{-1}$ ($d_{\rm helio} < 85$\,kpc)
					or $< 170$\,km\,s$^{-1}$ (beyond)
					& \texttt{RRL\_PTR\_SpecDist\_rv\_cut}
					& 116 \\
					\midrule
					
					\multirow{2}{*}{\textit{DESI}}
					& 4a & \texttt{DESI DR1 SpecDist} \newline \citep{Li2025SpecDis}
					& $> 4\times10^6$
					& distance error $< 25\%$; \texttt{BINARY\_FLAG} $= 1$;
					$\log g < 3.8$ (giants); RUWE $< 1.2$;
					\texttt{PHOT\_VARIABLE\_FLAG} $\neq$ VARIABLE
					& \texttt{DESI\_SpecDist\_cut}
					& $\sim$387,126 \\
					\noalign{\vskip 2pt}\cline{2-7}\noalign{\vskip 2pt}
					& 4b & \texttt{DESI DR1 MWS-VAC} \newline \citep{Koposov2024}
					& 625,688
					& $16 < r < 19$; $r_{\rm obs} < 20$; $g-r > 0.7$;
					\texttt{type} = PSF;
					$\epsilon_a < 3$;
					\texttt{duplicated\_source} = False;
					\texttt{brick\_primary} = True;
					$\varpi < \max(3\sigma_\varpi, 1\,\text{mas})$;
					$|\mu| < 7$\,mas\,yr$^{-1}$;
					\texttt{astrometric\_params\_solved} $= 31$
					& \texttt{DESI\_vac\_rv\_cut}
					& 25,994 (red giants); 961 (BHB) \\
					\midrule
					
					\multirow{3}{*}{\textit{SDSS}}
					& 5a & \texttt{BOSS-HALO MINESweeper} \newline \citep{2025Mainesweeper}
					& 8,777
					& $|b| > 15^\circ$ (plane exclusion);
					parallax cuts for nearby dwarfs;
					WISE infrared colours for giants;
					\textit{Gaia} parallax quality cuts as
					Section~\ref{sec:gaia_cuts}
					& \texttt{BOSS\_HALO\_PlxDist\_rv\_cut}
					& 1,333 \\
					\noalign{\vskip 2pt}\cline{2-7}\noalign{\vskip 2pt}
					& 5b & \texttt{SEGUE K-giants} \newline \citep{Xue2014SEGUE}
					& 6,036
					& relative distance error $< 20\%$; RV not null
					& \texttt{SEGUE\_KG\_SpecDist\_rv\_cut}
					& 5,038 \\
					\noalign{\vskip 2pt}\cline{2-7}\noalign{\vskip 2pt}
					& 5c & \texttt{SDSS-BHB DR8} \newline \citep{Xue2011KinematicSubstructure}
					& 12,530
					& $0.8 < (u-g) < 1.5$; $-0.5 < (g-r) < 0.0$; RV not null
					& \texttt{BHB\_SDSS\_SpecDist\_rv\_cut}
					& 4,981 \\
					\midrule
					
					\multirow{2}{*}{\textit{LAMOST}}
					& 6a & \texttt{LAMOST K-giants DR8} \newline \citep{Zhang2023}
					& 19,544
					& $4000 < T_{\rm eff} \leq 5600$\,K;
					$\log g < 3.5$ ($T_{\rm eff} < 4600$\,K) or
					$\log g < 4.0$ ($4600 \leq T_{\rm eff} \leq 5600$\,K);
					RC exclusion; HB cut: Eq.~(\ref{eq:HB_cut});
					$E(B-V) < 0.25$
					& \texttt{LAMOST\_KG\_SpecDist\_rv\_cut}
					& 19,543 \\
					\noalign{\vskip 2pt}\cline{2-7}\noalign{\vskip 2pt}
					& 6b & \texttt{LAMOST VMP DR9} \newline \citep{LAMOSTVMP}
					& 111,000
					& $[\text{Fe/H}] < -2.0$ (VMP, 32,631 stars);
					$[\text{Fe/H}] < -3.0$ (EMP, 702 stars);
					$[\text{Fe/H}] < -4.0$ (UMP, 30 stars);
					$[\text{Fe/H}] < +0.39$;
					$0.5 \leq (g-r)_0 \leq 1.4$
					& \texttt{LAMOST\_VMP\_SpecDist\_rv\_cut}
					& 9,207 \\
					\midrule
					
					\textit{LSDR9}
					& 7 & \texttt{Legacy Survey DR9 BHB}
					\newline \citep{Amarante2024_AnisotropicHalo}
					& 95,446
					& $-0.14 < grz < 0.07$; $-0.30 < (g-r) < -0.05$;
					Eq.~(\ref{eq:grz});
					BHB probability $> 70\%$;
					cross-matched by DESI source ID to obtain RV
					& \texttt{BHB\_LSDR9\_PhotDist\_rv\_cut}
					& 969 \\
					\midrule
					
					\multirow{2}{*}{\parbox{1.3cm}{\textit{Catalina/}\\\textit{LINEAR}}}
					& 8a & \texttt{AGB Candidates}
					\newline \citep{Mauron2019OxygenRichGiants}
					& 417
					& $E(B-V) < 0.3$; $|z| > 5$\,kpc (halo);
					$10.5 < (V_{\rm CSS})_0 < 17$;
					$2.6 < (V-K_s)_0 < 5.6$; $0.75 < (J-K_s) < 1.4$
					& \texttt{AGB}
					& 330 (90 with RV) \\
					\noalign{\vskip 2pt}\cline{2-7}\noalign{\vskip 2pt}
					& 8b & \texttt{Catalina Variables}
					\newline \citep{Drake2014Catalina}
					& 61,000
					& derived distances; cross-matched with
					\texttt{SOS\_I\_rv\_cut} and \texttt{GAIA\_rv\_BPRP\_cut}
					& \texttt{CAT\_Variables\_PhotDist\_rv\_cut}
					& 16,451 \\
					\midrule
					
					\textit{NGVS}
					& 9 & \texttt{NGVS RRLyrae} \newline \citep{Feng2024}
					& 180
					& QSO/AGN exclusion; RV not null
					& \texttt{NG\_Virgo\_SpecDist\_rv\_cut}
					& 154 \\
					\midrule
					
					\parbox{1.3cm}{\textit{SEGUE/}\\\textit{LEGUE}}
					& 10 & \texttt{SEGUE+LEGUE RRL stars} \newline \citep{Liu2020}
					& 5,290
					& distances from Table~\ref{tab:rrl_surveys}
					& \texttt{RRL\_SpecDist\_rv\_cut}
					& 3,345 \\
					\midrule
					
					\textit{SoS-I}
					&  & \texttt{SoS-I} \newline \citep{Tsantaki2022SurveyI}
					& $\sim$11M
					&$(\texttt{quality\_flags}~\&~32) = 0$;
					$1.0 + 0.015(G_{\rm BP}-G_{\rm RP})^2 < E <
					1.3 + 0.06(G_{\rm BP}-G_{\rm RP})^2$;
					$u < 1.2\max(1, e^{-0.2(G-19.5)})$,
					where
					$u = \sqrt{\chi^2/\nu}$ and
					$\nu = \texttt{astrometric\_n\_good\_obs\_al}-5$.
					& \texttt{SOS\_I\_rv\_cut}
					& $\sim$1M; 1,595 with $R>30$\,kpc \\
					
					\hline\hline
				\end{tabular}
			\end{adjustbox}
		\end{table*}
		
		\medskip
		
		\noindent\textbf{\texttt{DESI\_SpecDist\_cut}} (No.~4a): We use the DESI
		Data Release~1 spectrophotometric distance catalogue of
		\citet{Li2025SpecDis}, which employs a neural network trained on
		\textit{Gaia} parallaxes. The original catalogue is pre-filtered by the
		authors for non-variable, astrometrically well-behaved stars via
		$\texttt{PHOT\_VARIABLE\_FLAG} \neq \texttt{VARIABLE}$ and
		$\texttt{RUWE} < 1.2$. We then apply the following additional quality cuts:
		\begin{align}
			\texttt{DISTERR}/\texttt{DIST} &< 0.25, \\
			\texttt{BINARY\_FLAG} &= 1, \\
			\log g &< 3.8,
		\end{align}
		where \texttt{DISTERR}/\texttt{DIST} is the fractional distance uncertainty
		(retaining only sources with distances precise to better than 25\%),
		\texttt{BINARY\_FLAG} $= 1$ flags sources identified as likely single stars
		by the catalogue pipeline, and $\log g < 3.8$ selects giants. These cuts
		yield $\approx$387,126 giants extending to $\approx$100\,kpc, which we
		cross-match with \texttt{GAIA\_PlxDist\_cut}.
		
		\medskip
		
		\noindent\textbf{\texttt{BOSS\_HALO\_PlxDist\_rv\_cut}} (No.~5a): We use
		the MINESweeper catalogue \citep{2025Mainesweeper}, which simultaneously
		fits BOSS spectra \citep{BOSS2013}, broad-band photometry, and
		\textit{Gaia} parallaxes to MIST stellar isochrones \citep{Choi2016},
		naturally blending astrometric and spectroscopic distance information.
		\textit{Gaia} parallaxes dominate within $\approx$10\,kpc, while isochrone
		distances take over at larger distances. We select sources satisfying
		Galactic latitude $|b| > 15^\circ$ (to reduce the impact of dust
		extinction and crowding near the plane), the \textit{Gaia} astrometric
		quality cuts of Section~\ref{sec:gaia_cuts} and
		Eqn. \ref{eq:tension}, WISE colour refinement to confirm giant
		status, and exclusion of confirmed quasars
		\citep{VeronCettyV2010, Schneider2010}. Of 8,777 initial sources, 1,333
		survive, including 379 Sagittarius stream members, 381 substructure
		candidates, and metal-poor populations.
		
		\medskip
		
		\noindent\textbf{\texttt{PDR1\_PGS\_PhotDist\_cut}} (No.~2): We use the
		Pristine survey photometric distance catalogue
		\citep{Viswanathan2025Pristine, Viswanathan2024PristineXXVII}, which
		derives distances for red giant branch candidates by fitting three
		independent sets of stellar isochrones MIST \citep{Choi2016}, PARSEC
		\citep{Bressan2012, Pietrinferni2004}, and BaSTI \citep{Hidalgo2018}to
		provide a cross-check on systematic uncertainties in the stellar models.
		The typical distance uncertainty is 12\%, reaching $\approx$100\,kpc for
		PDR1 and $\approx$70\,kpc for PGS. We retain sources satisfying
		\begin{align}
			-4.0 < [\mathrm{Fe/H}] &< +0.1, \\
			\log g &< 4.2, \\
			0.5 < (G_{\rm BP} - G_{\rm RP})_0 &< 1.5, \\
			\sigma_\varpi / |\varpi| &< 0.10,
		\end{align}
		where $(G_{\rm BP} - G_{\rm RP})_0$ is the dereddened blue-to-red
		photometer colour and $\sigma_\varpi/|\varpi|$ is the fractional parallax
		uncertainty. We additionally apply a piecewise dwarf--giant luminosity
		separator: sources are classified as giants if their absolute $G$-band
		magnitude satisfies $M_G < M_G^{\rm thresh}(C)$, where
		$C \equiv (G_{\rm BP} - G_{\rm RP})_0$ is the dereddened colour and
		\begin{equation}
			M_G^{\rm thresh}(C) =
			\begin{cases}
				0.9,          & C < 0.8, \\
				6.25C - 1.7,  & 0.8 \leq C \leq 0.952, \\
				4.25,         & C > 0.952.
			\end{cases}
			\label{eq:pristine_MG}
		\end{equation}
		This colour-dependent threshold rises steeply through the subgiant--giant
		transition region ($0.8 < C < 0.952$), cleanly separating the dwarf and
		giant sequences in the colour--magnitude diagram. We recover missing radial
		velocities by cross-matching with \texttt{DESI\_vac\_rv\_cut} and
		\texttt{SOS\_I\_rv\_cut}, yielding 1,406 sources in total.
		
		\begin{figure*}[htbp]
			\centering
			\includegraphics[width=0.95\linewidth]{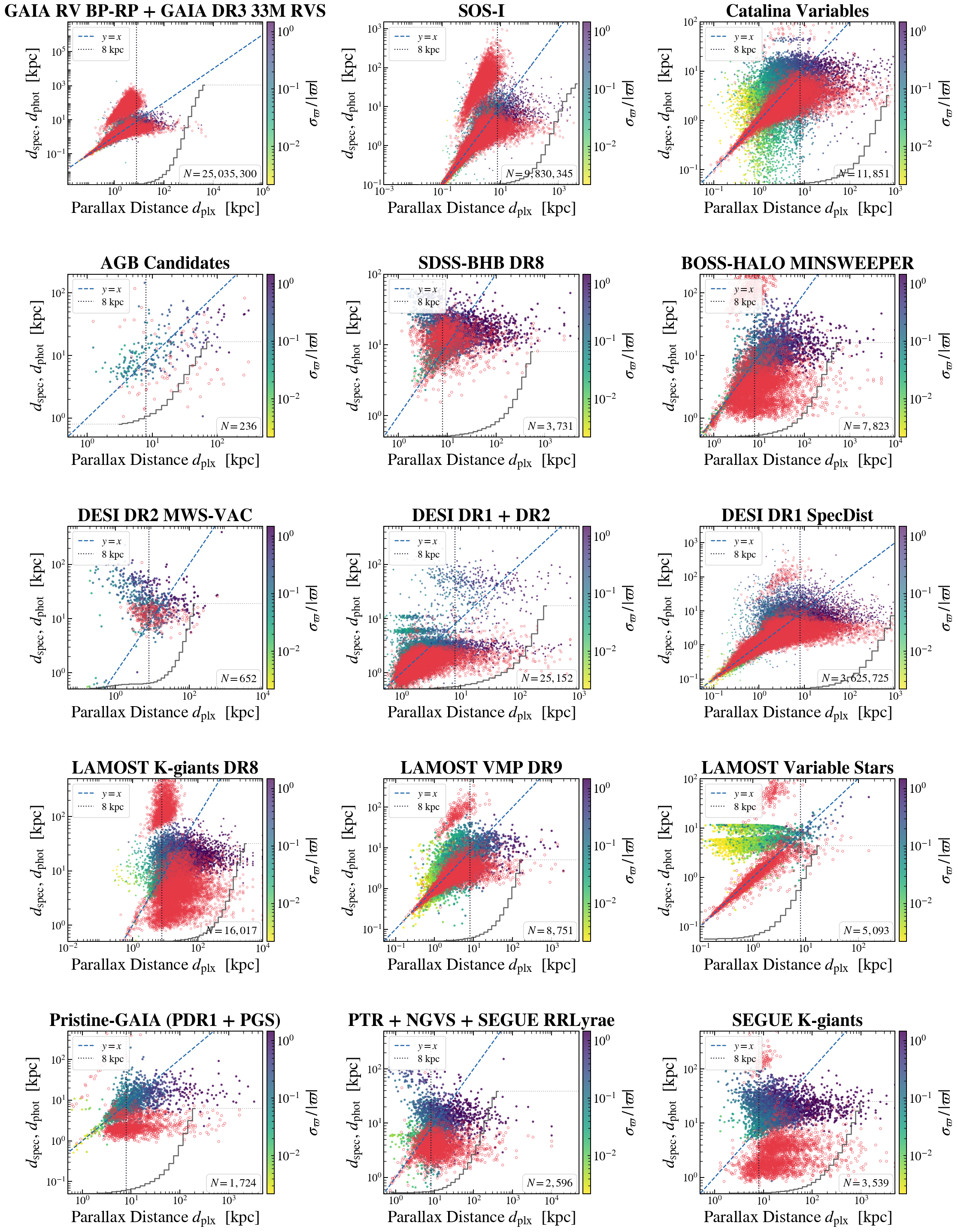}
			\caption{\small Comparison of spectroscopic and photometric distances with
				inverse-parallax distances for the derived sub-catalogues. In each panel,
				$d_{\rm spec}$ and $d_{\rm phot}$ are plotted against $d_{\rm plx}$, with
				points colour-coded by relative parallax precision, $\sigma_\varpi/|\varpi|$.
				The dashed line marks $y=x$, and the vertical dotted line marks
				$8\,\mathrm{kpc}$. Good agreement is seen at nearby distances, while at
				larger distances the inverse-parallax estimates become increasingly scattered
				and biased, whereas the spectroscopic and photometric distances remain more
				stable. The level of agreement varies across tracer populations, illustrating
				the complementary role of non-parallax distance estimates in the outer halo.}
			\label{fig:spectophoto}
		\end{figure*}
		
		\medskip
		
		\noindent\textbf{Red Clump Separation}: Across all red giant samples, we
		identify and exclude red clump (RC) stars core helium-burning giants that
		are overluminous relative to the RGB and would bias spectrophotometric
		distance estimates using the criteria of
		\citet{Huang..2015RAA....15.1240H}. A source is flagged as RC if it
		simultaneously satisfies
		\begin{align}
			&1.8 \leq \log g \leq
			0.0009\,\mathrm{dex\,K^{-1}}(T_{\rm eff} - T_{\rm eff}^{\rm Ref}) + 2.5,
			\label{eq:RC_logg}\\
			&T_{\rm eff}^{\rm Ref} =
			-876.8\,\mathrm{K\,dex^{-1}}\,[\mathrm{Fe/H}] + 4431\,\mathrm{K},
			\label{eq:RC_Teff}\\
			&1.21\,(J\!-\!K_s)_0^9 - 0.0859 < Z_{\rm met} <
			2.58\,(J\!-\!K_s)_0^3 - 0.4003,
			\label{eq:RC_Z}
		\end{align}
		where $T_{\rm eff}^{\rm Ref}$ is a metallicity-dependent reference
		temperature defining the centre of the RC locus in the $(\log g,
		T_{\rm eff})$ plane, $(J-K_s)_0$ is the dereddened near-infrared colour,
		and $Z_{\rm met}$ is the stellar metal mass fraction (dimensionless ratio
		of metal mass to total mass) obtained via
		$\log Z_{\rm met} = 0.977\,[\mathrm{Fe/H}] - 1.699$
		\citep{Bertelli..1994A&AS..106..275B}. Note that $Z_{\rm met}$ is distinct
		from the Galactic height coordinate $z$ (in kpc) used elsewhere in this
		work. For \textit{Gaia}-only sources lacking spectroscopic parameters, we
		instead follow \citet{CantatGaudin2024_data} and flag RC stars via
		\begin{equation}
			|M_G - \mathcal{P}(G - G_{\rm RP})| < 0.4,
			\label{eq:RC_Gaia}
		\end{equation}
		where $\mathcal{P}$ is a polynomial fit to the RC locus in the
		$(G-G_{\rm RP}, M_G)$ colour--magnitude diagram,
		\begin{equation}
			\mathcal{P}(x) = 0.2274x^4 + 0.2983x^3 + 1.1574x^2 + 1.0753x - 0.8317.
			\label{eq:RC_poly}
		\end{equation}
		We additionally exclude horizontal branch (HB) contamination by removing
		sources near the locus
		\begin{equation}
			(g\!-\!r)_0^{\rm HB} =
			0.087\,[\mathrm{Fe/H}]^2 + 0.39\,[\mathrm{Fe/H}] + 0.96,
			\label{eq:HB_cut}
		\end{equation}
		which defines the HB ridge line in the dereddened $(g-r)_0$
		vs.\ $[\mathrm{Fe/H}]$ plane.
		
		\medskip
		
		\noindent\textbf{\texttt{SEGUE\_KG\_SpecDist\_rv\_cut}} (No.~5b): We use
		the Bayesian spectrophotometric distance catalogue of 6,036 SEGUE K~giants
		\citep{Xue2014SEGUE}, calibrated on globular-cluster fiducials (absolute
		magnitude vs.\ colour). The distances achieve 16\% median precision and
		reach 125\,kpc, with 283 sources beyond 50\,kpc. We retain the 5,038
		sources with relative distance error below 20\%.
		
		\medskip
		
		\noindent\textbf{\texttt{LAMOST\_KG\_SpecDist\_rv\_cut}} (No.~6a): We use
		LAMOST Data Release~8 K~giants \citep{Zhang2023}, supplemented with
		Pan-STARRS1 photometry. We select giants via
		\begin{equation}
			4000 < T_{\rm eff} \leq 5600\,\mathrm{K}, \
			\log g <
			\begin{cases}
				3.5, & T_{\rm eff} < 4600\,\mathrm{K}, \\
				4.0, & 4600 \leq T_{\rm eff} \leq 5600\,\mathrm{K},
			\end{cases}
		\end{equation}
		following \citet{Liu2014}, with $E(B\!-\!V) < 0.25$ (colour excess due
		to interstellar dust reddening) to limit extinction systematics. Bayesian distances \citep{Xue2014} achieve 11\% precision,
		and the 19,543 K~giants span 4 to 126\,kpc. We exclude red clump stars
		using Eqns. \ref{eq:RC_logg}--\ref{eq:HB_cut}.
		
		\medskip
		
		\noindent\textbf{\texttt{LAMOST\_VMP\_SpecDist\_rv\_cut}} (No.~6b): We use
		LAMOST Data Release~9 \citep{LAMOSTVMP}, which provides 111,000 metal-poor
		stars: 32,631 very metal-poor ($[\mathrm{Fe/H}] < -2.0$), 702 extremely
		metal-poor ($[\mathrm{Fe/H}] < -3.0$), and 30 ultra metal-poor
		($[\mathrm{Fe/H}] < -4.0$). We retain sources satisfying
		$[\mathrm{Fe/H}] < +0.39$ and $0.5 \leq (g\!-\!r)_0 \leq 1.4$
		\citep{Zhang2023}, yielding 9,207 sources.

		\subsection{Radial Velocity Catalogues}\label{sec:rv_cats}

		\noindent\textbf{\texttt{GAIA\_rv\_BPRP\_cut}} (No.~1b): We use radial velocities derived from \textit{Gaia} BP/RP spectra
		\citep{Verberne2024}. Of the $\approx219$~million XP spectra analysed,
		radial-velocity estimates were obtained for $\approx125$~million sources. We reject sources
		with
		\begin{equation}
			\left|\frac{\Delta v_r}{\sigma_{v_r}}\right| > 3,
			\label{eq:rv_quality}
		\end{equation}
		where $\Delta v_r$ is the difference between the BP/RP-derived velocity and
		an independent reference measurement, and $\sigma_{v_r}$ is the combined
		uncertainty; this standard $3\sigma$ outlier cut removes sources with
		unreliable velocities due to template mismatches or contaminated spectra,
		while retaining $>$99.7\% of well-measured sources assuming Gaussian
		residuals. We additionally require
		\begin{align}
			E(B\!-\!V) &< 0.5, \\
			\texttt{rv\_err} &< 300\,\mathrm{km\,s^{-1}}, \\
			\texttt{CMD\_outlier\_fraction} &< 0.1, \\
			\texttt{bad\_measurement} &< 0.1,
		\end{align}
		where \texttt{CMD\_outlier\_fraction} is the fraction of Monte Carlo
		realisations placing the source outside the expected colour--magnitude
		locus, and \texttt{bad\_measurement} is the fraction of individual epoch
		measurements flagged as unreliable. These cuts yield 6,367,355 sources,
		providing $\approx$1.5~million new velocities in the faint regime ($G > 16$)
		lacking Radial Velocity Spectrometer (RVS) coverage.
		
		\medskip
		
		\noindent\textbf{\texttt{GAIA\_rv\_RVS\_cut}} (No.~1c): We use the
		\textit{Gaia} RVS catalogue \citep{Katz2023} of $\approx$33~million
		sources, which provides the most precise \textit{Gaia} radial velocities
		($\sigma_{v_r} \approx 1$--$10\,\mathrm{km\,s^{-1}}$) for bright stars
		($G \lesssim 16$). We filter to
		\begin{align}
			\varpi/\sigma_\varpi &> 2.3, \\
			\texttt{RUWE} &< 1.4, \\
			\sigma_\mu &< 1.0\,\mathrm{mas\,yr^{-1}},
		\end{align}
		where $\sigma_\mu$ denotes the proper-motion uncertainty, retaining
		2,314,443 sources for validation and supplementary radial velocities.
		
		\medskip
		
		\noindent\textbf{\texttt{SOS\_I\_rv\_cut}} (SoS-I): We use the Survey of
		Surveys \citep[SoS-I;][]{Tsantaki2022SurveyI}, which homogenises radial
		velocities from APOGEE, GALAH, Gaia-ESO, RAVE, and LAMOST onto the
		\textit{Gaia} reference frame ($\approx$11~million sources). We apply the
		photometric and astrometric quality filters of
		\citet{Evans2018, Arenou2018}:
		\begin{align}
			1.0 + 0.015(G_{\rm BP} - G_{\rm RP})^2
			&< E <
			1.3 + 0.06(G_{\rm BP} - G_{\rm RP})^2,
			\label{eq:sosi_E}\\
			u &< 1.2\max\!\bigl(1,\, e^{-0.2(G-19.5)}\bigr),
			\label{eq:sosi_u}
		\end{align}
		where $E \equiv \texttt{phot\_bp\_rp\_excess\_factor}$ is the
		$BP{+}RP$-to-$G$ flux ratio (defined in
		\texttt{GAIA\_PlxDist\_RGBSubset\_cut} above), and
		$u \equiv \sqrt{\chi^2/\nu}$ is the astrometric unit-weight error, with
		$\chi^2=\texttt{astrometric\_chi2\_al}$ and
		$\nu=\texttt{astrometric\_n\_good\_obs\_al}-5$ the corresponding degrees of freedom. The lower bound on $E$
		rejects unphysical negative-excess artefacts; the upper bound removes
		blended sources. Large $u$ signals binarity or non-stellar morphology; the
		exponential factor relaxes the threshold smoothly for faint stars where
		$\chi^2/\nu$ is noisier. We additionally require
		$(\texttt{quality\_flags}~\&~32) = 0$, which rejects sources for which
		bit~5 of the SoS-I bitmask is set, indicating a failed RV
		cross-calibration. Approximately 1~million sources are retained, of which
		1,595 lie at $R > 30$\,kpc.
		
		\medskip
		
		\noindent\textbf{\texttt{DESI\_vac\_rv\_cut}} (No.~4b): We use the DESI
		Milky Way Survey Value-Added Catalogue \citep{Koposov2024}, which provides
		radial velocities ($\sigma_{v_r} \sim 1\,\mathrm{km\,s^{-1}}$),
		$T_{\rm eff}$ (200\,K), $\log g$ (0.3\,dex), $[\mathrm{Fe/H}]$ (0.15\,dex),
		and $[\alpha/\mathrm{Fe}]$ (0.1\,dex) from DESI spectra without photometric
		or \textit{Gaia} priors. From 625,688 sources, we apply general quality
		requirements:
		\begin{align}
			&16 < r_{\rm mag} < 19,\quad r_{\rm obs} < 20,\quad
			\texttt{type} = \texttt{PSF},
			\label{eq:desi_mag}\\
			&\epsilon_a < 3,\quad
			\texttt{duplicated\_source} = \texttt{False},
			\label{eq:desi_astrom}\\
			&\texttt{brick\_primary} = \texttt{True},\quad
			\texttt{nobs}_{g,r} > 0,
			\label{eq:desi_obs}\\
			&\texttt{flux}_{g,r} > 0,\quad
			\texttt{fracmasked}_{g,r} < 0.5,
			\label{eq:desi_phot}
		\end{align}
		where $r_{\rm mag}$ is the Legacy Survey $r$-band magnitude (we use
		$r_{\rm mag}$ here to distinguish it from the Galactocentric radius $r_{\rm gc}$
		used in Section~\ref{sec:bhb}), $r_{\rm obs}$ is the observed depth,
		$\texttt{type} = \texttt{PSF}$ selects unresolved point sources (rejecting
		galaxies), $\epsilon_a$ is the astrometric excess noise (defined in
		\texttt{GAIA\_PlxDist\_RGBSubset\_cut} above), $\texttt{duplicated\_source}$
		and $\texttt{brick\_primary}$ together ensure a unique primary detection
		per sky position, $\texttt{nobs}_{g,r} > 0$ ensures at least one
		observation in each band, and $\texttt{fracmasked}_{g,r} < 0.5$ ensures
		fewer than half the pixels are masked. For the \texttt{MWS\_MAIN\_RED}
		giant subsample, we additionally require $g - r \geq 0.7$ (selecting cool
		red giants), $\texttt{astrometric\_params\_solved} = 31$ (all five
		astrometric parameters solved: two position, two proper motion, and
		parallax), and
		\begin{align}
			\varpi &< \max(3\sigma_{\varpi},\, 1\,\mathrm{mas}),
			\label{eq:desi_kin}\\
			|\mu| &< 7\,\mathrm{mas\,yr^{-1}},
		\end{align}
		where $|\mu| \equiv \sqrt{\mu_{\alpha^*}^2 + \mu_\delta^2}$ is the total
		proper motion, the parallax constraint selects distant halo giants (parallax
		consistent with zero at $3\sigma$, or $d > 1$\,kpc), and the proper-motion
		cut removes nearby fast-moving disc stars.
		
		\subsection{Blue Horizontal Branch Catalogues}\label{sec:bhb}

		\noindent\textbf{\texttt{BHB\_LSDR9\_PhotDist\_rv\_cut}} (No.~7): We
		select BHB stars, which occupy $-0.3 < (g\!-\!r) < 0$
		\citep{Deason2011, Belokurov2016Streams}, from the Legacy Survey DR9.
		Following \citet{Li2019S5}, we use the combined colour index
		\begin{equation}
			\begin{aligned}
				grz =\ & 1.07163(g\!-\!r)^5 - 1.42272(g\!-\!r)^4
				+ 0.69476(g\!-\!r)^3 \\
				&- 0.12911(g\!-\!r)^2
				+ 0.66993(g\!-\!r) - 0.11368 - (r\!-\!z),
			\end{aligned}
			\label{eq:grz}
		\end{equation}
		where $grz$ combines $g$, $r$, and $z$ photometry to separate BHB stars
		from the overlapping blue straggler sequence; its density is modelled as a
		two-component Gaussian mixture. We retain sources with $-0.14 < grz < 0.07$,
		$-0.30 < (g\!-\!r) < -0.05$, and BHB classification probability above
		70\% \citep{Amarante2024_AnisotropicHalo}. We assign distances using the
		colour-dependent absolute magnitude relation of \citet{Belokurov2016Streams},
		\begin{equation}
			\begin{aligned}
				M_{g}^{\rm BHB} = 0.398 &- 0.392(g\!-\!r)_0 + 2.729(g\!-\!r)_0^2 \\
				&+ 29.113(g\!-\!r)_0^3 + 113.57(g\!-\!r)_0^4,
			\end{aligned}
			\label{eq:bhb_Mg}
		\end{equation}
		which exploits the narrow intrinsic luminosity range of BHB stars and its
		residual colour dependence. Cross-matching with DESI by source identifier
		yields 969 sources with radial velocities.
		
		\medskip
		
		\noindent\textbf{\texttt{BHB\_SDSS\_SpecDist\_rv\_cut}} (No.~5c): We use
		12,530 BHB candidates from SDSS Data Release~8
		\citep{Xue2011KinematicSubstructure}, selected via $0.8 < (u\!-\!g) < 1.5$
		and $-0.5 < (g\!-\!r) < 0.0$, which isolates the BHB locus in the
		$(u-g, g-r)$ two-colour diagram. \citet{Xue2011KinematicSubstructure} fit
		a S\'{e}rsic profile to the Balmer lines H$\delta$ and H$\gamma$ in each
		spectrum,
		\begin{equation}
			y(\lambda) = 1 - a\exp\!\left[-\!\left(
			\frac{|\lambda - \lambda_0|}{b}\right)^{\!c}\right],
			\label{eq:sersic}
		\end{equation}
		where $\lambda_0$ is the line centre, $a$ is the line depth, $b$ is the
		line width, and $c$ controls the line shape; the Balmer line widths provide
		a luminosity diagnostic separating BHB stars from blue-straggler and
		main-sequence contaminants. We retain sources with fractional fit variance
		$\leq 0.10$, yielding 4,981 high-probability BHB stars (4,625 at
		$|z| > 4\,\mathrm{kpc}$, where $z$ is the Galactic height). We compute
		heliocentric distances from the theoretical absolute magnitude range
		$M_g \sim 0.60$--0.80 as
		\begin{equation}
			d_{\rm helio} = 10^{0.2(g - M_g + 5)}\,\mathrm{pc}.
		\end{equation}
		Distance precision is $\approx$10\%
		(0.2\,mag scatter in $M_g$) and velocity errors are
		5--20\,km\,s$^{-1}$ \citep{Xue2008}, with BHB contamination below 10\%
		\citep{Xue2008, Xue2011KinematicSubstructure}.
		
		\subsection{Variable Star Catalogues}\label{sec:variable}
		
		\begin{table*}[htbp]
			\centering
			\caption{Large-scale photometric surveys of RR~Lyrae used in this work.}
			\label{tab:rrl_surveys}
			\renewcommand{\arraystretch}{0.9}
			\small
			\begin{ruledtabular}
				\begin{tabular}{lcrl}
					Survey & $V$ range & Sources & Reference \\
					\hline
					Catalina   & 12--20     & 23,306 & \cite{Drake2013a}, \cite{Drake2014Catalina} \\
					QUEST      & 13.5--19.7 &  1,857 & \cite{Vivas2004}, \cite{Mateu2012, Zinn2014} \\
					NSVS       & $<$14      &  1,304 & \cite{Kinemuchi2006}, \cite{Hoffman2009} \\
					LINEAR     & 14--17     &  5,684 & \cite{Sesar2013} \\
					LONEOS     & $<$18      &    838 & \cite{Miceli2008} \\
					SDSS Str82 & 15--21     &    601 & \cite{Watkins2009}, \cite{Sesar2010, Suveges2012} \\
					GCVS       & \ldots     &  7,954 & \cite{Samus2009} \\
				\end{tabular}
			\end{ruledtabular}
		\end{table*}
		
		\noindent\textbf{\texttt{RRL\_SpecDist\_rv\_cut}} (No.~10): We use 5,290
		RR~Lyrae from \citet{Liu2020}, of which 3,642 have systemic radial
		velocities from LAMOST LEGUE and SDSS SEGUE. We obtain distances using
		period--luminosity--metallicity (PLZ) standard-candle relations of the form
		\begin{equation}
			M_V = a\,[\mathrm{Fe/H}] + b, \qquad
			M_{K_s/W1} = c\log P + d\,[\mathrm{Fe/H}] + e,
			\label{eq:PMZ}
		\end{equation}
		where $M_V$, $M_{K_s}$, and $M_{W1}$ are the absolute magnitudes in the $V$, $K_s$ (2MASS near-infrared), and WISE $W1$ ($3.4\,\mu$m) bands respectively, $P$ is the pulsation period, and $a$--$e$ are empirically calibrated coefficients; the near-infrared relations exploit the tighter period--luminosity correlation at longer wavelengths. We adopt the
		mid-infrared calibration of \citet{Klein2014},
		\begin{equation}
			M_{W1} = -1.64\log(P/0.32) - 0.231.
			\label{eq:MW1_rrl}
		\end{equation}
		where $M_{W1}$ is the absolute $W1$ magnitude derived from the
		period--luminosity relation.
		We derive systemic velocities from phased single-epoch spectra via the
		radial velocity template
		\begin{equation}
			\mathrm{RV}_{\rm obs}(\Phi) = A_{\rm rv}\,T(\Phi) + \mathrm{RV}_{\rm sys},
			\label{eq:RV_template}
		\end{equation}
		where $\Phi$ is the pulsation phase, $T(\Phi)$ is a standardised velocity-curve template, $\mathrm{RV}_{\rm sys}$ is the systemic radial velocity to be recovered, and $A_{\rm rv}$ is the velocity amplitude
		($A_{\rm rv} = 111.9,\, 90.9,\, 82.1$\,km\,s$^{-1}$ for H$\alpha$,
		H$\beta$, H$\gamma$ respectively; \citealt{Sesar2012VelocityCurves}).
		This yields 3,345 sources with infrared distances showing 1\% mean
		relative difference and 5\% dispersion.
		
		\medskip
		
		\noindent\textbf{\texttt{RRL\_PTR\_SpecDist\_rv\_cut}} (No.~3): We use 446 RRab stars from \citet{Cohen2017} at $d_{\rm helio} \geq
		50$\,kpc with classification probability above 0.70, excluding quasars
		and sources within $|b_{\rm Sgr}| < 9^\circ$, where $b_{\rm Sgr}$ is
		the latitude measured from the Sagittarius orbital plane. We derive
		distances from extinction-corrected mean $R$-band magnitudes
		($M_R \approx +0.6$\,mag, where $M_R$ is the absolute $R$-band magnitude, with a period-dependent correction
		$\Delta M \simeq -1.39\log P$; $\approx$5\% precision) and systemic
		velocities via the H$\alpha$ template of \citet{Sesar2012VelocityCurves},
		with velocity errors of 17--20\,km\,s$^{-1}$. We remove outliers by
		requiring
		\begin{equation}
			|v_r| <
			\begin{cases}
				200\,\mathrm{km\,s^{-1}}, & d_{\rm helio} < 85\,\mathrm{kpc}, \\
				170\,\mathrm{km\,s^{-1}}, & d_{\rm helio} \geq 85\,\mathrm{kpc},
			\end{cases}
		\end{equation}
		retaining 116 RRab stars.
		
		\medskip
		
		\noindent\textbf{\texttt{NG\_Virgo\_SpecDist\_rv\_cut}} (No.~9): We use
		180 high-confidence RR~Lyrae (139 RRab, 41 RRc) from NGVS, identified via template light-curve fitting, calibrated
		on 483 SDSS Stripe~82 sources \citet{Feng2024}. We obtain distances using the PS1-calibrated $i'$-band period--luminosity relation,
		\begin{equation}
			M_{i'} = -1.77\log_{10}(P/0.6) + 0.46,
			\label{eq:NGVS_PL}
		\end{equation}
		where $M_{i'}$ is the absolute magnitude in the PS1 $i'$ band and $P$
		is the pulsation period in days.
		with RRc periods fundamentalised as $\log_{10}P_F = \log_{10}P_{\rm RRc}
		+ 0.128$ to place them on the RRab sequence. The distance precision is found to be 5\%, and after excluding AGN and
		requiring valid radial velocities, we retain 154 sources.
		
		\medskip
		
		\noindent\textbf{LAMOST variable stars:} We cross-match the 80,702
		radial velocity variables of \citet{Tian2020} (probability above 60\%)
		with the 631,769 candidates of \citet{Xu2022} (probability above 95\%,
		distances from PLZ relations), yielding 3,138 sources with both distance
		and velocity measurements.
		
		\medskip
		
		\noindent\textbf{\texttt{CAT\_Variables\_PhotDist\_rv\_cut}} (No.~8b): We
		obtain distances using $\sim$61,000 periodic variables from
		\citet{Drake2014Catalina} using class-specific period--luminosity relations
		for Classical Cepheids, Type~II Cepheids, RR~Lyrae, Miras, $\delta$~Scuti,
		and SX~Phe stars (calibrations summarised in
		Appendix Table~\ref{tab:pl_relations_variables}); overtone pulsators are
		fundamentalised to the fundamental mode period where required. We draw
		radial velocities from cross-matches with \texttt{SOS\_I\_rv\_cut} and
		\texttt{GAIA\_rv\_BPRP\_cut}, yielding 16,451 sources.
		
		\medskip
		
		\noindent\textbf{\texttt{AGB}} (No.~8a): We use 417 non-carbon Mira and
		SRa variables from Catalina and LINEAR surveys \citep{Mauron2019OxygenRichGiants}.
		We retain sources satisfying
		\begin{align}
			E(B\!-\!V) &< 0.3, \\
			10.5 &< (V_{\rm CSS})_0 < 17, \\
			2.6 &< (V\!-\!K_s)_0 < 5.6, \\
			0.75 &< (J\!-\!K_s) < 1.4,
		\end{align}
		where $(V_{\rm CSS})_0$ is
		the extinction-corrected Catalina Sky Survey $V$-band magnitude, and the
		near-infrared colour cuts $(V-K_s)_0$ and $(J-K_s)$ select oxygen-rich AGB
		stars while excluding carbon stars and foreground dwarfs. Of 330 sources
		passing these cuts, 90 have radial velocities
		\citep{Feast2000, Smak1965, Luo2016, West2011} with errors of
		5--15\,km\,s$^{-1}$.

		\subsection{Catalogue Assembly and Distance Validation}\label{sec:assembly}
		
		Combining all distance and radial
		velocity sub-catalogues described in Sections~\ref{sec:dist_cats} and
		\ref{sec:rv_cats}, the raw assembled catalogue contains $N =
		52{,}112{,}379$ entries across 213 columns. These columns include the full 6D phase-space information for each Galactic tracer (comprising three position coordinates: right ascension $\alpha$,
		declination $\delta$, and heliocentric distance $d_{\rm helio}$, and
		three velocity coordinates: proper motions in right ascension and
		declination, $\mu_{\alpha^*} \equiv \dot{\alpha}\cos\delta$ and
		$\mu_\delta$, and line-of-sight radial velocity $v_r$, along with
		their associated uncertainties). These six quantities jointly determine
		the full phase-space position of each star. Detailed descriptions of all 213
		columns are provided in Fig.~\ref{fig:column_header_appendix} in the appendix.
		
		The catalogue assembly retains sources with distance and radial velocity,
		cross-matched by ($\alpha$, $\delta$, $\mu_{\alpha^*}$, $\mu_\delta$, $\varpi$) to flag duplicates, and applies the
		selections described in the previous sections. Bright disc tracers ($G < 17$, $d_{\rm helio} < 10$\,kpc) are dominated
		by the \texttt{GAIA\_PlxDist\_cut} ($\sim 90M$). The G broad-band magnitude in range $\sim (13, 16)$ provides the lever arm for inter-survey calibration
		(Section~\ref{sec:pipeline}). The deduplication pipeline consolidates these
		into $N_\star = 32{,}552{,}876$ unique stars as described in Section~\ref{sec:pipeline}.
		
		Figure~\ref{fig:dist_combined} shows how fractional distance uncertainty changes with heliocentric distance. At small distances, high-significance \textit{Gaia} parallaxes remain competitive, but their fractional uncertainty grows rapidly beyond roughly $d_{\rm helio} \sim 5$--$7$\,kpc. The spectrophotometric distances and specialised photometric channels therefore become increasingly important at larger distances. For kinematic work, we recommend $\texttt{DISTERR}/\texttt{DIST}<0.3$. Figure~\ref{fig:spectophoto} compares these estimates with the simple parallax inversion $1/\varpi$ for 3.7 million overlap sources. This comparison is informative only where the parallax has adequate signal-to-noise; for the better-constrained spectrophotometric samples the scatter around the one-to-one relation is typically $15$--$35\%$ at $d<20$\,kpc, while the LAMOST and variable-star samples are broader. Beyond $\sim30$\,kpc many \textit{Gaia} parallaxes are noise dominated, hence $1/\varpi$ is no longer a usable external distance reference, and large offsets should not be interpreted as failures of the spectrophotometric scale. Horizontal structures in some photometric samples arise from the discrete or fixed absolute-magnitude relations used for these tracers.
		
		\subsection{Spectrophotometric Distance Estimation}\label{sec:specphot}
		
		\subsubsection{The need for spectrophotometric distances}
		
		\textit{Gaia} provides direct parallaxes for a large fraction of the stars in
		our catalogue, but its astrometric precision degrades rapidly at faint
		magnitudes and large distances. For stars in the outer disc and halo, the
		fractional parallax uncertainty $\sigma_{\varpi}/\varpi$ is often comparable
		to unity or larger, which means that simply inverting the parallax to obtain
		a distance produces an extremely uncertain and biased estimate. To probe distances 
		in this faint regime, we use an {\it ansatz} that predicts  a star's \emph{would be} parallax
		from its observed atmospheric parameters and broad-band
		magnitudes. The resulting prediction, the \emph{spectrophotometric parallax}
		$\varpi_{\rm phot}$, is then inverted to give a heliocentric distance
		\emph{spectrophotometric distance}, $d_{\rm phot}$.
		
		The underlying physical assumption is that a star's intrinsic luminosity is
		encoded in its atmospheric parameters (effective temperature, surface
		gravity, metallicity) and its colours. If we know the intrinsic luminosity
		and we measure the apparent brightness, we know the distance. The ansatz we fit below is, in effect, a high-dimensional generalisation of
		the classical bolometric or absolute-magnitude relation, learned directly
		from data rather than imposed from stellar models. The methodology is the
		multi-band extension of the data-driven approach introduced by
		\citet{Eilers2019}, applied here to a substantially higher-dimensional
		feature set.
		
		The key idea here is to use \textit{Gaia} stars with \emph{precise} parallaxes (($\sigma_{\varpi}/\varpi < 0.1$)) as
		the calibration anchor; these are used to learn the relation between the
		observables and the parallax, and we then apply the knowledge gained to every
		remaining star including those whose own \textit{Gaia} parallax is too
		noisy to be used directly in constructing the tracer phase-space.
		
		\subsubsection{The empirical model}
		
		For each star $i$, we collect a vector of observables $\mathbf{A}_i \in
		\mathbb{R}^{19}$ (described in detail below). We model the spectrophotometric
		parallax as
		\begin{equation}
			\varpi_{\mathrm{phot},i}
			\;=\;
			\exp\!\bigl(\mathbf{A}_i \cdot \boldsymbol{\theta}\bigr),
			\label{eq:specphot_model}
		\end{equation}
		where $\boldsymbol{\theta} \in \mathbb{R}^{19}$ is the vector of model
		coefficients to be determined from the calibration data, and the dot product
		$\mathbf{A}_i \cdot \boldsymbol{\theta} = \sum_{k} A_{ik}\theta_k$ is the
		ordinary scalar product of the two vectors.
		
		The exponential function on the right-hand side plays two roles: (i) it
		guarantees $\varpi_{\rm phot} > 0$ by construction\footnote{Parallax is a
			strictly positive quantity, and we want our model to respect that without
			needing to impose constraints during the fit.}, and  (ii) taking the logarithm of
		equation~(\ref{eq:specphot_model}) gives a linear relation between
		$\ln\varpi_{\rm phot}$ and the observables, which is the natural form when
		one uses observables including apparent magnitudes (themselves logarithmic in
		flux).
		
		\subsubsection{The 19-dimensional feature vector}
		
		The feature vector $\mathbf{A}_i$ for star $i$ is built from three groups of
		observables, plus an intercept:
		
		\begin{enumerate}[label=(\roman*)]
			\item \textbf{Atmospheric parameters} (3 features): the effective
			temperature $T_{\mathrm{eff}}$, the surface gravity $\log g$, and the
			metallicity $[\mathrm{Fe}/\mathrm{H}]$. Each of these is
			\emph{standardised} across the training sample, i.e.\ the sample mean
			is subtracted and the result is divided by the sample standard
			deviation, so that each parameter contributes on a comparable numerical
			scale during the fit.
			
			\item \textbf{Apparent magnitudes} (8 features): these are the optical
			\textit{Gaia} bands $G$, $BP$, and $RP$, the near-infrared 2MASS bands
			$J$, $H$, and $K_s$, and the mid-infrared WISE bands $W1$ and $W2$.
			Apparent magnitudes carry the distance information, since for a given
			intrinsic luminosity a more distant star will be fainter.
			
			\item \textbf{Colour indices} (7 features): the differences
			$BP - RP$, $G - RP$, $J - K_s$, $H - K_s$, $J - H$, $G - J$, and
			$G - K_s$. Colours are distance-independent quantities,
			so they primarily encode the star's intrinsic spectral energy
			distribution and therefore its luminosity class.
			
			\item An \textbf{intercept} term, which absorbs any overall
			normalisation.
		\end{enumerate}
		
		Counting these gives $3 + 8 + 7 + 1 = 19$ features per star, hence the
		dimensionality of $\mathbf{A}_i$ and of the coefficient vector
		$\boldsymbol{\theta}$. The full design matrix $\mathbf{A}$ is built by
		stacking the row vectors $\mathbf{A}_i$ for all usable stars, so that its
		$(i,k)$th element $A_{ik}$ is the value of the $k$th feature for the $i$th
		star.
		
		\subsubsection{Required versus optional inputs}
		
		We treat the input data carefully so as to keep as many stars as possible.
		A star is \emph{excluded entirely} from the pipeline only when one, or more, of its
		{\it strictly required} quantities are missing or unreliable, namely:
		\begin{itemize}
			\item the \textit{Gaia} parallax $\varpi$ and its uncertainty
			$\sigma_{\varpi}$;
			\item the \textit{Gaia} optical magnitudes $G$, $BP$, $RP$;
			\item the spectroscopic atmospheric parameters $T_{\rm eff}$, $\log g$,
			and $[\mathrm{Fe}/\mathrm{H}]$.
		\end{itemize}
		The infrared bands ($J$, $H$, $K_s$, $W1$, $W2$) are treated as
		\emph{optional}. If one or more of these is missing or flagged as
		unreliable, the star is still kept in the catalogue. The missing magnitude
		is set to zero, and any colour index whose computation depends on the
		missing band is also set to zero. Setting the corresponding feature to zero
		makes that feature contribute nothing to
		$\mathbf{A}_i \cdot \boldsymbol{\theta}$ for that particular star, so the
		prediction is effectively driven by the features that \emph{are} available
		for it. This way, faint or partially observed stars still receive a
		prediction, even if it relies on a smaller subset of the features.
		
		\subsubsection{Training: objective function}
		
		The model coefficients $\boldsymbol{\theta}$ are obtained by minimising the
		penalised objective
		\begin{equation}
			H(\boldsymbol{\theta})
			\;=\;
			\underbrace{\frac{1}{2}\sum_{i \in \mathcal{T}}
				\frac{\bigl(\varpi_i - \varpi_{\mathrm{phot},i}\bigr)^{2}}
				{\sigma_{\varpi,i}^{2}}}_{\text{data term}}
			\;+\;
			\underbrace{\sum_{j=1}^{18}\lambda_j\,|\theta_j|}_{\text{penalty term}},
			\label{eq:objective}
		\end{equation}
		where $\mathcal{T}$ is the training set (defined in
		Section~\ref{sec:specphot_quality_cuts} below), $\varpi_i$ is the observed
		\textit{Gaia} parallax, and $\sigma_{\varpi,i}$ is its uncertainty.
		
		The two terms in Equation~\ref{eq:objective} play distinct roles. The first term is a chi-squared residual and quantifies the fit of the model-predicted parallax to the observed \textit{Gaia}
		parallax for each training star. Minimising this term alone is akin to standard weighted nonlinear least squares. The second term adds the
		sum of the absolute values of the coefficients, each multiplied by a
		strength $\lambda_j \geq 0$. This is the so-called \emph{$L_1$ penalty}, used in the well-known LASSO method
		\citep{Tibshirani1996}. The role of this term is to discourage a model
		from fitting large coefficients to features that do not help the fit\footnote{The absolute-value penalty has a sharp ``corner'' at zero, the
			optimum will tend to set weakly informative coefficients exactly to zero.}.
		This produces a \emph{sparse} solution, a small number of features carry
		real predictive power, while the rest are switched off. Sparsity reduces the risk
		of overfitting and the
		intercept $\theta_0$ is exempted from the penalty (we set $\lambda_0 = 0$),
		since penalising the intercept would distort the overall normalisation
		without serving any regularisation purpose.
		
		\subsubsection{Training: gradient and minimisation}
		
		To minimise $H$ efficiently, we provide the optimiser with the analytically
		computed gradient of the smooth term and an $L_1$ subgradient for the penalty. For
		$k = 1, \dots, 18$,
		\begin{equation}
			\frac{\partial H}{\partial \theta_k}
			\;=\;
			-\sum_{i \in \mathcal{T}}
			\frac{\varpi_i - \varpi_{\mathrm{phot},i}}{\sigma_{\varpi,i}^{2}}
			\,\varpi_{\mathrm{phot},i}\,A_{ik}
			\;+\; \lambda_k\,\operatorname{sgn}(\theta_k),
			\label{eq:gradient}
		\end{equation}
		where the factor of $\varpi_{\rm phot,i}$ in the first sum comes from
		differentiating the exponential link in
		equation~(\ref{eq:specphot_model}), and $\operatorname{sgn}(\theta_k)$ denotes
		the usual $L_1$ subgradient away from zero; at $\theta_k=0$, where the absolute-value
		penalty is not differentiable, we use the admissible subgradient value 0. For the intercept $\theta_0$,
		which is unpenalised, only the data term, i.e. the first weighted-residual term in Eq.~(\ref{eq:gradient})\footnote{Note, the second term is the regularisation penalty.} is present.
		
		Supplying the analytic gradient/subgradient, rather than relying on finite-difference
		approximations, is important here for two reasons: it is numerically more
		stable in the high-dimensional regime, and it is much faster to evaluate
		on the large training samples used in this work.
		
		\subsubsection{Quality cuts on the training set}\label{sec:specphot_quality_cuts}
		
		The model is only as good as the parallaxes it is calibrated against, so we
		restrict the training set $\mathcal{T}$ to stars whose \textit{Gaia}
		parallaxes are demonstrably trustworthy. A star is admitted to $\mathcal{T}$
		only if all four of the following criteria are satisfied:
		
		\begin{enumerate}[label=(\roman*)]
			\item \textbf{Small absolute parallax uncertainty:}
			$\sigma_{\varpi} < 0.1\,\mathrm{mas}$.
			
			\item \textbf{High parallax signal-to-noise ratio:}
			$\varpi/\sigma_{\varpi} > 20$. Together with criterion (i), this
			selects stars whose \textit{Gaia} parallax is determined to better than
			$5\%$.
			
			\item \textbf{Sufficient \textit{Gaia} visibility periods:}
			$N_{\rm per} \geq 8$. The number of visibility periods is the number of
			distinct epochs at which \textit{Gaia} observed the star; a higher
			number means the astrometric solution is constrained by more independent
			observations and is therefore more reliable.
			
			\item \textbf{A clean astrometric solution:} either the empirical
			astrometric fit-quality statistic $\chi^{2}_{\mathrm{AL}}/\sqrt{N_{\mathrm{AL}}-5} \leq 35$, or, when
			that quantity is unavailable, the renormalised unit-weight error $\mathrm{RUWE} < 1.4$. This criterion is adopted from the spectrophotometric-parallax training selection of \citet{Hogg2019}. The subtraction of five reflects the five fitted astrometric parameters in the \textit{Gaia} solution used by that selection, while 35 is the empirical screening threshold used in that work rather than a universal chi-square critical value. RUWE is a single-number summary of how well the \textit{Gaia} single-star astrometric model fits the data; values much
			above $1.4$ flag binarity, blending, or other model misfits, and are
			routinely excluded in \textit{Gaia} science analyses.
		\end{enumerate}
		
		It is essential that \emph{these cuts apply only to the training set
			$\mathcal{T}$, and not to the catalogue at large.} The pipeline still
		\emph{predicts} $\varpi_{\rm phot}$ for stars that fail the cuts; it only
		refuses to use them as training anchors. This is exactly the regime in
		which spectrophotometric distances are most valuable: stars with poor
		\textit{Gaia} parallaxes are the ones for which we aim for an
		independent, photometry-driven distance estimate.
		
		If, in a given cross-validation fold (one of the $K$ disjoint subsets), the strict training cuts yield too few stars for a
		numerically stable fit, the parallax signal-to-noise threshold (criterion
		(ii)) is relaxed in a second optimisation step. This means that the new relaxed fit is initialised with the coefficient vector obtained from the strict-cut optimisation so the second step changes only the training-set acceptance threshold rather than restarting the optimisation from an unrelated point. This ensures
		that the calibration always converges, while preferring the stricter
		solution whenever the data allows.
		
		\subsubsection{$K$-fold cross-validation}
		
		A naive approach would be to fit the coefficients on the full training set
		and then use the same coefficients to predict $\varpi_{\rm phot}$ for every
		star. This would, however, give optimistically biased predictions for stars
		that were used during training, because the model has effectively seen
		them. The standard remedy is \emph{$K$-fold cross-validation}, described below.
		
		The usable stars are partitioned at random into $K$ disjoint subsets, called
		\emph{folds}, of approximately equal size. For each of the $K$ folds in
		turn, the model is trained using only the other $K-1$ folds, and is then
		applied to predict $\varpi_{\rm phot}$ for the stars in the held-out fold.
		Crucially, every star is held out exactly once, so every star ultimately
		receives a prediction made by a model that was \emph{not} trained on that
		particular star. The bias from learning and predicting on the same data is
		therefore eliminated by construction.
		
		In our pipeline, the within-fold quality cuts of
		Section~\ref{sec:specphot_quality_cuts} are applied independently inside
		each of the $K-1$ training folds at every iteration, so that the calibration
		sample is always composed of high-quality astrometric standards while
		predictions are produced for the entire usable catalogue.
		
		\subsubsection{From parallax to distance}
		
		Once $\varpi_{\rm phot,i}$ has been obtained for each star, the
		heliocentric distance follows from the standard inversion
		\begin{equation}
			d_{\mathrm{phot}}\,[\mathrm{kpc}]
			\;=\;
			\frac{1}{\varpi_{\mathrm{phot}}\,[\mathrm{mas}]},
			\label{eq:dist_conversion}
		\end{equation}
		equivalent to $d_{\mathrm{phot}}\,[\mathrm{pc}] =
		1000/\varpi_{\mathrm{phot}}\,[\mathrm{mas}]$. Since the model
		parametrisation in equation~(\ref{eq:specphot_model}) guarantees
		$\varpi_{\rm phot} > 0$, the inversion in
		equation~(\ref{eq:dist_conversion}) is always well-defined and produces a
		finite, positive distance for every star. This is a
		significant practical advantage over directly inverting noisy
		\textit{Gaia} parallaxes, where small or negative values, for large $\sigma_{\varpi}/\varpi$,  make the inversion ill-defined. Note, that a measured astrometric parallax can be negative because the \textit{Gaia} solution is a noisy estimator whose errors are approximately symmetric about the true value; the physical parallax itself is, of course, non-negative.
		
		\subsubsection{Validation}
		
		The fidelity of the resulting distances is assessed through a suite of
		diagnostic comparisons. We compare the predicted spectrophotometric
		parallaxes $\varpi_{\rm phot}$ against the observed \textit{Gaia}
		parallaxes $\varpi$ for the full usable sample, focusing in particular on
		the high-S/N regime where the \textit{Gaia} value is most trustworthy.
		Where available, we also compare the derived distances $d_{\rm phot}$
		against catalogue-based geometric distances (e.g.\ Bayesian distance
		posteriors using \textit{Gaia} parallaxes as input).

		\section{From Heterogeneous Surveys to a Unified Catalogue}\label{sec:pipeline}
		
		The final unified catalogue described in this paper is built by combining $20$ parent catalogues (see Table~\ref{tab:survey_summary_20}) into a single, internally consistent table
		of stars. We start with an initial assembled catalogue containing 
		$N = 52{,}112{,}379$ rows, on which the procedures listed below are applied, resulting in a final catalogue containing $N_{\star} = 32{,}552{,}876$ \emph{unique} stars.
		The reduction from $\sim 52$M rows to $\sim 32.5$M stars is the result of four logical
		distinct operations, each having its own purpose:
		
		\begin{enumerate}
			\item \textbf{Error normalisation} (Section~\ref{sec:error_norm}). The
			$\sigma_{\rm RV}$ reported by each survey pipeline typically omits calibration systematics and pipeline-level correlations, so that the quoted
			uncertainty underestimates or overestimates the true measurement scatter by a
			survey-dependent factor. Since the final catalogue uncertainty assumes $\sigma_{\rm RV}$ is the
			true $1\sigma$ scatter, we must first rescale each survey's
			uncertainties to match the empirical scatter measured from repeated
			observations.
			
			\item \textbf{Deduplication} (Section~\ref{sec:dedup}). The same physical
			star may appear many times in our raw input table, either observed
			repeatedly within a single survey or observed independently
			by several surveys. These repeated observations need to be identified and grouped so that the final catalogue contains only distinct, unique stars.
			
			\item \textbf{Zero-point calibration} (Section~\ref{sec:zp_cal}). Even when
			two surveys  have similarly small random uncertainties, their measured velocity scales may nevertheless be systematically offset; one survey's velocity scale can be offset
			from another's by a survey-dependent constant (and sometimes by smooth
			trends in stellar parameters). We remove these offsets by tying every
			survey to a common reference frame defined by \textit{Gaia}.
			
			\item \textbf{Combination} (Section~\ref{sec:combination}). Once the RV uncertainties have been normalised, and duplicate rows assigned to physical stars and survey-dependent RV zero points calibrated, we combine the multiple measurements of each star into
			a single weighted record in the final unified catalogue.
		\end{enumerate}
		
		A natural question is why a simple weighted average of all measurements would
		not, by itself, be sufficient in producing the final catalogue? The reason lies in the fact that the uncertainties
		quoted by survey pipelines are not directly comparable, and the values
		themselves carry survey-dependent systematic offsets. A naive weighted average
		would therefore (i) over-trust surveys with optimistic error bars and
		(ii) inherit any zero-point bias as a hidden shift in the combined value.
		Steps~1 and~3 fix these two problems separately. Step~1 recalibrates the
		errors, and step~3 recalibrates the values.  Only then
		is the final combination in step~4 statistically well-founded.
		
		For the radial-velocity (RV) analysis, we regroup the $20$ parent catalogues
		into $14$ \emph{surveys}. Four surveys are large and play a central
		role: \texttt{GAIA} (the Gaia RV channels), \texttt{DESI} (the DESI
		spectroscopic-distance and stellar value-added catalogue channels),
		\texttt{LAMOST} (the K-giant and very metal-poor channels), and \texttt{SDSS}
		(the BOSS-HALO, SEGUE K-giant, and SDSS-BHB channels). Four further auxiliary
		families  \texttt{APOGEE}, \texttt{GALAH}, \texttt{GES}, and \texttt{RAVE}
		are smaller but contribute to high-precision spectroscopy.
		
		\begin{figure*}[htbp]
			\centering
			\includegraphics[width=0.8\linewidth]{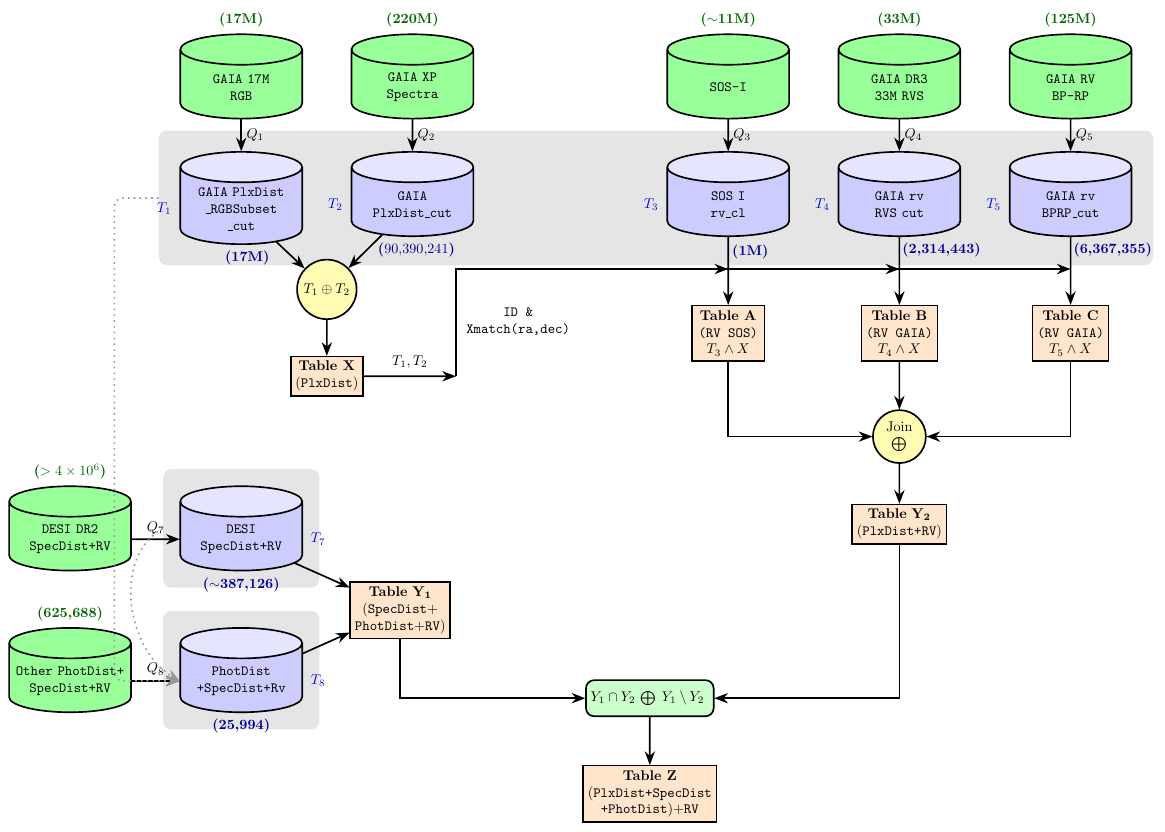}
			\caption{\small Database-merging procedure used to construct the final
				catalogue. The Gaia parallax-distance samples are first merged into a
				common \texttt{PlxDist} table and then cross-matched with the Gaia and
				SoS-I radial-velocity catalogues to form Table~$Y_2$ (\texttt{PlxDist+RV}).
				Separately, the DESI spectroscopic-distance and auxiliary
				photometric/spectroscopic-distance catalogues are combined into
				Table~$Y_1$ (\texttt{SpecDist+PhotDist+RV}). The final catalogue,
				Table~$Z$, is produced from the overlap $Y_1 \cap Y_2$ together with the
				unique component $Y_1 \setminus Y_2$. Here, $\oplus$ denotes the
				concatenation of the matched $Y_1 \cap Y_2$ records with the unmatched
				$Y_1 \setminus Y_2$ records.}
			\label{fig:catalogue_flowchart}
		\end{figure*}
		
		\subsection{Reading the catalogue-merging flowchart}
		\label{subsec:catalogue_flowchart_explained}
		
		The workflow in building the final homogeneous catalog is pictorially shown in Figure~\ref{fig:catalogue_flowchart}.
		Below, we briefly describe the different parts of this process:
		\begin{enumerate}
			\item We start by compressing the catalogue assembly into two
			parallel data branches. The first branch (top-left block)  begins with the \textit{Gaia}-based
			parallax-distance samples. 
			\item These are brought together in the intermediate
			\texttt{PlxDist} table and then cross-matched with the \textit{Gaia} and SoS-I
			radial-velocity catalogues (top-right block). 
			\item The resulting table, denoted by $Y_2$, therefore
			collects the parallax-distance and RV information available for the same
			source. 
			\item The second branch (bottom-left block)collects the DESI spectroscopic-distance catalogue
			and the auxiliary spectroscopic- and photometric-distance catalogues, together
			with their available RV measurements. These entries form the intermediate
			table $Y_1$. This second branch is especially important for stars for which a
			useful distance is supplied by a spectroscopic or photometric estimator rather
			than by direct parallax inversion.
			\item The two intermediate tables are then compared at the level of source identity.
			The intersection $Y_1\cap Y_2$ denotes stars represented in both branches, so
			that complementary measurements from the parallax-based and
			spectroscopic/photometric channels can be associated with the same physical
			star. The component $Y_1\setminus Y_2$ denotes stars present in $Y_1$ but with
			no corresponding entry in $Y_2$; retaining this component prevents the merge
			from discarding tracers that are supplied only by the spectroscopic or
			photometric-distance branch \footnote{The set notation in the flowchart therefore
				refers to matching and bookkeeping of stellar identities, rather than to any
				arithmetic operation on the measured quantities themselves.}.
		\end{enumerate}
		It is also important to distinguish this database-level schematic from the
		statistical calibration described in the following sections. A source appearing
		in more than one input catalogue is not simply averaged at the point where the
		flowchart branches meet. Its quoted RV uncertainty is first placed on the common
		empirical scale of Section~\ref{sec:error_norm}, repeated detections are grouped
		by the deduplication procedure of Section~\ref{sec:dedup}, and survey-dependent
		RV zero points are tied to the \textit{Gaia} reference frame in
		Section~\ref{sec:zp_cal}. Only after performing these steps, are the available measurements
		combined as described in Section~\ref{sec:combination}. Thus,
		Figure~\ref{fig:catalogue_flowchart} should be read as a map of how the input
		data streams are brought together, while the subsequent sections specify how those matched measurements are calibrated and reduced to the single record per
		physical star in the final catalogue. 
		The RV error-normalisation schematic shown in Figure~\ref{fig:rv_normalization_flowchart}, is introduced at the start of Section~\ref{sec:error_norm} where the calibration steps are described in detail.

		\textsc{Notation:} In the following sections, we use $\mathcal{S}$ for a survey family. For star $i$
		observed by survey $\mathcal{S}$, $v_{{\rm RV},\mathcal{S},i}$ denotes the
		reported radial velocity and $\sigma_{{\rm RV},\mathcal{S},i}$ its reported
		uncertainty. A primed symbol, such as $\sigma'$, denotes a normalised
		uncertainty (after Section~\ref{sec:error_norm}). A pairwise difference between
		two measurements of the same star is denoted by
		$\Delta v_{{\rm RV,pw}}$, and the additive zero-point correction by
		$\Delta v_{{\rm RV,ZP}}$.

		\section{Putting Radial-Velocity Uncertainties on a Common Scale}\label{sec:error_norm}
		
		This section addresses radial-velocity error normalisation  so that quoted RV uncertainties from different survey pipelines can be interpreted on a common empirical scale.
		
		Each survey's pipeline turns a spectrum into an RV measurement using a particular set of templates, continuum models, and noise assumptions. As a result, the uncertainty $\sigma_{{\rm RV},\mathcal{S}}$ produced by the pipeline only captures \emph{some} of the true error budget. Calibration systematics, template mismatches, and pipeline-level correlations are usually not accounted. The relevant question in this section is therefore whether each survey's quoted $\sigma_{{\rm RV}}$ correctly represents its random measurement scatter; additive shifts in the RV values are handled separately by the zero-point calibration. To calibrate the uncertainty scale, we multiply each survey's quoted uncertainty by an
		\emph{rescaling factor} $f_{\mathcal{S}}^{\rm RV}$, so that the rescaled
		uncertainty
		\begin{equation}
			\sigma'_{{\rm RV},\mathcal{S}}
			\;=\;
			f_{\mathcal{S}}^{\rm RV} \,\times\, \sigma_{{\rm RV},\mathcal{S}}
			\label{eq:rv_corr}
		\end{equation}
		matches the empirical scatter actually observed when the same star is
		measured more than once. Note, in Eq.~(\ref{eq:rv_corr}), $\sigma_{{\rm RV},\mathcal S}$ and $\sigma'_{{\rm RV},\mathcal S}$ are per-measurement $1\sigma$ uncertainties which are given in the input catalogue. The ``empirical scatter'' is a property of the ensemble of repeat-measurement residuals; it is used only to determine the multiplicative factor $f_{\mathcal S}^{\rm RV}$ that calibrates those individual uncertainties. Thus, if a survey pipeline's quoted errors are already
		correct, we would obtain $f_{\mathcal{S}}^{\rm RV} = 1$. Departures from
		unity indicate miscalibrated uncertainties.
		
\begin{figure*}[htbp]
			\centering
			\resizebox{0.5\textwidth}{!}{\includegraphics[width=\linewidth]{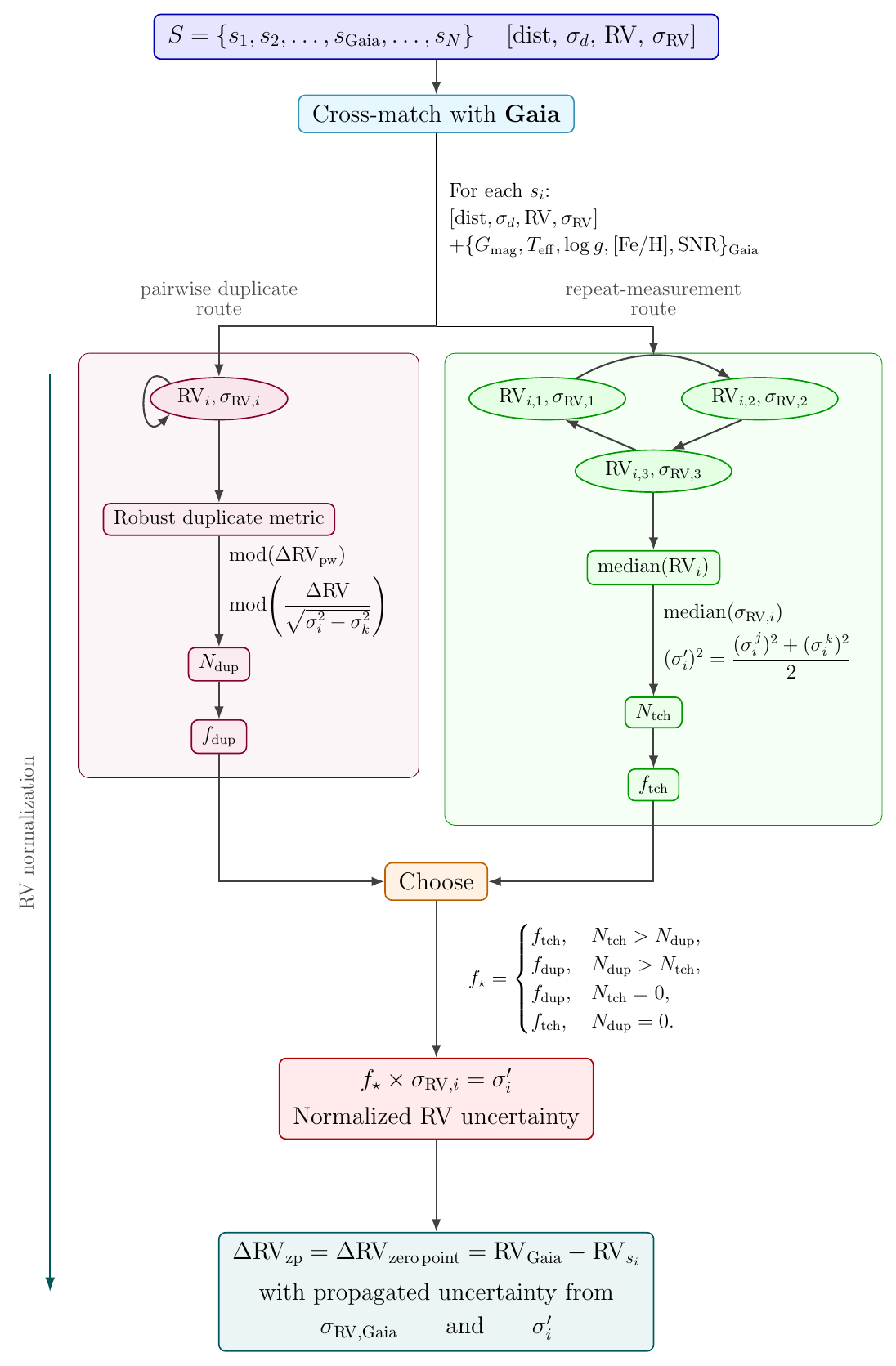}}
			\caption{\small Schematic of the radial-velocity normalisation procedure.
				Each source is cross-matched with \textit{Gaia} and routed through either
				the within-survey duplicate branch (left) or the cross-survey TCH
				branch (right). The resulting factors $f_{\rm DUP}$ and $f_{\rm TCH}$ are
				compared to define the adopted normalisation factor
				$f_{\star} \equiv f_{\mathcal{S}}^{\rm RV}$, which rescales the RV
				uncertainty and is then used in the construction of the zero-point
				correction relative to \textit{Gaia}.}
			\label{fig:rv_normalization_flowchart}
		\end{figure*}

		We compute $f_{\mathcal{S}}^{\rm RV}$ in two complementary ways: from
		\emph{repeated measurements (`duplication') within a survey} and from the
		\emph{three-cornered-hat} (TCH) comparison \emph{across surveys}. The two
		methods probe different aspects of the error budget. Repeated measurements
		test internal repeatability; TCH tests external consistency between
		independent surveys. The flowchart in Figure~\ref{fig:rv_normalization_flowchart} shows the
		overall flow.
		
		\subsection{Method 1: repeated measurements within a survey ($f_{\rm DUP}$)}
		
		Consider the same star being observed twice by the same survey, yielding
		velocities $v_{{\rm RV},1}$ and $v_{{\rm RV},2}$ with reported uncertainties
		$\epsilon_1$ and $\epsilon_2$. The pairwise difference is
		\begin{equation}
			\Delta v_{{\rm RV,pw}} \;=\; v_{{\rm RV},1} - v_{{\rm RV},2},
			\label{eq:delta_v}
		\end{equation}
		and, if the two measurements are independent, the standard deviation of this
		difference is
		\begin{equation}
			\sigma_{{\rm RV,pw}} \;=\; \sqrt{\epsilon_1^2 + \epsilon_2^2}.
			\label{eq:sigma_pw}
		\end{equation}
		The \emph{normalised} pairwise difference is therefore
		\begin{equation}
			\frac{\Delta v_{{\rm RV,pw}}}{\sigma_{{\rm RV,pw}}}.
		\end{equation}
		If the reported per-measurement uncertainties correctly describe the repeat-measurement scatter, then this quantity should be drawn from a unit
		Gaussian, $\mathcal{N}(0,1)$. Its width is therefore a direct diagnostic of whether the quoted uncertainty scale is calibrated, and we use its robust width below to determine the rescaling factor.
		
		Real difference distributions are rarely perfectly Gaussian, and commonly show non-Gaussian wings. To estimate a width that is not dominated by these outliers, we use the normalised median absolute deviation (normMAD) instead of the standard deviation. We first compute the median absolute deviation (MAD), and then convert it to normMAD, which is the quantity used as the Gaussian-equivalent width henceforth. The MAD is defined as the median of $|x_i - \mathrm{median}(x_i)|$. We multiply it by $1.4826$ so that, for a Gaussian distribution, the resulting normMAD equals the standard deviation.
		The repeated-measurement scaling factor for survey $\mathcal{S}$ is then
		\begin{equation}
			f_{{\rm DUP},\mathcal{S}}
			\;=\;
			\mathrm{normMAD}\!\left(
			\frac{\Delta v_{{\rm RV,pw}}}{\sigma_{{\rm RV,pw}}}
			\right),
		\end{equation}
		where the MAD is taken over all duplicate pairs available within survey
		$\mathcal{S}$.
		
		If a single star has
		$N$ repeated observations within the same survey, the number of distinct
		two-measurement pairs (duplicate pairs) that can be formed from those observations is
		$N_{\rm DUP} \;=\; \binom{N}{2}$. We use $N_{\rm DUP}$ to count how many internal pairs each survey contributes; the per-survey totals are listed in
		Table~\ref{tab:rv_norm_verified}.
		
\begin{figure}[htbp]
			\centering
			\includegraphics[width=\linewidth]{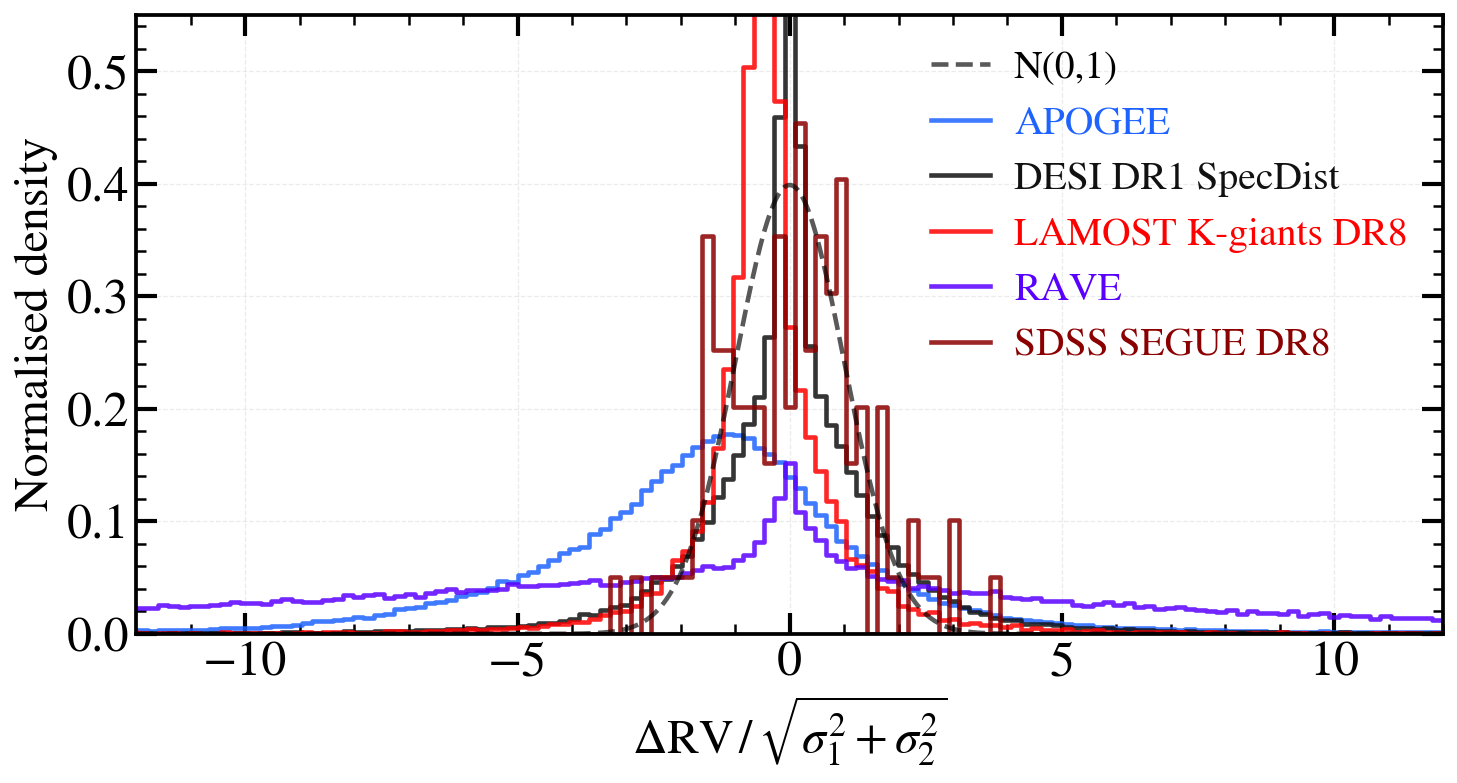}
			\caption{\small Normalised radial-velocity differences,
				$\Delta \mathrm{RV}/\sqrt{\sigma_1^2+\sigma_2^2}$, formed from duplicate
				observations of the same star within a single survey. Each coloured
				histogram corresponds to one survey; the dashed black curve is the
				$\mathcal{N}(0,1)$ reference. A histogram broader (narrower) than the
				reference indicates that the survey's quoted RV uncertainties are
				underestimated (overestimated). The width of each histogram, measured
				via the normalised MAD, defines the rescaling factor
				$f_{{\rm DUP},\mathcal{S}}$.}
			\label{fig:sigma_rv}
		\end{figure}

		Figure~\ref{fig:sigma_rv} shows the resulting normalised difference
		distributions. A perfectly calibrated survey would track the dashed
		$\mathcal{N}(0,1)$ reference curve. In practice, the histograms are
		narrower than the reference for some surveys (over-estimated quoted errors,
		$f < 1$) and broader for others (under-estimated quoted errors, $f > 1$).

		\subsection{Method 2: three-cornered hat across surveys ($f_{\rm TCH}$)}
		
		The repeated-measurement test probes internal repeatability within one survey. It can miss additional variance that appears only when that survey is compared with independent external data. To estimate this externally inferred variance, we also use the three-cornered-hat (TCH) method, which combines pairwise differences among independent surveys. Additive RV zero-point offsets are not what $f_{\rm TCH}$ is designed to measure; those are fitted separately in Section~\ref{sec:zp_cal}.\footnote{It is widely used in metrology
			to calibrate atomic clocks against one another \citep{Sjoberg2021}.} The central idea
		is as follows:
		
		Suppose three independent surveys measure the same physical
		quantity (for example, the radial velocity of a common star). Let their true
		measurement variances be $\sigma_1^2$, $\sigma_2^2$, and $\sigma_3^2$. The
		\emph{observed} pairwise difference variances are then
		\begin{equation}
			\sigma_{12}^2 = \sigma_1^2 + \sigma_2^2,\quad
			\sigma_{13}^2 = \sigma_1^2 + \sigma_3^2,\quad
			\sigma_{23}^2 = \sigma_2^2 + \sigma_3^2.
			\label{eq:tch_system}
		\end{equation}
		Solving the above gives
		\begin{equation}
			\sigma_i^2
			\;=\;
			\tfrac{1}{2}\!\left(\sigma_{ij}^2 + \sigma_{ik}^2 - \sigma_{jk}^2\right),
			\qquad i = 1, 2, 3,
			\label{eq:tch_solution}
		\end{equation}
		which isolates the per-survey variance from the pairwise differences alone,
		\emph{without ever needing to know the true RV of any individual star}. 
		
		For a target survey $\mathcal S$, we use every \emph{valid} pair $(j,k)_{\mathcal S}$ for which the required pairwise RV-difference variances can be estimated from duplicate measurements for that survey.  With $N$ available surveys, fixing the target survey leaves $N-1$ others, from which another survey is chosen. The maximum possible number of survey-pairs is therefore $\binom{N-1}{2}$, but only such survey-pairs that have  sufficient number of common stars are chosen. We denote the number of retained pairs by $N_{\rm TCH}$. For each valid pair, $(j,k)_{\mathcal S}$,  we compute $\sigma_{\mathcal S}^2$ from equation~(\ref{eq:tch_solution}) and average the valid estimates to obtain the survey-level TCH variance. Comparing this to the median of the
		RV uncertainties quoted for the same survey, we define the TCH scale factor
		\begin{equation}
			f_{{\rm TCH},\mathcal{S}}
			\;=\;
			\sqrt{\frac{\langle \sigma_{\mathcal{S}}^2 \rangle_{\rm TCH}}
				{\mathrm{median}\!\left(\sigma_{{\rm RV},\mathcal{S}}^2\right)}}.
		\end{equation}
		As before, $f_{{\rm TCH},\mathcal{S}} = 1$ would mean the quoted errors
		already match the externally measured precision.
		For the cross-matched stars from survey $\mathcal S$ that enter the TCH comparison, we square each pipeline-reported RV uncertainty and take the median of those squared values. The numerator is the TCH-inferred variance for the same survey; their ratio therefore compares externally inferred scatter with the typical quoted variance.
		
		\subsection{The normalisation factors for different surveys}
		
\begin{figure*}[htbp]
			\centering
			\includegraphics[width=1\linewidth]{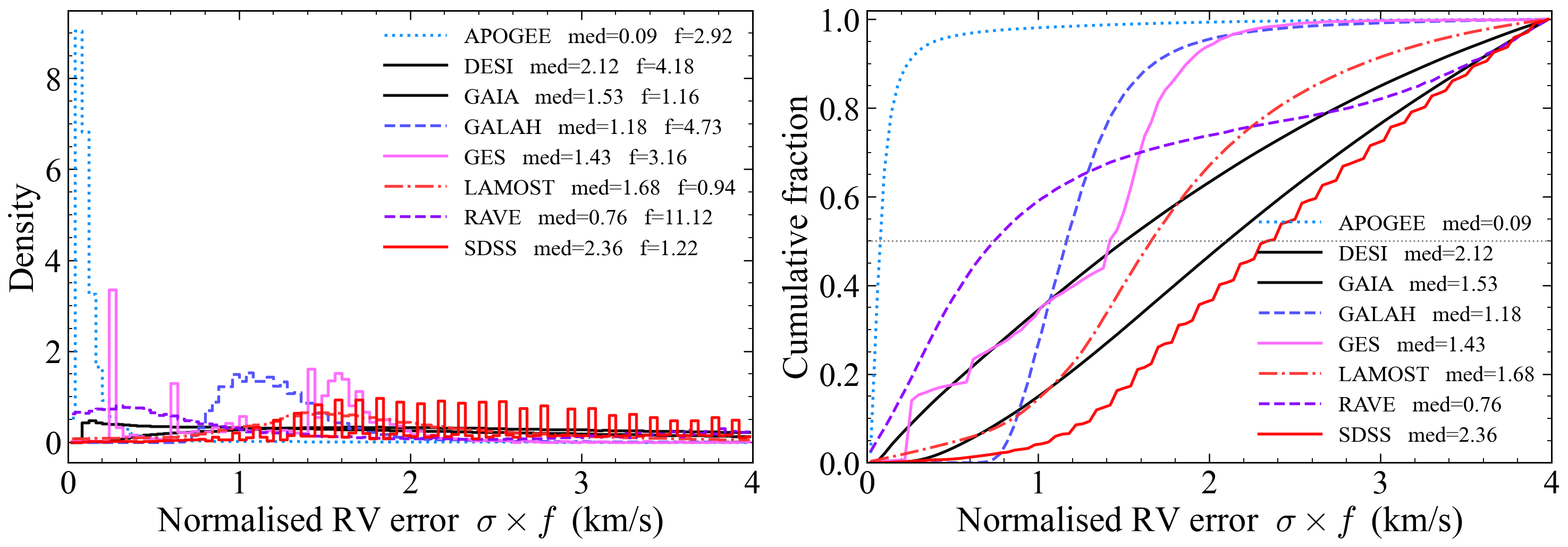}
			\caption{\small Distribution of the rescaled radial-velocity uncertainty
				$\sigma \times f$ for each survey. \emph{Left}: density distributions.
				\emph{Right}: cumulative distributions, with the dotted horizontal line
				marking the median. Per-survey medians and adopted factors are listed
				in the legends.}
			\label{fig:norm_RV_error}
		\end{figure*}

		After $f_{{\rm DUP},\mathcal{S}}$ and $f_{{\rm TCH},\mathcal{S}}$ are estimated, a survey may have one or both factors available, depending on its repeat observations and cross-survey overlap.
		We choose the factor that comes with a larger sample size\footnote{This choice is guided primarily by sample size, rather than by selecting whichever of \(f_{{\rm DUP},\mathcal{S}}\) or \(f_{{\rm TCH},\mathcal{S}}\) lies closer to unity, since an estimate based on a larger number of independent comparisons is generally expected to be more stable. This is a pragmatic selection rule rather than a formally optimal statistical estimator. In particular, duplicate pairs and TCH triplets may share individual measurements, so \(N_{\rm DUP}\) and \(N_{\rm TCH}\) cannot be interpreted as strictly independent information counts. A fully data-driven optimal estimator would therefore require an explicit treatment of these correlations and of the effective number of independent comparisons. In the absence of such a covariance model, we use the available sample size as a practical indicator of statistical stability. When both factors are available, their mutual agreement is used as an additional consistency diagnostic. Thus, the adopted rule does not claim statistical optimality; rather, it favours the correction factor supported by the larger observational sample, while retaining the comparison between the two methods as a check on possible systematic differences.
		}, i.e.
		\begin{equation}
			f_{\mathcal{S}}^{\rm RV}
			=
			\begin{cases}
				f_{{\rm DUP},\mathcal{S}}, & N_{\rm DUP} > N_{\rm TCH},\\[3pt]
				f_{{\rm TCH},\mathcal{S}}, & N_{\rm TCH} \ge N_{\rm DUP},
			\end{cases}
			\label{eq:f_rv_adopted}
		\end{equation}
		The final normalised uncertainties are then computed using
		equation~(\ref{eq:rv_corr}). The
		resulting $\sigma' = f \sigma$ distributions for all surveys are shown in Figure~\ref{fig:norm_RV_error}, and summarized in Table~\ref{tab:rv_norm_verified}.
		
		Among the $14$ families analysed, internal duplicates or triplets are
		available for $8$. We highlight a few cases that illustrate the importance
		of doing the previous steps:
		
		\begin{itemize}
			\item \textbf{\texttt{APOGEE}} has the smallest median quoted RV
			uncertainty in our catalogue ($0.093\,\mathrm{km\,s^{-1}}$), yet its
			duplicate-based factor is $f_{\rm DUP} = 2.92$. \emph{A small quoted
				error does not by itself guarantee a well-calibrated error.} APOGEE has
			very many duplicate pairs and only a handful of TCH triplets, so we adopt
			the duplicate factor.
			
			\item \textbf{\texttt{LAMOST}} sits closest to unity, with
			$f_{\rm DUP} = 0.94$. Its quoted errors require only a minor rescaling.
			
			\item \textbf{\texttt{RAVE}} has the largest factor of all, $f = 11.1$,
			again from duplicates rather than TCH since the duplicate sample is
			much larger.
			
			\item \textbf{\texttt{DESI}} illustrates the complementarity of the two
			methods. Its duplicate factor is $0.96$ (very close to unity, suggesting
			excellent internal repeatability), but its TCH factor is $4.18$. The discrepancy means that DESI is internally repeatable, while the cross-survey differences imply substantially larger external scatter than is predicted by the quoted DESI errors alone. A constant RV zero-point offset is diagnosed and corrected separately in Section~\ref{sec:zp_cal}.
			
			\item \textbf{\texttt{SDSS}} returns similar duplicate and TCH factors
			($1.33$ and $1.22$), and we adopt the TCH factor, which rests on more
			triplets than there are duplicate pairs.
			
			\item \textbf{\texttt{GAIA}} has no usable internal duplicates in the
			relevant subsample, so we use the TCH factor only. Its TCH value is
			$1.155$, indicating that Gaia's  RV errors are already a
			reasonably faithful description of the external scatter; our finding agrees
			with \citet{Tsantaki2022SurveyI}.
		\end{itemize}
		
		The DESI comparison,  in particular, shows why both methods are needed.
		A factor close to unity from internal duplicates does not guarantee that a
		survey has the survey-specific uncertainty scale which is consistent with the rest of the surveys.

		\begin{table*}[htbp]
			\centering
			\caption{Radial-velocity error normalisation summary, per survey family.
				Columns: total number of stars with reported stellar parameters; total
				unique stars; number of duplicate stars (i.e., stars observed more than once
				within the survey); number of within-survey duplicate pairs $N_{\rm DUP}$;
				the duplicate-based and TCH-based scaling factors; the adopted factor; the
				median quoted RV uncertainty in km/s; and the method by which the adopted
				factor was chosen.}
			\label{tab:rv_norm_verified}
			\small
			\setlength{\tabcolsep}{3pt}
			\resizebox{\linewidth}{!}{%
				\begin{tabular}{lccccccccl}
					\toprule
					Survey & With Stellar Parameters & Unique Stars & Duplicate Stars & DUP Pairs & $f_{{\rm DUP},\mathcal{S}}$ & $f_{{\rm TCH},\mathcal{S}}$ & Adopted factor & Median Error & Method \\
					\midrule
					GAIA   & 24{,}767{,}966 & 31{,}554{,}531 & 0       & 0       & --     & 1.155  & 1.155  & 1.526 & TCH only \\
					DESI   &  3{,}356{,}610 &  3{,}897{,}147 & 101{,}063 & 114{,}240 & 0.959  & 4.179  & 4.179  & 2.123 & TCH preferred \\
					APOGEE &     274{,}660  &     395{,}607  & 259{,}376 & 267{,}658 & 2.922  & 17.816 & 2.922  & 0.093 & DUP only \\
					GALAH  &     133{,}439  &     180{,}316  & 0       & 0       & --     & 4.735  & 4.735  & 1.184 & TCH only \\
					GES    &       2{,}515  &       3{,}169  & 0       & 0       & --     & 3.158  & 3.158  & 1.431 & TCH only \\
					LAMOST &  1{,}749{,}032 &  2{,}106{,}595 &  23{,}243 &  23{,}795 & 0.943  & 2.347  & 0.943  & 1.678 & DUP only \\
					RAVE   &     216{,}556  &     368{,}718  & 114{,}615 & 131{,}116 & 11.117 & 3.346  & 11.117 & 0.763 & DUP only \\
					SDSS   &      13{,}334  &      18{,}504  &     124  &     105  & 1.332  & 1.217  & 1.217  & 2.360 & TCH preferred \\
					\bottomrule
			\end{tabular}}
		\end{table*}
		
		\section{Counting the Distinct Stars}\label{sec:exact_counting}
		Before describing how we identify duplicates, we first set up the bookkeeping
		that tells us, in principle, how many distinct stars our combined input
		contains. This is essentially a set-theoretic problem.
		
		First, consider a single survey family $i$ (out of $K$ families in total). Now, let $S_i$ be the
		set of distinct stars in that survey family, and let $R_i$ be
		the set of \emph{rows} that family contributes to the input. These are not
		the same: a single star can appear in $R_i$ multiple times if it was observed
		more than once. We split $S_i$ into two pieces:
		\begin{itemize}
			\item $n_i$, the number of stars in family $i$ that were observed
			\emph{exactly once} within that family;
			\item $m_i$, the number of stars in family $i$ that were observed
			\emph{two or more times} within that family.
		\end{itemize}
		The number of distinct stars contributed by family $i$ is therefore
		\begin{equation}
			N_i \;\equiv\; |S_i| \;=\; n_i + m_i \;.
			\label{eq:Ni_def}
		\end{equation}
		
		Next, consider multiple survey families where the same star may appear in more than
		one family. We define the cardinality of the various overlaps as:
		\begin{align}
			P_{ij}    &= |S_i \cap S_j|                       && \text{(pair overlap)},\\
			T_{ijk}   &= |S_i \cap S_j \cap S_k|              && \text{(triple overlap)},\\
			Q_{ijkl}  &= |S_i \cap S_j \cap S_k \cap S_l|     && \text{(quadruple overlap)},
		\end{align}
		and so on. Here $|\cdot|$ denotes the number of elements in the set.
		
		The exact number of distinct stars in the union of all $K$ families is given by the inclusion--exclusion formula,
		\begin{equation}
			\begin{split}
				N_{\rm unique}
				&= \sum_{i=1}^{K} N_i
				\;-\; \sum_{1 \le i < j \le K} P_{ij}
				\;+\; \sum_{1 \le i < j < k \le K} T_{ijk}\\
				&\quad - \sum_{1 \le i < j < k < l \le K} Q_{ijkl}
				\;+\; \cdots,
			\end{split}
			\label{eq:inclusion_exclusion_general}
		\end{equation}
		which, after substituting $N_i = n_i + m_i$, becomes
		\begin{equation}
			\begin{split}
				N_{\rm unique}
				&= \sum_{i=1}^{K} (n_i + m_i)
				\;-\; \sum_{i<j} P_{ij}
				\;+\; \sum_{i<j<k} T_{ijk}\\
				&\quad - \sum_{i<j<k<l} Q_{ijkl}
				\;+\; \cdots.
			\end{split}
			\label{eq:inclusion_exclusion_nm}
		\end{equation}
		
		If no star were ever
		shared by more than two families, equation~(\ref{eq:inclusion_exclusion_nm})
		would reduce to the much simpler form
		$N_{\rm unique} = \sum_{i}(n_i + m_i) - \sum_{i<j} P_{ij}$. However, this is
		\emph{not} the case for our catalogue. Higher-order overlaps are common: for
		example, $178{,}776$ stars are shared by \texttt{GAIA}, \texttt{DESI}, and
		\texttt{LAMOST}; $103{,}885$ are shared by \texttt{GAIA}, \texttt{APOGEE},
		and \texttt{LAMOST}; and $13{,}588$ are shared by \texttt{GAIA},
		\texttt{APOGEE}, and \texttt{RAVE}. Ignoring these triple (and higher)
		overlaps would over-count the number of unique stars.
		
\begin{figure*}[htbp]
			\centering
			\resizebox{0.9\textwidth}{!}{\includegraphics[width=\linewidth]{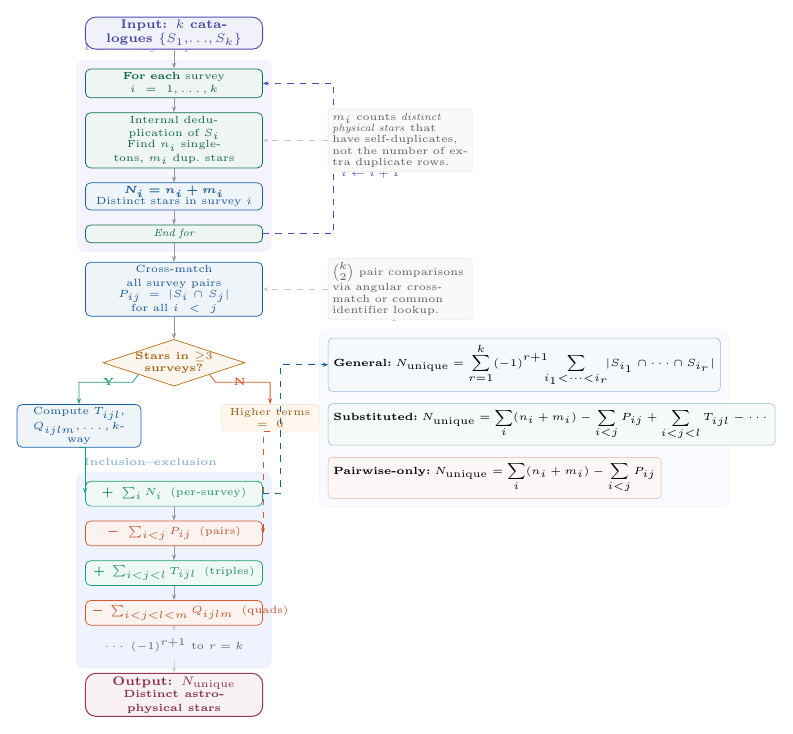}}
			\caption{\small Schematic of the inclusion--exclusion bookkeeping
				underlying the deduplication step. After internal deduplication within
				each survey ($N_i = n_i + m_i$), inter-survey overlaps are incorporated
				via pairwise terms $P_{ij}$ and, where present, higher-order terms
				$T_{ijk}$, $Q_{ijkl}$, etc. The figure also shows the simpler
				pairwise-only formula that would apply if higher-order overlaps were
				absent, which is \emph{not} the case for the present catalogue.}
			\label{fig:deduplicaton_diagram}
		\end{figure*}

		Note that although Eq.~\ref{eq:inclusion_exclusion_nm}
		is exact, when applied across $20$ parent catalogues, it requires evaluating an exponentially large number of intersection terms. In
		Section~\ref{sec:dedup}, we therefore implement a
		graph-based algorithm that produces an identical result without explicitly
		enumerating every higher-order overlap. The schematic in
		Figure~\ref{fig:deduplicaton_diagram} summarises the relationship between
		the inclusion--exclusion view and the graph-based implementation. 
		
		\section{Identifying Repeated Observations of the Same Star}\label{sec:dedup}
		
		After the radial velocity uncertainty-normalisation stage, we identify the rows in the input that refer to the same physical star. We use the term
		\emph{duplicate} for any pair of input rows that correspond to the same
		star, regardless of whether the two rows come from the same survey or from
		different surveys. A duplicate group is the set of all rows that, by our
		matching criteria (as discussed below), refer to one star.

		In principle, finding all duplicate groups across $20$ surveys could be done
		by directly applying the inclusion--exclusion formula in
		equation~(\ref{eq:inclusion_exclusion_nm}). In practice, this requires
		explicitly evaluating $2^{20}-1$ intersection terms, which is computationally
		prohibitive. The same logical content can be obtained much more efficiently
		with a graph-based approach.
		
		We build an undirected graph in which each \emph{node} represents one row
		(one measurement of one star in one parent catalogue), and we draw an
		\emph{edge} between two nodes if the two rows pass our matching criteria
		described below. Once all edges have been added, every \emph{connected
			component} of this graph (i.e.\ every set of nodes that can be reached from
		each other by following edges) corresponds to one physical star.
		
		The reason behind the graph-based method is transitivity by extension: if measurement $A$
		matches measurement $B$, and $B$ matches $C$, then $A$, $B$, and $C$ are all
		treated as the same star, even if $A$ and $C$ would not have been linked
		directly by themselves. This handles two-survey, three-survey, and
		higher-order overlaps automatically, with no explicit inclusion--exclusion
		expansion.
		
		\subsection{Matching criteria}
		
		In our algorithm, two rows are joined by an edge if and only if they pass two tests: a
		sky-position test and (when both rows have $\varpi$ and its reported uncertainty are finite and that $\sigma_\varpi>0$ parallax) a parallax
		test.
		
		To avoid comparing every row
		against every other row in the input, we partition the sky into
		\textsc{healpix} pixels \citep{Gorski2005}. For each row, we restrict the search for matching candidates to the row's own HEALPix pixel plus the neighbouring pixels returned by the HEALPix adjacency query (up to eight). The neighbour-pixel padding prevents us from missing genuine matches that
		straddle a pixel boundary.
		
		Within each local search region, we represent
		each row's sky position by its right ascension $\alpha$ and declination
		$\delta$, and convert these to a three-dimensional unit vector,
		\begin{equation}
			\hat{\boldsymbol{r}}
			=
			(\cos\delta\cos\alpha,\; \cos\delta\sin\alpha,\; \sin\delta).
		\end{equation}
		where the unit vectors are indexed with a $k$-dimensional tree
		(the SciPy implementation \texttt{cKDTree})\footnote{The $k$-d tree algorithm is a standard
			data structure that organises points in space such that all points within a
			given distance from a query point can be found rapidly.  We use \texttt{scipy.spatial.cKDTree}; the SciPy implementation is referenced by \citet{Virtanen2020}.}. 
		The tree operates on ordinary three-dimensional Euclidean distances, so the on-sky angular threshold $\theta_{\rm match}$ must be converted to the corresponding straight-line (chord) distance,
		\begin{equation}
			d_{\rm chord} = 2\sin\!\left(\tfrac{1}{2}\theta_{\rm match}\right).
		\end{equation}
		All pairs separated by less than $d_{\rm chord}$ are 
		selected as candidate matches. Each candidate is then required to satisfy
		the final on-sky tolerance: $\Delta\theta < 1''$.
		
		Next, if and only if, both candidate rows have a finite parallax with a finite, positive reported uncertainty, we apply a second filter. Let $\varpi_1, \varpi_2$ be the
		two parallaxes and $\sigma_{\varpi,1}, \sigma_{\varpi,2}$ their
		uncertainties. We accept the pair if it satisfies \emph{either} of two
		conditions:
		\begin{equation}
			\frac{|\varpi_1 - \varpi_2|}
			{\sqrt{\sigma_{\varpi,1}^{2} + \sigma_{\varpi,2}^{2}}}
			\;\le\; 3
			\quad\text{(statistical agreement of $3\sigma$)},
			\label{eq:plx_statistical}
		\end{equation}
		\emph{or}
		\begin{equation}
			\frac{|\varpi_1 - \varpi_2|}{\max(|\varpi_1|, |\varpi_2|)}
			\;<\; 0.2
			\quad\text{(fractional agreement of $20\%$)},
			\label{eq:plx_fractional}
		\end{equation}
		where the first condition selects the parallaxes that agree
		within $3\sigma$ of their combined uncertainty.
		The second condition is a safety net for cases where the first becomes
		needlessly strict; for example, when parallaxes are very precise (i.e., very small
		$\sigma_{\varpi}$), even small genuine differences  such as numerical
		rounding between catalogues  can blow up
		$|\varpi_1 - \varpi_2|/\sqrt{\sigma_{\varpi,1}^{2} + \sigma_{\varpi,2}^{2}}$
		beyond $3$, leading to spurious rejections. The fractional condition
		provides a sensible tolerance based on the size of the parallax difference
		itself, accepting pairs whose parallaxes agree to better than $20\%$.
		
		Note that if either or both candidate rows lack finite parallax with a finite, positive reported uncertainty, we do not enforce the parallax criteria, and such pairs are accepted as matches as long as they satisfy the on-sky tolerance $\Delta\theta < 1''$. This helps minimise candidate loss in faint or low-S/N
		regimes.
		
		\subsection{Graph construction, deduplications and final tally}
		
		Applying the above matching criteria to all candidate pairs yields $P_{\rm total} \;=\; 27{,}540{,}273$ matched pairs (i.e. edges in the graph). The set of nodes is the set of
		input rows, and the set of edges is this collection of validated matches.
		
		To extract the connected components from the graph, we use a
		\emph{union--find} (also called \emph{disjoint-set union})  data structure
		with path compression and union-by-rank (see \citet{cormen2009}). Union--find is a standard data structure that tracks the elements belonging to the same set as more edges are added one-by-one. It begins from each node in its own
		singleton group, and every new edge merges two groups. Queries regarding which
		node contains which group can then be answered in nearly constant amortised
		time. In our implementation, the cost of resolving the graph is essentially linear in
		the number of validated pairs.
		
		Applied to the $52{,}112{,}379$ input rows, the connected-component analysis
		yields $N_{\star} = 32{,}552{,}876$ unique stars. The deduplication
		efficiency (fraction of input rows removed as duplicates) is
		$37.53\%$, with a corresponding compression factor of $1.60$.
		Of these unique stars,
		\begin{itemize}
			\item $19{,}459{,}422$ ($59.8\%$) are \emph{singletons}  stars that
			appear in our input from only a single observation in a single survey;
			\item $13{,}093{,}454$ ($40.2\%$) are observed two or more times,
			either by repeated visits within one survey or by independent detections
			across different surveys.
		\end{itemize}
		
		The largest pairwise overlap is
		\texttt{GAIA}--\texttt{LAMOST}, with $2{,}050{,}645$ shared stars. Other
		significant pairwise overlaps include \texttt{APOGEE}--\texttt{GAIA}
		($359{,}005$), \texttt{GAIA}--\texttt{RAVE} ($320{,}919$),
		\texttt{GAIA}--\texttt{GALAH} ($169{,}847$), and
		\texttt{DESI}--\texttt{LAMOST} ($175{,}049$). The most populated triple
		overlaps are \texttt{GAIA}--\texttt{DESI}--\texttt{LAMOST} ($178{,}776$),
		\texttt{GAIA}--\texttt{APOGEE}--\texttt{LAMOST} ($103{,}885$), and
		\texttt{GAIA}--\texttt{APOGEE}--\texttt{RAVE} ($13{,}588$). Higher-order
		overlaps also exist among
		\texttt{GAIA}--\texttt{GALAH}--\texttt{RAVE},
		\texttt{GAIA}--\texttt{LAMOST}--\texttt{RAVE},
		\texttt{GAIA}--\texttt{GALAH}--\texttt{LAMOST},
		\texttt{DESI}--\texttt{APOGEE}--\texttt{LAMOST},
		\texttt{GAIA}--\texttt{APOGEE}--\texttt{GALAH}, and
		\texttt{APOGEE}--\texttt{LAMOST}--\texttt{RAVE}. These higher-order overlaps
		are precisely the terms that the graph algorithm absorbs automatically
		without our having to write each intersection term down. Interestingly, the maximum multiplicity for a single star in our catalogue is $18$
		detections.

		\section{Calibrating Radial Velocities to a Common Zero Point}\label{sec:zp_cal}
		
		After error normalisation (Section~\ref{sec:error_norm}) and deduplication
		(Section~\ref{sec:dedup}), every measurement carries a normalised (error-rescaled) uncertainty, and
		every duplicate group is identified. We can now examine the \emph{values} of the survey RV measurement.
		
		When the same star is measured by two surveys, the two RV values rarely
		agree exactly. Some of the disagreement is genuine random scatter, captured
		by the renormalised uncertainties. However, part of it is \emph{systematic}: the
		two surveys may report values shifted by a survey-dependent offset, and
		these offsets can themselves drift with stellar parameters, such as magnitude
		or temperature. This section removes such offsets by tying every survey to
		a single reference frame, which is the \textit{Gaia} radial-velocity scale.
		
		\subsection{Setup}
		
		We choose \textit{Gaia} as the common reference for two reasons: (i) it has by
		far the largest sky coverage of any RV surveys in our input-catalogue provenance, and (ii) its
		companion stellar parameters $(G_{\rm mag}, T_{\rm eff}, \log g,
		[\mathrm{Fe/H}], v_{\rm RV}, {\rm S/N}_{\rm RV})$ are available for every
		matched star, which allows us to look for parameter-dependent calibration
		trends. For each survey family $\mathcal{S}$ that we wish to calibrate, we
		work with the subset of stars that have RV measurements from both
		\textit{Gaia} and $\mathcal{S}$.
		
		For a given matched star $i$:
		\begin{itemize}
			\item $v_{{\rm RV},{\rm G},i}$ and $\sigma_{{\rm RV},{\rm G},i}$ are the
			\textit{Gaia} RV and its uncertainty;
			\item $v_{{\rm RV},\mathcal{S},i}$ and $\sigma_{{\rm RV},\mathcal{S},i}$
			are the radial velocity and its uncertainty from survey $\mathcal{S}$;
			\item the renormalised uncertainties (from
			Section~\ref{sec:error_norm}) are
			$\sigma'_{{\rm RV},{\rm G},i}    = f_{\rm G}^{\rm RV}\,\sigma_{{\rm RV},{\rm G},i}$
			and
			$\sigma'_{{\rm RV},\mathcal{S},i}= f_{\mathcal{S}}^{\rm RV}\,\sigma_{{\rm RV},\mathcal{S},i}$.
		\end{itemize}
		
		\subsection{The pairwise residual}
		
		The quantity of interest in the calibration is the \emph{pairwise residual}
		between Gaia and survey~$\mathcal{S}$ for star~$i$:
		\begin{equation}
			\Delta v_{{\rm RV,pw},i}
			\;=\;
			v_{{\rm RV},{\rm G},i} - v_{{\rm RV},\mathcal{S},i}.
			\label{eq:delta_rv_pw}
		\end{equation}
		If the two surveys  required no fitted systematic correction and had only random scatter, this
		quantity would be a random-measurement sample from a zero-mean Gaussian. Any systematic offset
		between the two surveys appears as a non-zero mean (or as a non-zero trend
		in $\Delta v_{{\rm RV,pw}}$ versus a stellar parameter).
		
		The renormalised uncertainty on $\Delta v_{{\rm RV,pw},i}$ is
		\begin{equation}
			\left(\sigma'_{{\rm RV,pw},i}\right)^{2}
			=
			\left(\sigma'_{{\rm RV},{\rm G},i}\right)^{2}
			+
			\left(\sigma'_{{\rm RV},\mathcal{S},i}\right)^{2},
			\label{eq:sigma_rv_pw_prime}
		\end{equation}
		which we will use as the inverse weight in the fits below.
		
		\subsection{The polynomial calibration model}
		
		For each survey $\mathcal{S}$ and each \textit{Gaia} stellar parameter
		we model the pairwise residual as a smooth, second-order polynomial in $x$:
		\begin{equation}
			\Delta v_{{\rm RV,ZP}}(x)
			\;=\;
			\alpha + \beta\,x + \gamma\,x^{2}.
			\label{eq:poly_model}
		\end{equation}
		where
		\begin{equation}
			x_i \in
			\bigl\{\,G_{\rm mag},\, T_{\rm eff},\, \log g,\,
			[\mathrm{Fe/H}],\, v_{\rm RV},\, {\rm S/N}_{\rm RV}\,\bigr\}_i,
			\label{eq:x_controls}
		\end{equation}
		Note, that a second-order polynomial allows for both a
		bulk offset (the constant $\alpha$) and a slow drift with $x$ (captured by
		$\beta$ and $\gamma$); higher-order terms are not statistically required by
		the data. We use this fitted curve as the survey's zero-point correction:
		the value of $\Delta v_{{\rm RV,ZP}}$ at a star's parameters is what must
		be added back to the survey RV to bring it onto the Gaia RV reference scale frame
		(see Section~\ref{sec:rv_corrected}).
		
		\subsection{Weighted least-squares solution}
		
		We fit equation~(\ref{eq:poly_model}) to the matched data using weighted
		least squares \citep{Bevington1969}. Each star contributes with weight
		\begin{equation}
			w_i \;=\; \frac{1}{\left(\sigma'_{{\rm RV,pw},i}\right)^{2}},
			\label{eq:weight_pw}
		\end{equation}
		i.e.\ inverse pairwise variance. Stars with smaller renormalised pairwise
		uncertainties carry more weight, as expected. We define the basis vector
		\begin{equation}
			\boldsymbol{\phi}(x_i)
			=
			\begin{bmatrix} 1 \\ x_i \\ x_i^{2} \end{bmatrix},
			\qquad
			\mathbf{c}
			=
			\begin{bmatrix} \alpha \\ \beta \\ \gamma \end{bmatrix}.
			\label{eq:basis_phi}
		\end{equation}
		where the normal equations are $\mathbf{M}\,\mathbf{c} = \mathbf{b}$, with
		\begin{align}
			M_{lk} &= \sum_{i=1}^{N} w_i\,\phi_l(x_i)\,\phi_k(x_i),
			\label{eq:M_matrix}\\
			b_k    &= \sum_{i=1}^{N} w_i\,\Delta v_{{\rm RV,pw},i}\,\phi_k(x_i).
			\label{eq:b_vector}
		\end{align}
		Inverting $\mathbf{M}$ gives the best-fit coefficients,
		\begin{equation}
			\mathbf{c} \;=\; \mathbf{M}^{-1}\,\mathbf{b},
			\label{eq:c_solution}
		\end{equation}
		which in turn defines the fitted zero-point correction
		$\Delta v_{{\rm RV,ZP}}(x)$.
		
		\subsection{Uncertainty on the fitted correction}
		
		The covariance matrix of the fitted coefficients $\mathbf{c}$ is given by
		$\mathbf{C} \;=\; \mathbf{M}^{-1}$.
		However, this expression is correct only if our renormalised pairwise uncertainties describe the data perfectly. They seldom do since the
		residuals retain real astrophysical scatter that no simple polynomial can
		absorb. We therefore rescale $\mathbf{C}$ by the reduced chi-squared of
		the fit,
		\begin{equation}
			\chi^{2}_{\nu}
			\;=\;
			\frac{1}{N - 3}
			\sum_{i=1}^{N}
			\frac{\bigl[\Delta v_{{\rm RV,pw},i} - \Delta v_{{\rm RV,ZP}}(x_i)\bigr]^{2}}
			{\left(\sigma'_{{\rm RV,pw},i}\right)^{2}},
			\label{eq:chi2_red}
		\end{equation}
		which adjusts the  coefficient errors so that the fit's per-degree-of-freedom
		$\chi^{2}$ equals unity. The rescaled covariance matrix is
		\begin{equation}
			\mathbf{C}'
			\;=\;
			\chi^{2}_{\nu}\,\mathbf{C}
			\;=\;
			\chi^{2}_{\nu}\,\mathbf{M}^{-1}.
			\label{eq:Cprime_matrix}
		\end{equation}
		Standard error propagation then gives the uncertainty of the fitted zero-point correction evaluated at $x_i$. For any value of $x$,
		\begin{eqnarray}
			\bigl(\sigma'_{{\rm RV,ZP},i}\bigr)^{2}  
			&=&  \boldsymbol{\phi}(x_i)^{\rm T}\,\mathbf{C}'\,\boldsymbol{\phi}(x_i)\\
			&=&  C'_{11} + x_i^{2}\,C'_{22} + x_i^{4}\,C'_{33}\\ \nonumber
			&&\quad + 2\bigl(x_i\,C'_{12} + x_i^{2}\,C'_{13} + x_i^{3}\,C'_{23}\bigr).
			\label{eq:sigma_rv_zp_prime_expanded}
		\end{eqnarray}
		Thus $\sigma'_{{\rm RV,ZP},i}$ is the propagated uncertainty of the fitted zero-point correction and is combined with the normalised measurement uncertainty when constructing the final RV uncertainty.
		
		\subsection{Stellar parameter dependence of calibration results}
		
		\begin{figure*}[!t]
			\centering
			\includegraphics[width=0.8\linewidth]{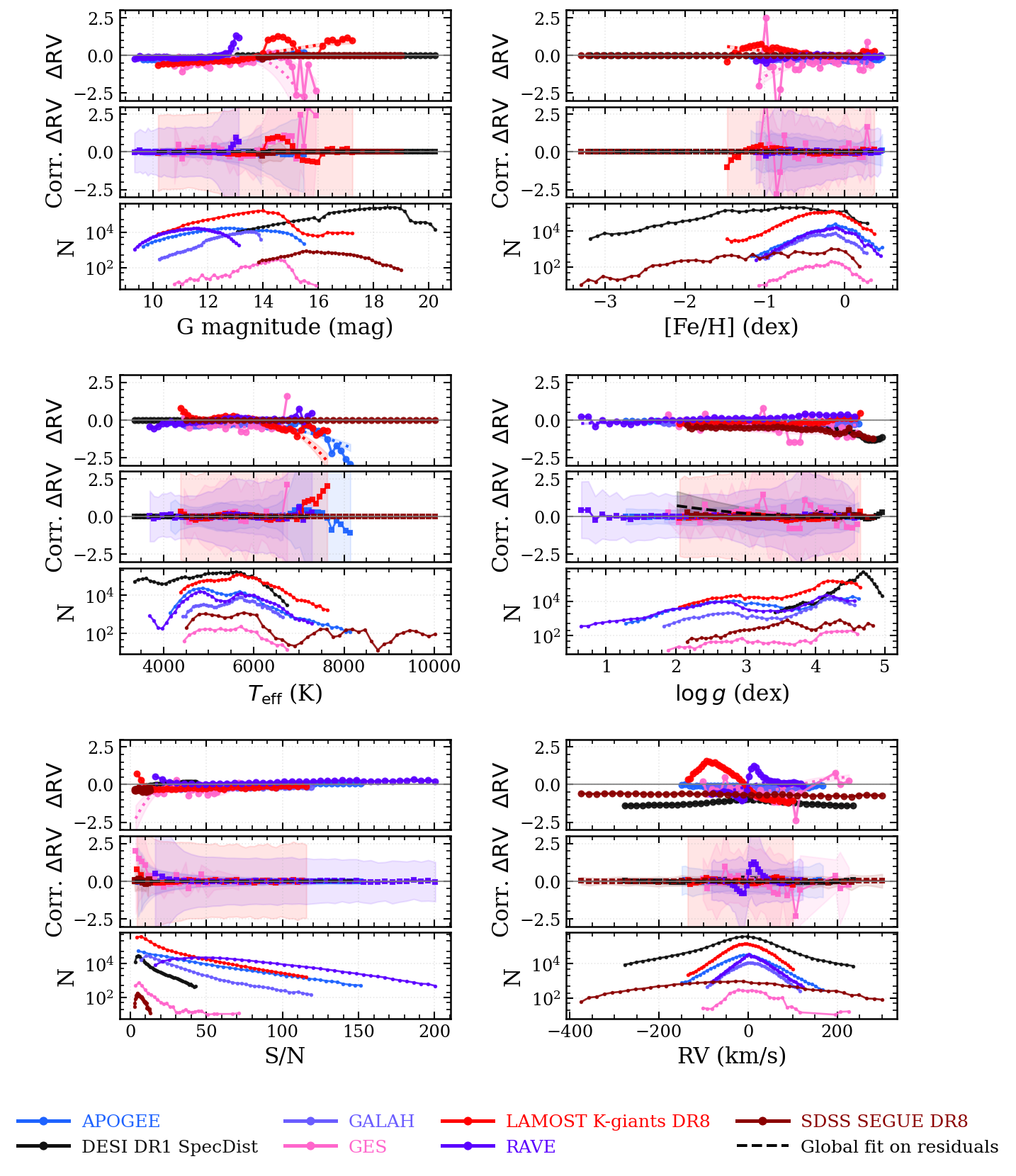}
			\caption{\small Median radial-velocity residual $\Delta\mathrm{RV}$
				relative to \textit{Gaia} as a function of six \textit{Gaia} stellar
				parameters, shown separately for each contributing survey. Each column
				is one parameter; within a column, the top panel shows the raw
				residual, the middle panel shows the residual after applying the
				polynomial zero-point correction, and the bottom panel shows the number
				of stars per bin. Dashed black curves in the middle panel are the polynomial fits used in
				the calibration, with grey-shaded $68\%$ confidence intervals.}
			\label{fig:raw_delta_rv}
		\end{figure*}
		
		Figure~\ref{fig:raw_delta_rv} shows the median residual $\Delta\mathrm{RV}$
		versus each of the six control parameters, separately for each survey. For each stellar or observational parameter, the top panel shows the raw residual $\Delta\mathrm{RV}=v_{{\rm RV},G}-v_{{\rm RV},\mathcal S}$, the middle panel shows the residual after the fitted zero-point correction, and the bottom panel gives the number of stars in each bin. The bin counts are important because excursions are influenced by the weighted average of the number of stars in each bin.
		The black dashed curve in
		the middle panel is the global polynomial fit; the grey shaded band is the
		$68\%$ confidence interval propagated from $\mathbf{C}'$.
		
		The dominant calibration term is generally a bulk RV offset, with weaker parameter-dependent structure superposed. The main features supported by the populated bins are as follows.
		\begin{itemize}
			\item \textbf{$G$ magnitude:} \texttt{LAMOST} shows the strongest magnitude dependence, including an excursion near $G\simeq15$--$16$ and a change toward the faint end; \texttt{RAVE} shows a weaker dependence.
			\item \textbf{$T_{\rm eff}$:} \texttt{LAMOST} shows structure at both the cool and hot ends of its populated range, while \texttt{RAVE} shows weaker low-temperature fluctuations.
			
			\item \textbf{S/N:} \texttt{LAMOST} shows the clearest low-S/N dependence, with a rapid change in $\Delta\mathrm{RV}$ below S/N $\sim10$--$15$; \texttt{RAVE} shows a milder feature.
			\item \textbf{$\log g$:} a weak population-level dependence is visible, with \texttt{DESI} contributing most coherently at high surface gravity. The \texttt{SDSS} sample is too sparse in this panel for a strong conclusion.
			\item \textbf{$[\mathrm{Fe/H}]$:} the large \texttt{GES} excursion at $[\mathrm{Fe/H}]\lesssim-2$ coincides with very small bin counts and is therefore treated as a small-$N$ effect. \texttt{LAMOST} shows a more modest sub-solar metallicity dependence.
			\item \textbf{$v_{\rm RV}$:} \texttt{LAMOST} shows an approximately monotonic residual trend across the well-populated RV range, with negative residuals at negative RV and positive residuals at positive RV. \texttt{DESI} is instead dominated by an approximately constant offset of $\simeq-0.4\,\mathrm{km\,s^{-1}}$ across the populated range; \texttt{SDSS} also shows a largely constant offset. \texttt{RAVE} has a smaller local feature around $v_{\rm RV}\sim0$, where its sample density is highest.
		\end{itemize}
		
\begin{figure}[htbp]
			\centering
			\includegraphics[width=1\linewidth]{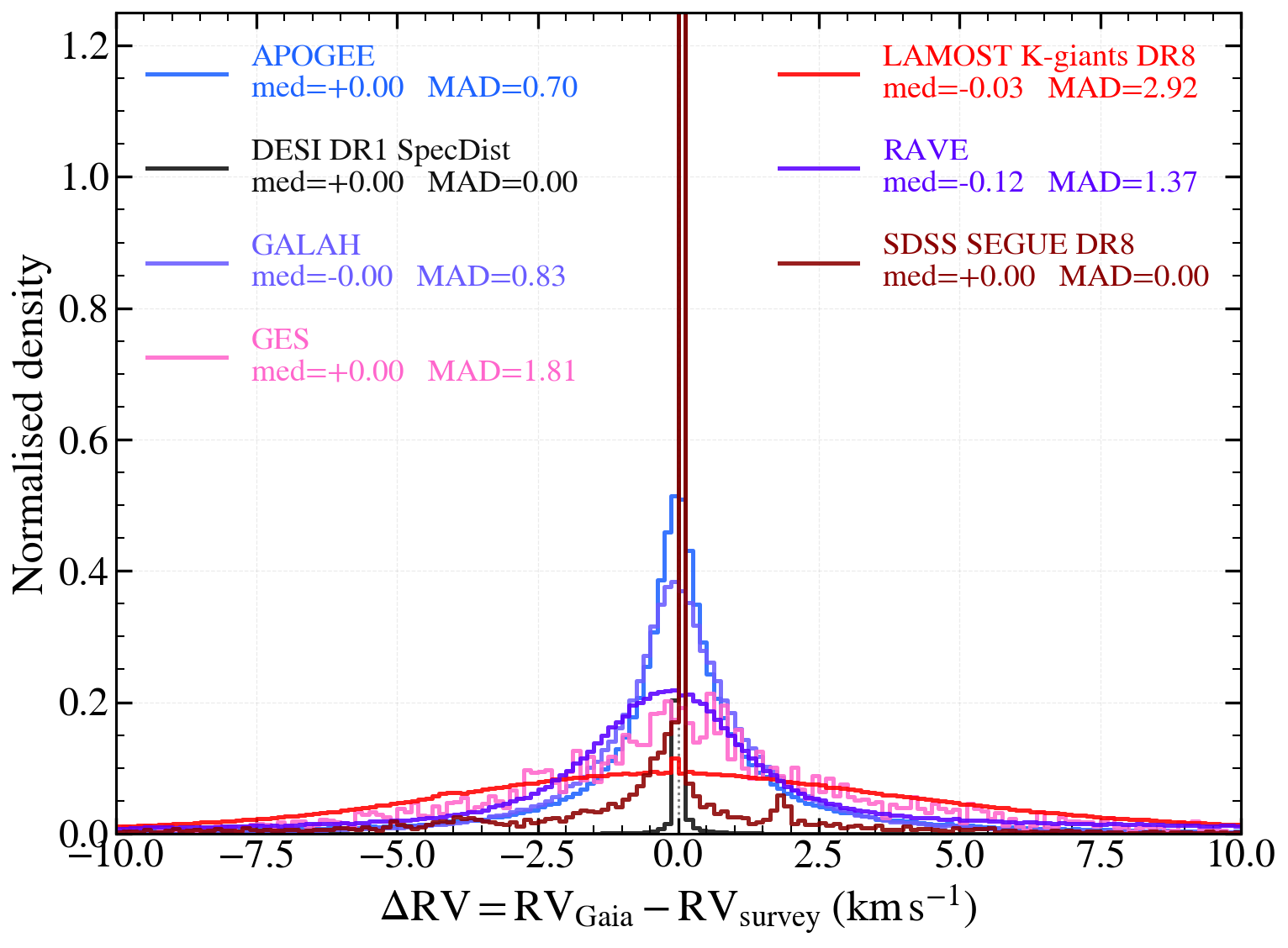}
			\caption{\small Distributions of the zero-point-corrected residual
				$\Delta\mathrm{RV} = v_{{\rm RV},{\rm G}} - v_{{\rm RV},\mathcal{S}}$
				for the contributing surveys. After calibration, all distributions are
				centred close to zero, indicating that the survey-to-survey systematic
				offsets have been removed; the remaining widths reflect the intrinsic
				measurement scatter of each survey.}
			\label{fig:drv_corrected}
		\end{figure}

		After applying the corrections, the residual centroids move closer to zero over the well-populated parameter ranges. The overall before--to--after change is summarised in Table~\ref{tab:rv_before_after} and Fig.~\ref{fig:drv_corrected}. The calibration removes coherent additive structure; it is not expected to eliminate the full residual width, which also contains measurement noise, non-Gaussian wings, and possible higher-dimensional correlations not represented by the one-parameter-at-a-time fits.
		
		\begin{table*}[htbp]
			\centering
			\scriptsize
			\setlength{\tabcolsep}{3pt}
			\caption{Per-survey $\Delta\mathrm{RV}$ diagnostics before and after the
				zero-point correction. Negative reductions indicate that the scatter
				estimator increased after the correction; this happens when the
				correction shifts the bulk of the distribution but redistributes the
				wings.Here, $\mathrm{Med}(|\Delta\mathrm{RV}|)$ denotes the median absolute
				residual.}
			\label{tab:rv_before_after}
			\begin{tabular}{l r rrrr rrrr rr}
				\toprule
				& & \multicolumn{4}{c}{Before} & \multicolumn{4}{c}{After}
				& \multicolumn{2}{c}{Reduction (\%)} \\
				\cmidrule(lr){3-6}\cmidrule(lr){7-10}\cmidrule(lr){11-12}
				Survey & $N$ & Med & MAD & Std & $\mathrm{Med}(|\Delta\mathrm{RV}|)$& Med & MAD & Std & $\mathrm{Med}(|\Delta\mathrm{RV}|)$ & MAD & Std \\
				\midrule
				APOGEE  &   395\,603 & $-$0.114 & 0.710 & 17.63 & 0.721 &    0.135 & 0.814 & 17.54 & 0.826 & $-$14.6 &   0.56 \\
				DESI    & 3\,897\,147 & $-$1.096 & 0.266 &  0.639 & 1.097 & $-$0.256 & 0.189 &  0.678 & 0.274 &   29.1 & $-$6.16 \\
				GALAH   &   180\,316 & $-$0.263 & 0.808 & 11.13 & 0.846 &    0.026 & 0.855 & 11.12 & 0.854 &  $-$5.90 &   0.03 \\
				GES     &     3\,169 & $-$0.411 & 2.052 & 23.43 & 2.080 &    0.597 & 2.249 & 23.45 & 2.343 &  $-$9.60 & $-$0.09 \\
				LAMOST  & 2\,106\,589 & $-$0.047 & 3.349 & 23.21 & 3.350 &    0.141 & 3.339 & 23.19 & 3.336 &    0.31 &   0.10 \\
				RAVE    &   368\,718 &    0.130 & 1.787 & 21.50 & 1.778 & $-$0.181 & 2.636 & 21.41 & 2.658 & $-$47.5 &   0.42 \\
				SDSS    &    18\,453 & $-$0.652 & 0.307 &  3.380 & 0.779 &    0.035 & 0.513 &  3.366 & 0.518 & $-$67.3 &   0.42 \\
				\bottomrule
			\end{tabular}
		\end{table*}

		\section{Combining Measurements into the Final Catalogue}\label{sec:combination}
		
		For any star $\star$ observed in survey $\mathcal{S}$, the preceding steps provide:
		\begin{itemize}
			\item the renormalised uncertainty
			$\sigma'_{{\rm RV},\mathcal{S}} = f_{\mathcal{S}}^{\rm RV}\,\sigma_{{\rm RV},\mathcal{S}}$
			is on a common error scale (Section~\ref{sec:error_norm});
			\item the duplicate group containing $\star$ has been identified
			(Section~\ref{sec:dedup});
			\item the survey RV has been brought onto the \textit{Gaia} reference
			frame by adding the polynomial zero-point correction
			(Section~\ref{sec:zp_cal}).
		\end{itemize}
		We now have everything we need to collapse all rows belonging to the same duplicate group into a single stellar record while retaining their survey provenance. The combination then proceeds in two weighted-average stages: first within each survey family, and then across families.
		
		\subsection{Calibrated survey measurements}\label{sec:rv_corrected}
		
		We first write down the calibrated survey RVs explicitly. Equation~(\ref{eq:delta_rv_pw}) defines $\Delta v_{\rm RV}=v_{{\rm RV},G}-v_{{\rm RV},\mathcal S}$, so the zero-point correction must be
		\emph{added} to the raw survey value to bring it onto the \textit{Gaia} frame:
		\begin{equation}
			\tilde{v}_{{\rm RV},\mathcal{S},i}
			\;=\;
			v_{{\rm RV},\mathcal{S},i}
			+
			\Delta v_{{\rm RV,ZP}}(x_i).
			\label{eq:rv_corrected}
		\end{equation}
		The calibrated pairwise residual is then
		\begin{equation}
			\widetilde{\Delta v}_{{\rm RV,pw},i}
			\;\equiv\;
			v_{{\rm RV},{\rm G},i} - \tilde{v}_{{\rm RV},\mathcal{S},i}
			\;=\;
			\Delta v_{{\rm RV,pw},i} - \Delta v_{{\rm RV,ZP}}(x_i),
			\label{eq:delta_rv_pw_tilde}
		\end{equation}
		This would vanish identically if the zero-point model captured the
		systematic offset perfectly.
		
		The uncertainty on the calibrated survey value combines two ingredients:
		the renormalised per-star measurement uncertainty and the propagated zero-point
		uncertainty $\sigma'_{{\rm RV,ZP},i}$. For the Gaia--survey residual used to diagnose the calibration, the calibrated pairwise variance is
		\begin{equation}
			\bigl(\tilde{\sigma}'_{{\rm RV,pw},i}\bigr)^{2}
			\;=\;
			\bigl(\sigma'_{{\rm RV,pw},i}\bigr)^{2}
			+
			\bigl(\sigma'_{{\rm RV,ZP},i}\bigr)^{2},
			\label{eq:sigma_rv_pw_tilde}
		\end{equation}
		whereas the uncertainty assigned to the calibrated survey measurement itself is
		\begin{equation}
			\bigl(\sigma'_{{\rm RV},\mathcal{S},i,{\rm cal}}\bigr)^{2}
			\;=\;
			\bigl(\sigma'_{{\rm RV},\mathcal{S},i}\bigr)^{2}
			+
			\bigl(\sigma'_{{\rm RV,ZP},i}\bigr)^{2}.
			\label{eq:sigma_rv_corrected}
		\end{equation}
		The Gaia measurement uncertainty enters the pairwise residual but is not added to the uncertainty of the survey measurement itself. Equations~\ref{eq:rv_corrected}--\ref{eq:sigma_rv_corrected}
		therefore fully define the calibrated radial velocity and its uncertainty for every
		star observed by the survey $\mathcal{S}$.
		
		\subsection{Within-family combination}\label{sec:rv_intra}
		
		If the same star has $n_{\mathcal{S},\star}$ calibrated measurements of
		$v'_{{\rm RV},\mathcal{S},j}$ (with $j = 1,\dots,n_{\mathcal{S},\star}$)
		within a single survey family $\mathcal{S}$, we combine them by
		inverse-variance weighting, with the weight of each measurement being simply
		\begin{equation}
			w_{\mathcal{S},j}
			\;=\;
			\frac{1}{\bigl(\sigma'_{{\rm RV},\mathcal{S},j,{\rm cal}}\bigr)^{2}},
			\label{eq:intra_weight}
		\end{equation}
		where $\sigma'_{{\rm RV},\mathcal{S},j,{\rm cal}}$ is the calibrated
		uncertainty from equation~(\ref{eq:sigma_rv_corrected}). The within-family
		weighted mean is
		\begin{equation}
			\bar{v}_{{\rm RV},\mathcal{S},\star}
			\;=\;
			\frac{\sum_{j=1}^{n_{\mathcal{S},\star}}
				w_{\mathcal{S},j}\, v'_{{\rm RV},\mathcal{S},j}}
			{\sum_{j=1}^{n_{\mathcal{S},\star}} w_{\mathcal{S},j}},
			\label{eq:intra_mean}
		\end{equation}
		with associated uncertainty
		\begin{equation}
			\sigma'_{{\rm RV},\mathcal{S},\star}
			\;=\;
			\biggl(\sum_{j=1}^{n_{\mathcal{S},\star}} w_{\mathcal{S},j}\biggr)^{-1/2}.
			\label{eq:intra_sigma}
		\end{equation}
		If a star has only one calibrated measurement within a given family, these
		expressions reduce trivially to the single calibrated value and its
		uncertainty.
		
		To suppress catastrophic outliers (e.g.\ a single bad spectrum among many
		good ones), we additionally perform a $3\sigma$ rejection step before the
		inverse-variance combination, where $\sigma$ here refers to the
		renormalised, calibrated uncertainty of each individual measurement.
		
		\subsection{Cross-family combination}\label{sec:rv_inter}
		
		After within-family combination, we now have  one calibrated, weighted-average RV per star for each
		survey. If this star is measured in $m_{\star}$
		distinct survey families, the remaining survey-level estimates are themselves combined by
		inverse-variance weighting. With per-survey weight
		\begin{equation}
			W_{\mathcal{S},\star}
			\;=\;
			\frac{1}{\bigl(\sigma'_{{\rm RV},\mathcal{S},\star}\bigr)^{2}},
			\label{eq:inter_weight}
		\end{equation}
		the final catalogue RV for the star is
		\begin{equation}
			v_{{\rm RV},\star}
			\;=\;
			\frac{\sum_{\mathcal{S}=1}^{m_{\star}}
				W_{\mathcal{S},\star}\,\bar{v}_{{\rm RV},\mathcal{S},\star}}
			{\sum_{\mathcal{S}=1}^{m_{\star}} W_{\mathcal{S},\star}},
			\label{eq:inter_mean}
		\end{equation}
		with the associated uncertainty
		\begin{equation}
			\sigma'_{{\rm RV},\star}
			\;=\;
			\biggl(\sum_{\mathcal{S}=1}^{m_{\star}} W_{\mathcal{S},\star}\biggr)^{-1/2}.
			\label{eq:inter_sigma}
		\end{equation}
		
		\subsection{Aggregation of non-RV columns}
		
		The combination procedure above is appropriate only for measurements that
		carry trustworthy uncertainties; in our case, these are the radial
		velocities and other kinematic quantities for which we have constructed
		calibrated $\sigma'$ values. However, for a given survey, there can be non-RV quantities that do not have an error reported for a given observed value; for a survey with duplicates, this means that a star can have multiple entries for a given non-RV quantity without associated errors. In such cases, the mean and median for each star, for that particular non-RV quantity, are kept in the final entry column.
		Let $x_1, \dots, x_n$ be the values
		contributed to each duplicate group of stars, i.e. all input rows identified as measurements of the same physical star. We can 
		compute the associated median absolute deviation given by
		\begin{equation}
			\mathcal{M}
			\;=\;
			\mathrm{median}\!\left(|x_i - \tilde{x}|\right),
			\label{eq:mad_def_section}
		\end{equation}
		where the median is $\tilde{x} \;=\; \mathrm{median}(x_i)$. Next, we estimate the
		Gaussian-equivalent scale via the standard factor
		\begin{equation}
			\hat{\sigma} \;=\; 1.4826\,\mathcal{M}.
			\label{eq:mad_gauss_equiv}
		\end{equation}
		Values lying further than $3\hat{\sigma}$ from $\tilde{x}$ are rejected, and the final aggregated
		value is the robust mean of the survivors \footnote{The factor $1.4826$ is the inverse of the $75$th-percentile of the
			standard normal distribution; it makes $\hat{\sigma}$ equal to the standard
			deviation of the data when those data are Gaussian.}. This protects the aggregated
		column against catastrophic outliers. This is how non-RV/non-distance numerical columns are collapsed during deduplication, thereby protecting the final aggregate against isolated bad values.
		
		For a few `special'  quantities, we deviate from this default. Astrometric positions are aggregated by a circular mean on the unit sphere rather than
		by direct arithmetic averaging of $\alpha$ and $\delta$. The spherical mean is defined here explicitly: map each $(\alpha,\delta)$ to the three-dimensional unit vector $(\cos\delta\cos\alpha,\cos\delta\sin\alpha,\sin\delta)$, average these vectors, renormalise the result, and convert it back to $(\alpha,\delta)$. This avoids the $0/360^\circ$ wrap discontinuity in right ascension while treating both sky coordinates consistently. For text metadata, we keep the distinct non-empty labels found among the rows of the same physical star. For list/array fields that cannot be meaningfully averaged, we keep the first valid non-empty value.
		
		\subsection{Improved precision from merging of radial velocities}
		
\begin{figure}[htbp]
			\centering
			\includegraphics[width=1.0\linewidth]{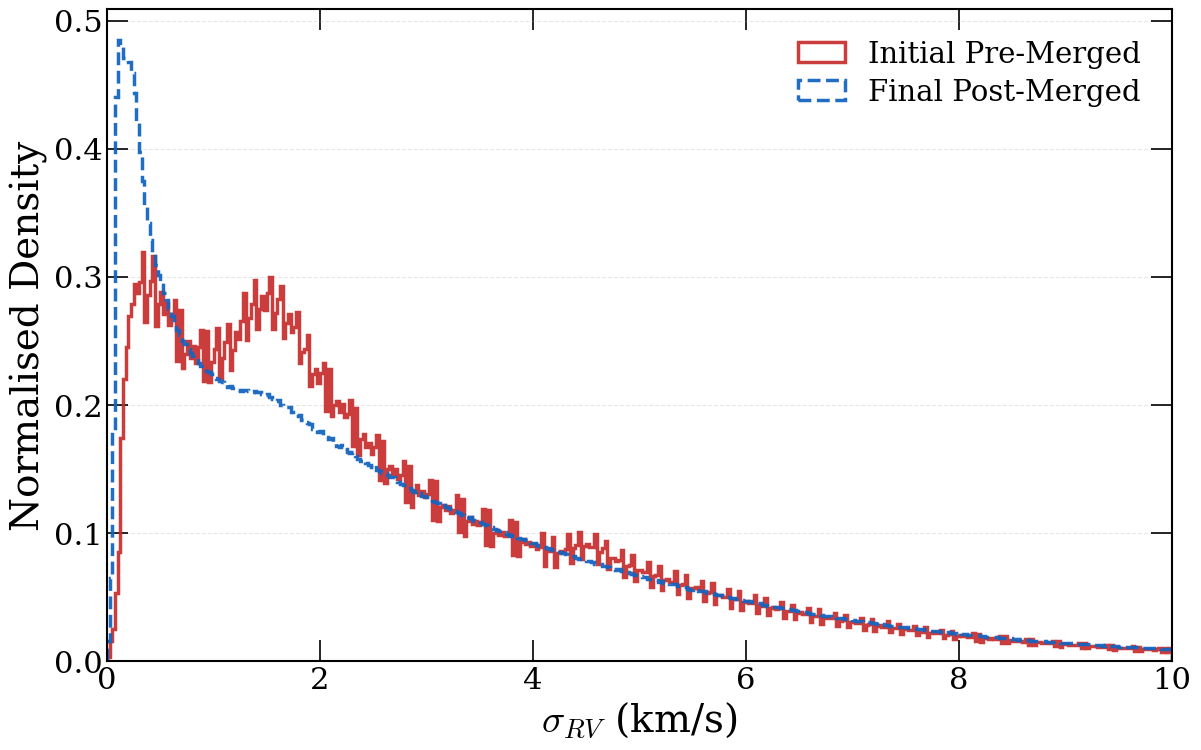}
			\caption{\small Distribution of the radial-velocity uncertainty before (red solid) and after (blue dashed) merging of duplicate measurements within and across survey families.}
			\label{fig:rv_merge_error}
		\end{figure}

		Figure~\ref{fig:rv_merge_error} compares the distribution of
		$\sigma_{\rm RV}$  before and after the merging that combines duplicate measurements of the same physical star within and across survey families as described above.
		The post-merged distribution is more strongly concentrated at low
		$\sigma_{\rm RV}$, reflecting the fact that combining several independent
		calibrated measurements of the same star yields a smaller 
		uncertainty than any single one of them.  The high-error tail is dominated by stars represented by only one usable RV measurement.
		
\begin{figure}[htbp]
			\centering
			\includegraphics[width=1\linewidth]{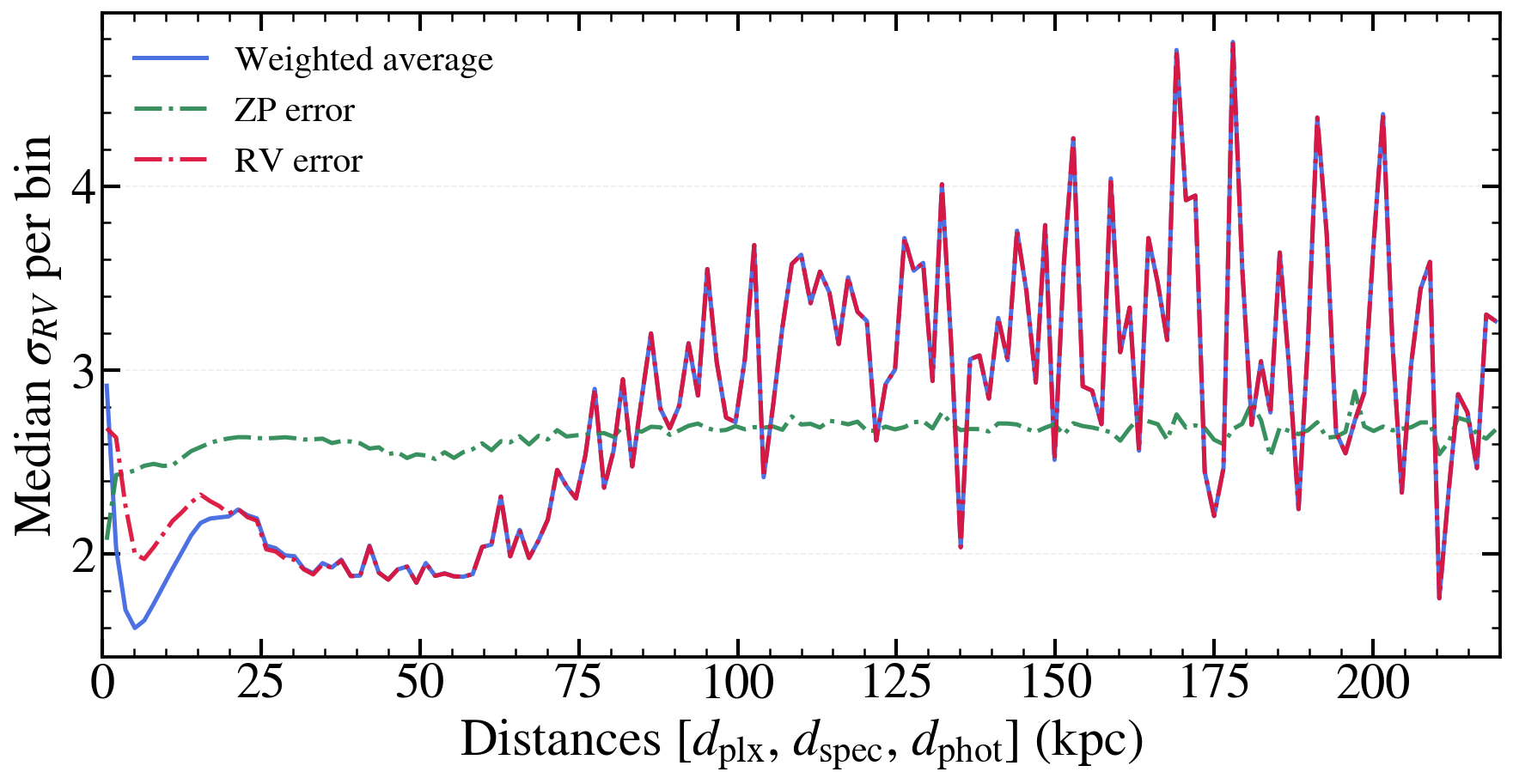}
			\caption{\small Binned-median RV uncertainty as a function of distance.
				The solid blue curve is the final catalogue weighted-average uncertainty;
				the dash-dotted green curve is the propagated zero-point contribution;
				the dash-dotted red curve is the raw quoted uncertainty.}
			\label{fig:ZP_error}
		\end{figure}

		Figure~\ref{fig:ZP_error} shows how the median catalogue-level RV
		uncertainty depends on heliocentric distance. The weighted-average and
		raw RV uncertainties track each other across most of the distance range,
		while the zero-point uncertainty term (the contribution from
		$\sigma'_{{\rm RV,ZP},i}$) remains comparatively smooth and nearly constant
		at $\sim 2.5$--$2.8\,\mathrm{km\,s^{-1}}$. Beyond a few tens of kpc, the
		median RV uncertainty rises and becomes increasingly noisy because the
		sample size in each distance bin drops sharply, and the catalogue is
		dominated by intrinsically bright distant tracers whose individual quoted
		errors are large.
		
		Below, we summarize the calibration--combination data flow that we performed. 
		The overall chain of logic, in Sections~\ref{sec:error_norm}--\ref{sec:combination},
		is as follows:
		\begin{enumerate}
			\item The per-survey factors $f_{{\rm DUP},\mathcal{S}}$ and
			$f_{{\rm TCH},\mathcal{S}}$ define the renormalised uncertainties
			$\sigma'_{{\rm RV},\mathcal{S},i}$.
			\item Duplicates are identified by the graph-based connected-component
			algorithm.
			\item The pairwise residuals
			$\Delta v_{{\rm RV,pw},i} = v_{{\rm RV},{\rm G},i} - v_{{\rm RV},\mathcal{S},i}$
			are fitted to obtain the additive zero-point correction
			$\Delta v_{{\rm RV,ZP}}(x_i)$, with propagated uncertainty
			$\sigma'_{{\rm RV,ZP},i}$.
			\item Each duplicate group is collapsed by inverse-variance weighting,
			first within each survey family and then across families.
		\end{enumerate}
		The end product is a survey-agnostic final catalogue in which each star
		appears exactly once, with a single combined RV $v_{{\rm RV},\star}$, a
		single combined uncertainty $\sigma'_{{\rm RV},\star}$, and an aggregated
		set of all other catalogue columns derived from every survey that observed
		the star.

		\section{External Validation: Cross-Match and Adaptive Membership Analysis}\label{sec:validation}
		\label{sec:membership}
		Throughout this section, the \emph{master catalogue} means the $32.55$ million-star unified catalogue constructed in the preceding sections, whereas a \emph{reference catalogue} is an external literature compilation that supplies known system centres, published members, or system-level parameters. The validation follows a fixed sequence: (1) select an external benchmark system and its literature information; (2) propagate the reference coordinates from their native epoch to J2016.0 when required; (3) cross-match those positions to stars in the master catalogue; (4) evaluate member and field likelihoods using measurements stored in the master catalogue, while externally compiled system parameters supply the benchmark/prior information; (5) iterate an expectation--maximisation (EM) model to obtain posterior membership probabilities and fitted system parameters; and (6) compare the recovered member distributions with the external benchmark values. Open-cluster distances are a special Gaia-anchored benchmark and are identified explicitly below.
		
		Validating a catalogue assembled from heterogeneous surveys requires checking whether it can recover known stellar systems without first selecting stars on the basis of the same final catalogue quantities. We therefore use four benchmark classes - globular clusters, open clusters, dwarf galaxies, and the Sagittarius stream and assign posterior membership probabilities to the master-catalogue cross-matches with an EM mixture model. The reference compilations provide the system definition and external benchmark parameters; the star-by-star astrometry, radial velocities, distances, photometry, and atmospheric parameters entering the likelihood are taken from the master catalogue. The discriminating power of the available likelihood terms changes naturally with distance: proper motion is particularly informative nearby, distance information becomes increasingly important at intermediate distances, and radial velocity/chemistry/gravity become essential in the distant halo as geometric parallaxes lose precision.
		
		For each source in a reference catalogue at coordinates $(\alpha_s,\delta_s)$ and epoch
		$J2000.0$,\footnote{Since the master catalogue stores Gaia-epoch positions at $J2016.0$,
			reference-catalogue coordinates listed at $J2000.0$ are propagated to the \textit{Gaia} reference epoch. J2000.0 is used only for reference catalogues whose tabulated coordinates are explicitly given at that epoch. More generally the propagation interval is $\Delta t=2016.0-t_{\rm ref}$ for the native reference epoch $t_{\rm ref}$; catalogues already at J2016.0 require no propagation.} are propagated to $J2016.0$ with \textsc{Astropy}
		\texttt{SkyCoord.apply\_space\_motion} when full proper-motion information is present, and otherwise with
		the linear approximation $\alpha_{J16}\simeq \alpha_{J2000}+\mu_{\alpha *}\Delta t/\cos\delta_{J2000}$
		and $\delta_{J16}\simeq \delta_{J2000}+\mu_\delta\Delta t$, with $\Delta t=16.0$~yr.
		We query the master catalogue using a $k$-d tree constructed on the three-dimensional unit-sphere Cartesian embedding of
		the sky positions.\footnote{The adopted matching tolerance is
			$\theta_\mathrm{match}=1^{\prime\prime}$, corresponding to the chord distance
			\begin{equation}
				d_\mathrm{chord}
				= 2\sin\!\left(\frac{\theta_\mathrm{match}}{2}\right)
				\approx 4.848 \times 10^{-6},
				\label{eq:chord_match}
			\end{equation}
			on the unit sphere. At high-latitude stellar surface densities of $\leq 500$~deg$^{-2}$, this implies a
			false-match probability
			$p_\mathrm{false} \approx \pi \theta_\mathrm{match}^2 \rho \lesssim 0.1\%$
			per query star. }
		\label{subsec:quality}
		
		Note that no additional hard cuts are imposed at the validation stage. Instead, all cross-matched stars are
		retained, and the influence of problematic measurements is  down-weighted through conservative error inflation.
		This preserves potentially useful photometric or spectroscopic information while preventing low-quality
		astrometric or radial-velocity measurements from dominating the likelihood.
		
		For stars with $\mathrm{RUWE} > 1.4$, the proper-motion uncertainties are inflated by an empirically
		chosen factor $f_\mathrm{RUWE}=3.0$:
		\begin{equation}
			\begin{aligned}
				\sigma_{\mu_\alpha}^\mathrm{adj}
				&= f_\mathrm{RUWE}\,\sigma_{\mu_\alpha}, \\
				\sigma_{\mu_\delta}^\mathrm{adj}
				&= f_\mathrm{RUWE}\,\sigma_{\mu_\delta}, \\
				f_\mathrm{RUWE}
				&=
				\begin{cases}
					3.0 & \mathrm{RUWE} > 1.4 \\
					1.0 & \text{otherwise.}
				\end{cases}
			\end{aligned}
			\label{eq:ruwe_inflate}
		\end{equation}
		where the threshold $\mathrm{RUWE}>1.4$ identifies stars whose astrometric residuals exceed the
		expectations of the Gaia single-star model. In this way, proper-motion errors that may underestimate the
		true external scatter are conservatively down-weighted rather than discarded, allowing genuine members to
		be retained while avoiding a hard RUWE cut that could remove them, especially in crowded regions.
		An analogous treatment is applied to low-significance radial velocities. Stars with $\mathrm{RV\;S/N}<3$
		receive an empirically chosen fivefold error inflation:
		\begin{equation}
			\sigma_v^\mathrm{adj} = f_\mathrm{RV}\,\sigma_v,
			\qquad
			f_\mathrm{RV}
			= \begin{cases}
				5.0 & \mathrm{RV\;S/N} < 3 \\
				1.0 & \text{otherwise,}
			\end{cases}
			\label{eq:rvsn_inflate}
		\end{equation}
		where $\sigma_v$ is the calibrated weighted-average radial-velocity uncertainty in the master catalogue.
		This renders such measurements effectively non-informative in the EM updates while preserving the
		remaining astrometric and photometric constraints. In practice, this approach retains more matched stars
		than equivalent hard cuts.

		\subsection{The Seven-Term Bayesian EM Likelihood}
		\label{subsec:em_likelihood}
		
		Each matched star $i$ is modelled as a draw from a two-component mixture consisting of a member
		population and a foreground/background field population from the master catalogue. The member likelihood can be estimated as the
		product of seven terms,
		\begin{equation}
			\mathcal{L}_i^\mathrm{mem}
			= \mathcal{L}_i^\mathrm{sp}
			\times \mathcal{L}_i^\mathrm{pm}
			\times \mathcal{L}_i^\mathrm{rv}
			\times \mathcal{L}_i^\mathrm{dist}
			\times \mathcal{L}_i^\mathrm{cmd}
			\times \mathcal{L}_i^{[\mathrm{Fe/H}]}
			\times \mathcal{L}_i^{\log g}.
			\label{eq:full_likelihood}
		\end{equation}
		An analogous product $\mathcal{L}_i^\mathrm{field}$ is evaluated for the field component. The mixture coefficient
		$\eta \in [0.01,0.99]$ gives the member fraction within the matched sample (details in Appendix A). 
		
		For globular clusters and dwarf galaxies, the projected member distribution is modelled by an elliptical
		Plummer profile. For a star, the angular offset $(\xi,\eta_\mathrm{sky})$ from the system centre, obtained
		through the gnomonic projection \footnote{\begin{align}
				\xi &= \frac{\cos\delta\,\sin(\alpha - \alpha_0)}
				{\sin\delta_0\,\sin\delta
					+ \cos\delta_0\,\cos\delta\,\cos(\alpha-\alpha_0)},
				\label{eq:xi} \\
				\eta_\mathrm{sky} &= \frac{\cos\delta_0\,\sin\delta
					- \sin\delta_0\,\cos\delta\,\cos(\alpha-\alpha_0)}
				{\sin\delta_0\,\sin\delta
					+ \cos\delta_0\,\cos\delta\,\cos(\alpha-\alpha_0)},
				\label{eq:eta_sky}
		\end{align}}.
		The elliptical radius is
		$r_\mathrm{ell}=\sqrt{(u/q)^2+v^2}$,
		where $(u,v)$ are the coordinates rotated by the position angle $\theta_\mathrm{PA}$ and $q=1-\epsilon$
		is the axis ratio. The member spatial likelihood is
		\begin{equation}
			\mathcal{L}_i^\mathrm{sp,mem}
			= \frac{a^2 + R_\mathrm{max}^2}{\pi a^2 R_\mathrm{max}^2 q}
			\cdot \frac{1}{[1 + (r_{\mathrm{ell},i}/a)^2]^2},
			\qquad a = r_h,
			\label{eq:plummer}
		\end{equation}
		where $r_h$ is the projected half-light radius; for the projected Plummer profile written above, its scale radius is therefore $a=r_h$. $R_\mathrm{max}$ is the maximum angular offset within
		the matched sample. The field spatial term is uniform, given by $
		\mathcal{L}_i^\mathrm{sp,field} = (\pi R_\mathrm{max}^2 q)^{-1}.
		$
		For open clusters and Sagittarius stream bins, a single Plummer profile is not an appropriate spatial model: open clusters span a wide range of angular morphologies and the Sagittarius stream is intrinsically elongated. Setting both spatial terms to unity deliberately makes position non-discriminating, so membership for these classes is determined by the remaining kinematic, distance, CMD, and chemical information rather than by a mis-specified spatial prior.
		
		The intrinsic proper-motion distribution of each component is described by a bivariate Gaussian with mean
		$\bm{\mu}_k=(\mu_{\alpha *,k},\mu_{\delta,k})$
		and covariance $\bm{\Sigma}_k$, where $k\in\{\mathrm{mem},\mathrm{field}\}$. The cross-match aperture contains both genuine system members and unrelated foreground/background stars. The field component models those contaminants, allowing the EM algorithm to compare a member hypothesis with an explicit non-member hypothesis rather than forcing every matched source to be a member. For star $i$, the effective
		covariance is
		\begin{equation}
			\bm{\Sigma}_k^{\mathrm{eff},i}
			= \bm{\Sigma}_k + \mathbf{C}_i + \epsilon_\mathrm{reg}\,\mathbf{I},
			\mathbf{C}_i
			= \begin{pmatrix}
				(\sigma_{\mu_\alpha}^{\mathrm{adj}})^2
				& \rho_i\,\sigma_{\mu_\alpha}^\mathrm{adj}\,
				\sigma_{\mu_\delta}^\mathrm{adj} \\
				\rho_i\,\sigma_{\mu_\alpha}^\mathrm{adj}\,
				\sigma_{\mu_\delta}^\mathrm{adj}
				& (\sigma_{\mu_\delta}^\mathrm{adj})^2
			\end{pmatrix},
			\label{eq:pm_cov}
		\end{equation}
		where $\rho_i$ is the Gaia proper-motion correlation coefficient and
		$\epsilon_\mathrm{reg}=10^{-6}$~$(\mathrm{mas\,yr}^{-1})^2$ is a regularisation term. The corresponding
		likelihood is
		\begin{equation}
			\mathcal{L}_i^{\mathrm{pm},k}
			= \frac{1}{2\pi\sqrt{|\bm{\Sigma}_k^{\mathrm{eff},i}|}}
			\exp\!\left(
			-\tfrac{1}{2}\,\Delta\bm{\mu}_i^T
			\bigl(\bm{\Sigma}_k^{\mathrm{eff},i}\bigr)^{-1}
			\Delta\bm{\mu}_i
			\right),
			\label{eq:pm_likelihood}
		\end{equation}
		where $\Delta\bm{\mu}_i = \bm{\mu}_i - \bm{\mu}_k,$
		which is evaluated in Mahalanobis form for numerical stability.
		
		The member radial-velocity distribution is represented by a Gaussian centred on the systemic velocity
		$v_\mathrm{sys}$ with total variance equal to the quadrature sum of the intrinsic dispersion
		$\sigma_\mathrm{int}$ and the adjusted measurement uncertainty $\sigma_{v,i}^\mathrm{adj}$:
		\begin{equation}
			\mathcal{L}_i^{\mathrm{rv,mem}}
			= \frac{1}{\sqrt{2\pi\bigl(\sigma_\mathrm{int}^2
					+ (\sigma_{v,i}^\mathrm{adj})^2\bigr)}}
			\exp\!\left(
			-\frac{(v_{r,i} - v_\mathrm{sys})^2}
			{2[\sigma_\mathrm{int}^2 + (\sigma_{v,i}^\mathrm{adj})^2]}
			\right).
			\label{eq:rv_mem}
		\end{equation}
		The intrinsic member
		dispersion is updated by maximising
		\begin{equation}
			\begin{split}
				\mathcal{L}_\sigma
				&=
				-\frac{1}{2}\sum_{i:\,v_{r,i}\,\mathrm{finite}} w_i
				\left[
				\log(2\pi s_i^2)
				+ \frac{(v_{r,i} - v_\mathrm{sys})^2}{s_i^2}
				\right], \\
				s_i^2
				&= \sigma_\mathrm{int}^2 + (\sigma_{v,i}^\mathrm{adj})^2,
			\end{split}
			\label{eq:mle_sigma}
		\end{equation}
		over the bounded interval
		$[\sigma_\mathrm{int,floor},200]$~km\,s$^{-1}$ with
		$\sigma_\mathrm{int,floor}=0.5$~km\,s$^{-1}$. When too few stars have finite radial velocities, the
		dispersion defaults to
		$\sigma_\mathrm{int,init}=10$~km\,s$^{-1}$. The field radial-velocity distribution is also Gaussian, with mean $v_\mathrm{field}$ and dispersion
		$\sigma_\mathrm{field}$ estimated iteratively from the non-member component. 
		
		Whenever an external reference distance $d_\mathrm{ref}$ is available, the member distance likelihood is
		\begin{equation}
			\begin{split}
				\mathcal{L}_i^{\mathrm{dist,mem}}
				&= \mathcal{N}\!\left(
				d_i;\, d_\mathrm{ref},\,
				\sqrt{\sigma_{d_\mathrm{ref}}^2 + \sigma_{d,i}^2}
				\right), \\
				\sigma_{d,i}
				&= \max\!\left(
				\sigma_{d,i}^\mathrm{cat},\, 0.5~\mathrm{kpc}
				\right).
			\end{split}
			\label{eq:dist_mem}
		\end{equation}
		where the $0.5$~kpc floor prevents unrealistically narrow distance posteriors at very small heliocentric
		distances. The field distance term is taken to be a broad Gaussian centred on $10$~kpc with
		$\sigma_\mathrm{field,dist}=\max(0.5\,d_\mathrm{ref},20~\mathrm{kpc})$.
		If no reliable reference distance is available, the distance term is set to unity for both components.
		
		A kernel-density estimate of the reference member sequence is constructed in the de-reddened
		colour--magnitude plane
		$(M_G,\,G_\mathrm{BP}-G_\mathrm{RP})$
		using a fixed bandwidth of $h=0.15$~mag, provided that at least 15 reference members have valid
		photometry. Absolute magnitudes are computed using
		$M_{G,i}=G_i-A_{G,i}-5\log_{10}(d_\mathrm{ref}/10\,\mathrm{pc})$.
		The member CMD likelihood, $\mathcal{L}_i^\mathrm{cmd, mem}$, is the KDE value at the de-reddened position of each matched star. 
		We use the standard term ``de-reddened'' consistently here, meaning that the photometry has been corrected for interstellar extinction/reddening before evaluating the CMD likelihood.
		The field CMD likelihood is derived from an annular KDE built from stars at projected radii $r>5\,r_h$. If fewer than 15 reference stars are available, both CMD terms are set to unity.
		
		When a reference metallicity prior is available, the member metallicity likelihood is
		\begin{equation}
			\begin{split}
				\mathcal{L}_i^{[\mathrm{Fe/H}],\mathrm{mem}}
				&= \mathcal{N}\!\left(
				[\mathrm{Fe/H}]_i;\,
				[\mathrm{Fe/H}]_\mathrm{sys},\,
				\sigma_\mathrm{[Fe/H],eff}
				\right), \\
				\sigma_\mathrm{[Fe/H],eff}
				&= \max\!\left(
				\sigma_\mathrm{[Fe/H]},\,
				0.05~\mathrm{dex}
				\right).
			\end{split}
			\label{eq:feh_mem}
		\end{equation}
		where $[\mathrm{Fe/H}]_\mathrm{sys}$ and $\sigma_\mathrm{[Fe/H]}$ are taken from the external reference
		catalogue. The field metallicity likelihood is modelled as
		\[
		\mathcal{L}_i^{[\mathrm{Fe/H}],\mathrm{field}}
		= \mathcal{N}([\mathrm{Fe/H}]_i;\,-1.0,\,0.8),
		\]
		which approximates the broad halo-field metallicity distribution.
		
		For globular clusters and dwarf galaxies (but not for open clusters or Sagittarius stream bins), a soft giant-branch requirement is imposed through
		\begin{equation}
			\begin{split}
				\mathcal{L}_i^{\log g,\mathrm{mem}}
				&= \exp\!\left[
				-\frac{\max(\log g_i-\log g_\mathrm{thr},\,0)^2}
				{2\,\sigma_{\log g}^2}
				\right], \\
				\log g_\mathrm{thr} &= 3.5, \\
				\sigma_{\log g} &= 0.5~\mathrm{dex}.
			\end{split}
			\label{eq:logg_mem}
		\end{equation}
		so that stars below the giant threshold receive unit weight while probable dwarf contaminants are
		progressively suppressed. The field term is represented by a broad mixture centred on the main sequence,
		\[
		\mathcal{L}_i^{\log g,\mathrm{field}}
		= 0.3 + 0.7\,\mathcal{N}(\log g_i;\,4.2,\,0.8).
		\]
		Both terms are set to unity when no $\log g$ measurement is available.

		\begin{figure*}[!t]
			\centering
			\includegraphics[width=\linewidth]{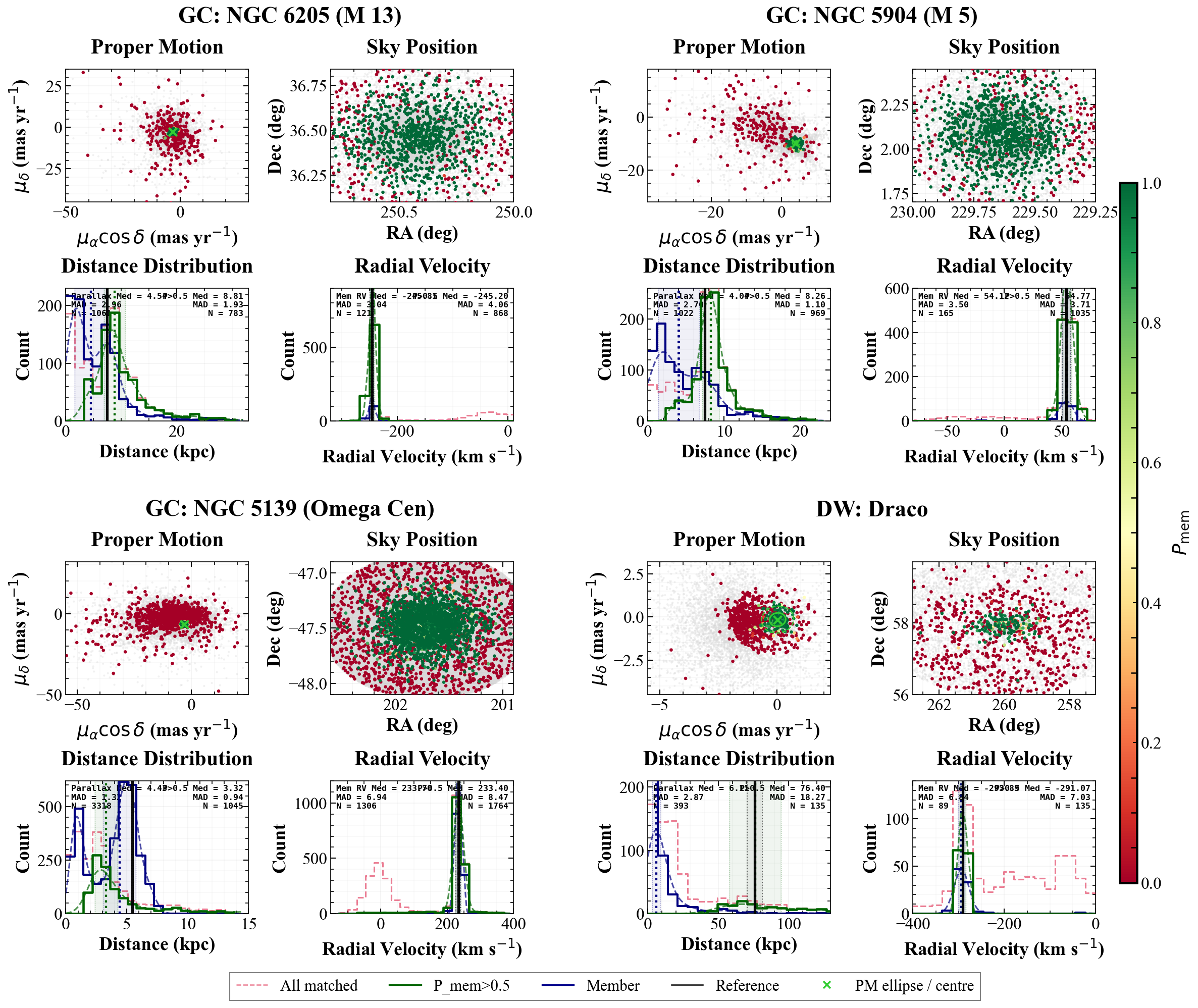}
			\caption{\small Representative validation cases spanning the nearby, intermediate-distance, and
				distant-halo regimes. Each $2\times 2$ block shows proper motion, sky position, distance distribution,
				and radial velocity, with the upper panels colour-coded by membership probability $P_{\rm mem}$. Black
				lines and grey bands denote the adopted external reference values and uncertainties.}
			\label{fig:specific_validation}
		\end{figure*}
		
		(i) \sf{Globular Clusters - }The 189 globular clusters, used for cross-validation, are processed with the full seven-term likelihood of Eq.~\eqref{eq:full_likelihood}, including the elliptical Plummer spatial model defined by the literature half-light radius, ellipticity, and position angle. Reference distances are taken from a compilation of 162 unique literature distance priors (\citet{baumgardt2021}). The selection criterion is defined uniformly throughout this
		section as $N_{P>0.5}\geq 10$, where $N_{P>0.5}$ denotes the number of stars with posterior membership
		probability $P_{\rm mem}>0.5$; this yields 49 globular clusters. The five top member-rich systems after
		quality adjustment  are NGC\,5139 ($\omega$\,Cen; 1,822 members in the $J2016$ implementation), NGC\,104
		(47\,Tuc; 1,347), NGC\,5904 (M\,5; 1,059), NGC\,3201 (985), and NGC\,6205 (M\,13; 886).\footnote{Unless
			stated otherwise, all member counts quoted in this section refer to the post-quality-adjustment matched
			sample used by the EM algorithm. This avoids ambiguity with larger raw cross-match counts quoted
			elsewhere in the paper.} The down-weighting rules applied before the EM fit are those defined above, including the RUWE-based proper-motion uncertainty inflation and the low-significance RV uncertainty inflation.
		
		(ii) \sf{Open Clusters - }The 1,481 open clusters are analysed with a five-term likelihood using the spatial, proper-motion,
		radial-velocity, distance, and CMD terms. The metallicity and $\log g$ terms are omitted because the wide
		age range and chemical diversity of the open-cluster population make a single-valued prior inappropriate
		in the present framework. For the open-cluster benchmark only, reference distances are estimated from the \textit{Gaia} parallax-averaged distances of stars in the external literature member lists,

		$d_\mathrm{ref}=\langle 1/\varpi\rangle$
		over stars with $\varpi>0$. With the same selection criterion,
		$N_{P>0.5}\geq 10$,
		the open-cluster selected set contains 785 systems.
		
		(iii) \sf{Sagittarius stream - }The Sagittarius stream is divided into heliocentric-distance bins spanning $15$--$80$~kpc in steps of
		$5$~kpc. Of the 13 bins, 9 contain at least 5 reference members and are processed without the Plummer
		spatial term but with the radial-velocity, distance, metallicity, and $\log g$ terms active. For each
		bin, the reference distance is the bin centre,
		$d_\mathrm{ref}=(d_\mathrm{lo}+d_\mathrm{hi})/2$,
		with uncertainty $\sigma_{d_\mathrm{ref}}=2.5$~kpc. All 9 populated bins converge within at most 4 EM
		iterations and typically return $\eta \approx 0.99$, consistent with the high purity of the input
		Sagittarius sample. The richest bin is the $25$--$30$~kpc interval, which contains 893 high-probability
		members. The $55$--$60$~kpc bin yields zero cross-matches, so the Sagittarius contribution is reduced to
		8 bins with a combined total of 2,669 high-probability members. The membership counts are reproduced exactly
		between the $J2016$ and $J2000$ implementations, indicating that epoch propagation introduces no
		measurable bias for the Sagittarius validation set.
		
		(iv) \sf{Dwarf Galaxies - }At larger heliocentric distances, the 52 dwarf galaxies constitute the most demanding validation class because their true parallaxes,
		$\varpi \approx d_\mathrm{ref}^{-1}$,
		are far smaller than the Gaia parallax uncertainty floor,
		$\sigma_\varpi \sim 0.1$--$0.5$~mas.
		To suppress any Milky Way foreground contamination before the EM stage, we reject stars satisfying
		$ \varpi - 3\sigma_\varpi > 0.10~\mathrm{mas}$, 
		which corresponds to a maximum heliocentric distance of approximately $10$~kpc. This removes nearby disc and
		inner-halo stars that would otherwise bias the mixture model toward a spurious foreground solution. The
		systemic velocity, intrinsic dispersion, and metallicity priors are taken from the reference dwarf-galaxy
		compilation. The numerical priors are applied to the \emph{member component of the EM likelihood for master-catalogue stars}; their parameter values are taken from the external reference system. They are not re-estimated from the same master-catalogue measurements before the validation fit.
		
		Of the 52 dwarf galaxies, 25 produce no cross-matches; these are mostly ultra-faint systems with $M_V \gtrsim -5$
		whose stellar populations fall below the effective magnitude limits of the contributing surveys. Only 3
		systems satisfy this $N_{P>0.5}\ge10$ minimum-member selection criterion: Draco
		($N_{P>0.5}=139$, $\eta=0.176$, 172 foreground stars removed),
		Sextans
		($N_{P>0.5}=133$, $\eta=0.150$, 132 foreground stars removed),
		and Ursa Major\,II
		($N_{P>0.5}=11$, $\eta=0.063$, 65 foreground stars removed).
		Sculptor
		($N_\mathrm{match}=28$)
		survives the parallax foreground cut but converges to the lower bound
		$\eta=0.010$ with $N_{P>0.5}=0$, indicating that the available matched sample is too small for a stable
		decomposition. Ursa Minor and Willman\,1 fail the post-cut minimum-sample requirement. Five additional
		systems, Bootes\,III, Canes Venatici\,I, Coma Berenices, Canes Venatici\,II, and Hercules complete the EM
		procedure but return $N_{P>0.5}\leq 8$ and therefore fall below this threshold. These, however, are
		retained in our final released membership catalogue with quality flags.
		
		\begin{table*}[htbp]
			\centering
			
			\captionsetup{width=\textwidth}
			\caption{Best-agreement spectrophotometric distance validation subset. For each object, the epoch with the smaller relative distance offset $|d_{\rm match}-d_{\rm ref}|/d_{\rm ref}$ is retained. $N_{\rm match}$ is the number of master-catalogue stars cross-matched to the benchmark system, and $d_{\rm match}$ is the median spectrophotometric distance of the $P_{\rm mem}>0.5$ subsample with stored finite distances. The rows shown here are the systems selected by the explicit numerical support cuts stated below and then ranked by relative distance offset within each benchmark class, using minimum-support cuts of $N_{\rm hi}\geq20$ for open clusters, $N_{\rm hi}\geq15$ for globular clusters, $N_{\rm hi}\geq5$ for dwarf galaxies, and $N_{\rm hi}\geq20$ for Sagittarius bins.}

			\label{tab:dist_validation_best_subset}
			\setlength{\tabcolsep}{4pt}
			\renewcommand{\arraystretch}{1.15}
			\begin{tabular*}{\textwidth}{@{\extracolsep{\fill}}
					l @{\hspace{3pt}} r @{\hspace{3pt}} c @{\hspace{3pt}} c
					@{\hspace{10pt}}
					l @{\hspace{3pt}} r @{\hspace{3pt}} c @{\hspace{3pt}} c
				}
				\hline\hline
				\noalign{\smallskip}
				\multicolumn{8}{c}{\textsc{Open Clusters}} \\
				\multicolumn{8}{c}{\footnotesize $d_{\rm ref}$ source: mean inverse-parallax distances from \textit{Gaia} parallaxes of stars in the external open-cluster member lists.} \\
				\noalign{\smallskip}
				\hline
				\noalign{\smallskip}
				Name & $N_{\rm match}$ & $d_{\rm ref}$\,(kpc) & $d_{\rm match}$\,(kpc) & Name & $N_{\rm match}$ & $d_{\rm ref}$\,(kpc) & $d_{\rm match}$\,(kpc) \\
				\noalign{\smallskip}
				\hline
				\noalign{\smallskip}
				Trumpler\,29 & 76 & $1.549^{+0.500}_{-0.500}$ & $1.547^{+0.252}_{-0.321}$ & Alessi\,9 & 65 & $0.208^{+0.500}_{-0.500}$ & $0.207^{+0.013}_{-0.013}$ \\
				IC\,1434 & 90 & $3.540^{+0.664}_{-0.664}$ & $3.535^{+1.624}_{-0.435}$ & Platais\,9 & 25 & $0.185^{+0.500}_{-0.500}$ & $0.184^{+0.012}_{-0.019}$ \\
				ASCC\,6 & 48 & $1.644^{+0.500}_{-0.500}$ & $1.639^{+0.150}_{-0.269}$ & Tombaugh\,1 & 80 & $2.631^{+0.500}_{-0.500}$ & $2.616^{+0.976}_{-0.444}$ \\
				NGC\,7245 & 65 & $3.636^{+0.970}_{-0.970}$ & $3.648^{+2.670}_{-0.937}$ & Trumpler\,22 & 48 & $2.657^{+0.500}_{-0.500}$ & $2.679^{+2.294}_{-0.649}$ \\
				\noalign{\smallskip}
				\hline
				\noalign{\smallskip}
				\multicolumn{8}{c}{\textsc{Globular Clusters}} \\
				\multicolumn{8}{c}{\footnotesize $d_{\rm ref}$ source: literature/mean cluster distances from the adopted globular-cluster compilation.} \\
				\noalign{\smallskip}
				\hline
				\noalign{\smallskip}
				Name & $N_{\rm match}$ & $d_{\rm ref}$\,(kpc) & $d_{\rm match}$\,(kpc) & Name & $N_{\rm match}$ & $d_{\rm ref}$\,(kpc) & $d_{\rm match}$\,(kpc) \\
				\noalign{\smallskip}
				\hline
				\noalign{\smallskip}
				NGC\,6205 (M\,13) & 1258 & $7.427^{+0.079}_{-0.078}$ & $7.734^{+4.755}_{-6.265}$ & NGC\,5634 & 67 & $25.990^{+0.630}_{-0.615}$ & $20.714^{+18.629}_{-16.111}$ \\
				NGC\,5904 (M\,5) & 1320 & $7.471^{+0.066}_{-0.065}$ & $7.853^{+2.438}_{-4.676}$ & NGC\,104 (47\,Tuc) & 1739 & $4.512^{+0.040}_{-0.039}$ & $3.371^{+2.349}_{-2.374}$ \\
				NGC\,5053 & 172 & $17.515^{+0.235}_{-0.232}$ & $14.706^{+4.592}_{-13.808}$ & NGC\,5272 (M\,3) & 523 & $10.167^{+0.085}_{-0.084}$ & $7.333^{+7.912}_{-6.602}$ \\
				\noalign{\smallskip}
				\hline
				\noalign{\smallskip}
				\multicolumn{4}{c}{\textsc{Dwarf Galaxies}} & \multicolumn{4}{c}{\textsc{Sagittarius Stream Bins}} \\
				\multicolumn{4}{c}{\footnotesize $d_{\rm ref}$: external dwarf-galaxy reference distances.} & \multicolumn{4}{c}{\footnotesize $d_{\rm ref}$: adopted 5-kpc bin centres.} \\
				\noalign{\smallskip}
				\hline
				\noalign{\smallskip}
				Name & $N_{\rm match}$ & $d_{\rm ref}$\,(kpc) & $d_{\rm match}$\,(kpc) & Bin & $N_{\rm match}$ & $d_{\rm ref}$\,(kpc) & $d_{\rm match}$\,(kpc) \\
				\noalign{\smallskip}
				\hline
				\noalign{\smallskip}
				Draco & 944 & $75.800^{+5.400}_{-5.400}$ & $76.396^{+40.189}_{-21.659}$ & SGR 35--40\,kpc & 126 & $37.500^{+2.500}_{-2.500}$ & $35.134^{+8.312}_{-8.838}$ \\
				Bootes III & 73 & $46.500^{+2.000}_{-2.000}$ & $44.900^{+9.096}_{-4.563}$ & SGR 45--50\,kpc & 372 & $47.500^{+2.500}_{-2.500}$ & $44.487^{+16.193}_{-12.224}$ \\
				Sextans & 981 & $92.500^{+2.500}_{-2.500}$ & $84.084^{+47.433}_{-21.978}$ & SGR 40--45\,kpc & 179 & $42.500^{+2.500}_{-2.500}$ & $39.798^{+10.017}_{-9.318}$ \\
				Canes Venatici I & 24 & $210.000^{+6.000}_{-6.000}$ & $179.308^{+244.595}_{-38.016}$ & SGR 50--55\,kpc & 255 & $52.500^{+2.500}_{-2.500}$ & $47.085^{+16.024}_{-10.156}$ \\
				\noalign{\smallskip}
				\hline
			\end{tabular*}
		\end{table*}

		\begin{figure*}[htbp]
			\centering
			\includegraphics[width=1\linewidth]{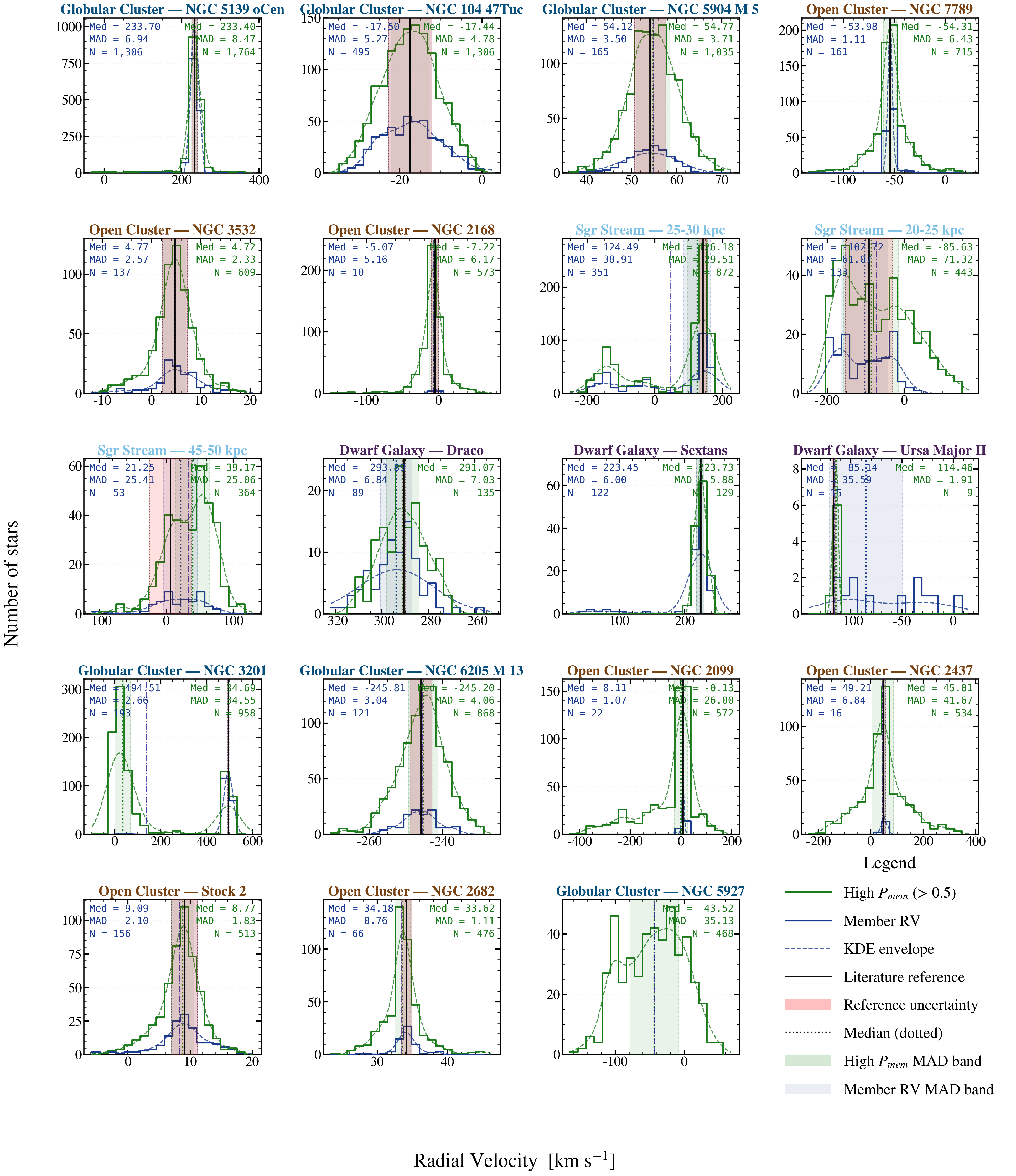}
			\caption{\small Examples of radial-velocity distributions for a selection of individual stellar systems. In each panel, the blue histogram denotes
				member-catalogue radial velocities and the green histogram denotes high-membership ($P_{\rm mem}>0.5$)
				master-catalogue radial velocities; dashed curves show the corresponding KDE envelopes. The black
				vertical line and shaded band mark the literature reference RV and its uncertainty, while dotted lines
				and lightly shaded bands indicate the median and MAD range of each distribution. The annotation boxes
				list the median, MAD, and sample size for the member-catalogue (blue) and high-$P_{\rm mem}$ (green)
				samples.}
			\label{fig:ext_valid_rv}
		\end{figure*}
		
		\begin{figure*}[htbp]
			\centering
			\includegraphics[width=1\linewidth]{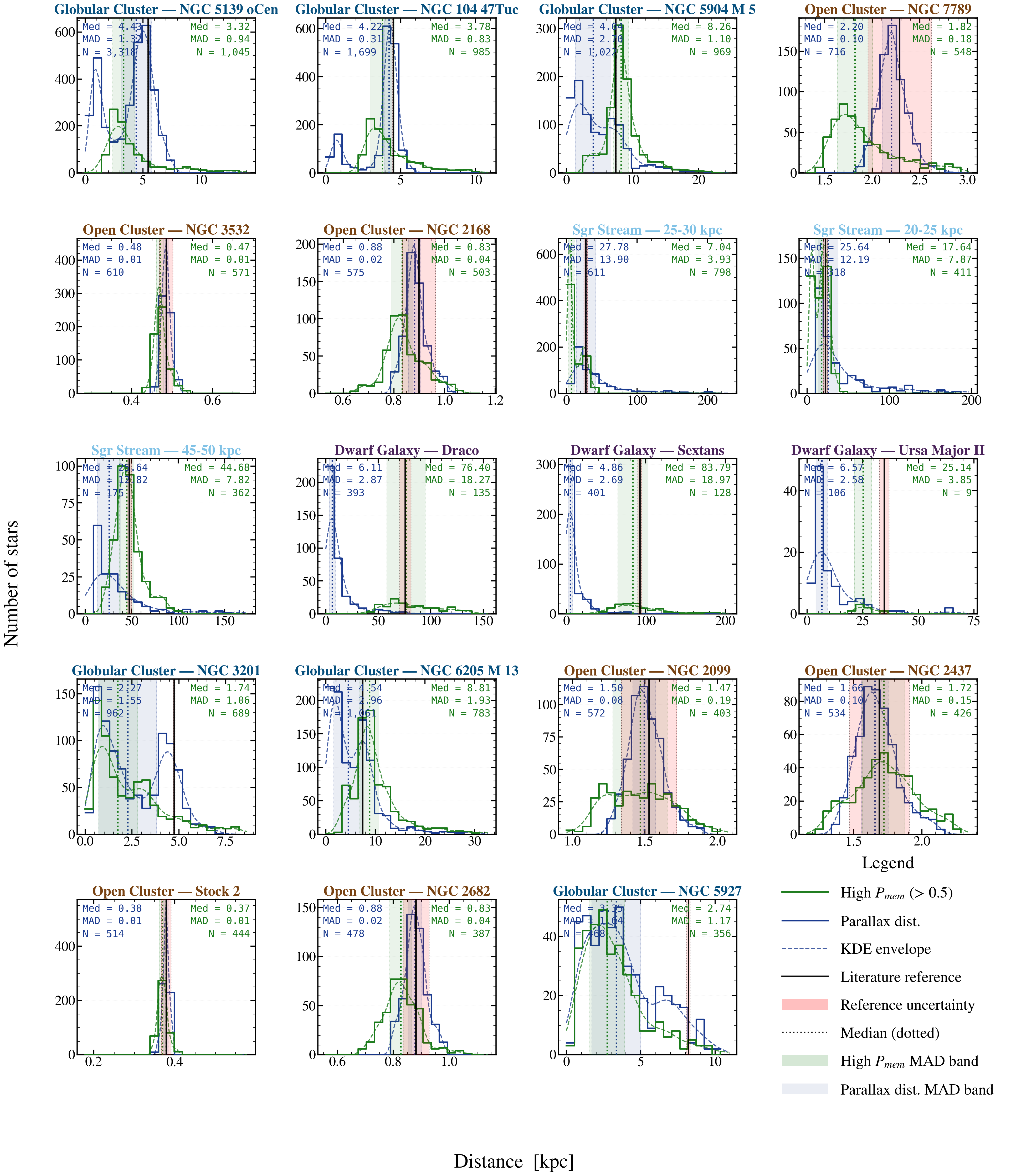}
			\caption{\small Examples of distance distributions for the external validation sample, shown as a
				multi-panel compilation of individual stellar systems. In each panel, the blue histogram denotes
				parallax distances and the green histogram denotes high-membership ($P_{\rm mem}>0.5$)
				spectrophotometric distances; dashed curves show the corresponding KDE envelopes. The black vertical
				line and shaded band mark the literature reference distance and its uncertainty, while dotted lines and
				lightly shaded bands indicate the median and MAD range of each distribution. The annotation boxes list
				the median, MAD, and sample size for the parallax (blue) and high-$P_{\rm mem}$ (green) samples.
			}
			\label{fig:ext_valid_dist}
		\end{figure*}
		
		\subsection{Results and Validation}
		\label{subsec:mem_results}
		
		The full validation pipeline is applied to 1,722 discrete stellar systems
		in two independent epoch realisations, $J2000$ and $J2016$, for a total of
		3,444 EM runs. The Sagittarius stream is analysed separately. Only 16 runs, corresponding to $0.46\%$ of the total,
		fail to converge within the 60-iteration limit. All such failures occur in the open-cluster class and are
		associated with small matched samples in high-background-density fields, an intrinsically degenerate
		regime for mixture modelling. The overwhelming majority of converged solutions require only 4--7
		iterations. Agreement between the two epoch implementations is excellent: for example the recovered
		member counts differ by only $1$--$2\%$ for representative rich systems such as $\omega$\,Cen (1,822 in
		the $J2016$ implementation versus 1,806 in the $J2000$ implementation) and 47\,Tuc (1,347 versus 1,335).
		
		With the uniform criterion $N_{P>0.5}\geq 10$, the validation set contains 845 systems in total,
		comprising 49 globular clusters, 785 open clusters, 3 dwarf galaxies, and 8 Sagittarius stream bins.
		Stable seven-term decompositions are obtained most robustly once the matched sample contains
		of order $200$--$500$ stars; below this range, the solutions increasingly tend toward the lower mixture
		boundary $\eta \rightarrow 0.01$.

		The adaptive distance dependence of the framework is illustrated by Fig.~\ref{fig:specific_validation}
		and quantified in Table~\ref{tab:dist_validation_best_subset}, which compares the median
		spectrophotometric distance $d_\mathrm{match}$ of the $P_\mathrm{mem}>0.5$ subsample with the adopted
		reference distance $d_\mathrm{ref}$ for all four member classes. The broader system-by-system
		behaviour is summarised in Figs.~\ref{fig:ext_valid_rv} and \ref{fig:ext_valid_dist}, which show that the
		framework performs as intended across the nearby, intermediate-distance, and distant-halo regimes.
		
		At $d \lesssim 5$~kpc, the spectrophotometric scale remains accurate for nearby systems. For the Pleiades
		($0.136$~kpc), $\alpha$\,Per ($0.175$~kpc), and IC\,2602 ($0.152$~kpc), the recovered $d_\mathrm{match}$
		agrees with $d_\mathrm{ref}$ to better than $3\%$. This nearby regime is not one in which spectrophotometric distances are expected to outperform high-S/N parallaxes in every system. For example, $\omega$\,Cen in Fig.~\ref{fig:specific_validation} illustrates a case where the parallax-based member distance is closer to the adopted reference value, where both member-distance estimates lie below the adopted reference
		distance and the parallax-based member distribution remains the closer of the two. This behaviour is
		physically reasonable and should be stated explicitly to avoid over-claiming a universal gain from
		spectrophotometric distances at small $d$. For a few nearby systems, both the median spectrophotometric distance and the median parallax-inversion distance of the selected members lie below the adopted system distance, and the parallax-based median can be closer. This is expected when the Gaia parallax has high signal-to-noise. We therefore do not claim that spectrophotometric distances universally improve nearby systems; their main advantage appears when the parallax precision degrades with distance.
		
		At intermediate distances, the transition from parallax to spectrophotometric dominated
		inference becomes clear. For M\,13 (NGC\,6205; $d_\mathrm{ref}=7.4$~kpc), the parallax-distance
		distribution peaks at $\tilde{d}_\varpi=6.94$~kpc with $\mathrm{MAD}=2.40$~kpc, whereas the
		spectrophotometric distribution yields $\tilde{d}=7.734$~kpc with $\mathrm{MAD}=1.93$~kpc, corresponding
		to a reduction in scatter of approximately $25\%$. The recovered radial-velocity distribution peaks at
		$\tilde{v}_r=-245.0$~km\,s$^{-1}$ with $\mathrm{MAD}=4.21$~km\,s$^{-1}$, in excellent agreement with the
		independently derived member-catalogue median of $-245.8$~km\,s$^{-1}$. For M\,5 (NGC\,5904;
		$d_\mathrm{ref}=7.5$~kpc), the improvement is stronger: the median parallax-inferred distance deviates
		from the adopted reference scale by approximately $2$~kpc, whereas the spectrophotometric median,
		$7.853$~kpc, has a median absolute deviation narrower by a factor of approximately 2.5 ($1.09$ versus
		$2.74$~kpc). These intermediate-distance systems mark the regime in which spectrophotometric distances
		begin to carry cleaner astrophysical information than direct parallax inversion.

		For globular clusters beyond approximately $15$~kpc, the spectrophotometric-distance posteriors broaden
		substantially, as seen in Table~\ref{tab:dist_validation_best_subset} and Fig.~\ref{fig:ext_valid_dist}.
		This broadening reflects the intrinsic growth of photometric distance uncertainty with heliocentric
		distance rather than a systematic failure of the method. Even in this regime, the median distance remains
		informative because the metallicity and CMD terms stabilise the member selection. The Sagittarius stream
		bins likewise show $d_\mathrm{match}$ values systematically $5$--$15\%$ below the adopted bin centres,
		consistent with the spectrophotometric distance floor at large heliocentric distance. The RV montages in
		Fig.~\ref{fig:ext_valid_rv} further show that, for most systems, the high-$P_{\rm mem}$ distributions are
		sharply peaked and centred close to the literature systemic velocities, whereas broader or offset cases
		flag either residual contamination or genuine internal kinematic complexity.
		
		The RUWE-based proper-motion error inflation of Eq.~\eqref{eq:ruwe_inflate} is important in crowded
		systems. In $\omega$\,Cen, $6.7\%$ of the matched stars require proper-motion error inflation owing to
		$\mathrm{RUWE}>1.4$. Without this adjustment, the proper-motion likelihood would artificially suppress
		genuine members in the crowded inner region.
		
		{\it The most striking validation occurs in the dwarf-galaxy regime}. For Draco, with
		$d_\mathrm{ref}=75.8$~kpc, the true parallax ($\varpi \approx 0.013$~mas) is overwhelmed by the Gaia
		uncertainty floor, and the raw parallax-distance distribution peaks at $\tilde{d}_\varpi=6.11$~kpc with
		$\mathrm{MAD}=2.87$~kpc, which is physically uninformative. After removing 172 foreground stars and applying the seven-term EM model, the spectrophotometric distance
		distribution converges to $\tilde{d}=76.5$~kpc with $\mathrm{MAD}=18.27$~kpc, more than $70$~kpc away
		from the naive parallax estimate and centred on the adopted reference distance. The recovery is driven
		jointly by the metallicity term, which isolates the $[\mathrm{Fe/H}] \approx -2$ population, the
		surface-gravity term, which favours evolved giants over dwarf contaminants, and the radial-velocity term,
		which anchors the solution at $v_\mathrm{sys,EM}=-290.8$~km\,s$^{-1}$, consistent with the literature
		value of $-291$~km\,s$^{-1}$ to within $1$~km\,s$^{-1}$. For Sextans ($d_\mathrm{ref}=92.5$~kpc), the
		pipeline recovers $v_\mathrm{sys,EM}=+224.5$~km\,s$^{-1}$ with $\sigma_\mathrm{int}=8.7$~km\,s$^{-1}$,
		while for Ursa Major\,II ($d_\mathrm{ref}=34.7$~kpc) it yields $v_\mathrm{sys,EM}=-115.8$~km\,s$^{-1}$
		with $\sigma_\mathrm{int}=3.5$~km\,s$^{-1}$. In this distant regime, the validation is driven not by
		geometric parallax but by the combined leverage of radial velocity, chemistry, gravity, and CMD
		consistency.

		\section{Science Usage of the Master Catalogue}
		\label{subsec:cp_guidance}
		
		A central requirement of the Milky Way tracer catalogue presented here is high internal reliability without discarding the rare, faint, and distant tracers needed for outer-halo science. Any quality cut therefore creates a trade-off: a permissive threshold retains more stars but also more poorly constrained measurements, whereas a restrictive threshold increases reliability at the cost of sample size. Because the catalogue combines several distance and RV channels whose precision varies with distance, we quantify this trade-off explicitly rather than imposing a single uniform hard cut across all tracer populations.
		
		We lay out some representative purity and completeness diagnostics in Appendix \ref{sec:completeness_purity}. However, it is up to the users to choose their own selection criterion, depending on the scientific objectives, when working with the final catalogue. For all applications, the \texttt{RUWE} $< 1.4$  criterion 
		is recommended to be applied as a first step.  Below, we provide a few explicit
		recommendations for fruitful use of the data:
		
		\begin{enumerate}

			\item Studies employing the Jeans equations, action-based modelling, or
			distribution-function fitting to constrain the Milky Way potential
			require precise line-of-sight velocities and accurate distances for
			large samples of halo tracers. We recommend the purity-prioritised
			regime ($q_d < 0.25$, $q_v < 0.15$) together with a heliocentric
			distance lower bound of $d_{\rm helio} > 5$\,kpc to exclude disc
			contaminants. At $d \gtrsim 20$\,kpc, where $q_d$ broadens
			substantially, we further recommend restricting the analysis to the spectrophotometric-distance channels
			(\texttt{DESI\_SpecDist\_cut}, \texttt{BOSS\_HALO\_PlxDist\_rv\_cut},
			\texttt{SEGUE\_KG\_SpecDist\_rv\_cut},
			\texttt{BHB\_SDSS\_SpecDist\_rv\_cut}) and applying
			$\texttt{DISTERR}/\texttt{DIST} < 0.20$ in lieu of the global $q_d$
			threshold.
			
			\item Searches for kinematic substructure via clustering in velocity or
			integral-of-motion space benefit from large sample sizes and broad sky
			coverage at the expense of individual measurement precision. We
			recommend the completeness-prioritised regime ($q_d < 0.75$,
			$q_v < 0.5$). The membership probabilities $P_{\rm mem}$ released for
			the 845 validated systems in Section~\ref{sec:validation} provide an
			independent fidelity indicator. Additionally, restricting to
			$P_{\rm mem} > 0.5$ for confirmed streams raises the
			sample purity for the same diagnostic thresholds.
			
			\item Determinations of cluster systemic velocities, internal velocity dispersions and, where the spatially resolved member sampling is sufficient, rotation signatures require both high radial-velocity
			precision and reliable membership assignment. We recommend the
			purity-prioritised regime ($q_v < 0.15$), supplemented by the
			seven-term expectation-maximisation membership probabilities of
			Section~\ref{sec:validation} with $P_{\rm mem} > 0.7$. The RUWE-based
			proper-motion error inflation described in
			Eq.~\eqref{eq:ruwe_inflate} automatically down-weights
			crowding-affected sources in cluster cores without requiring a
			supplementary hard cut on those stars.
			
			\item Studies using $[\mathrm{Fe/H}]$ or $[\alpha/\mathrm{Fe}]$ as tracers
			of Galactic chemical evolution are typically less sensitive to distance
			or velocity precision than to metallicity completeness. In this regime
			neither $q_d$ nor $q_v$ is the primary selection criterion. We
			recommend applying only $\texttt{RUWE} < 1.4$ and retaining all entries
			with valid atmospheric parameters from the contributing surveys,
			supplemented by the photometric metallicity columns from the
			\textit{Gaia}~XP and Pristine-Gaia channels.
			
			\item In the solar neighbourhood and inner halo, \textit{Gaia} geometric
			parallaxes dominate and spectrophotometric distances are not the primary
			distance estimator. Users should apply the astrometric quality criteria
			of Section~\ref{sec:gaia_cuts} and restrict to sources with parallax
			signal-to-noise above five, rather than relying on $q_d$, which is
			calibrated to the spectrophotometric regime and is not meaningful for
			parallax-supported samples.
			
			\item At large heliocentric distances the parallax-based distance column is
			entirely noise-dominated and the $q_d$ diagnostic should not be applied.
			Instead, users should restrict to the specialised outer-halo channels
			(\texttt{RRL\_PTR\_SpecDist\_rv\_cut},
			\texttt{NG\_Virgo\_SpecDist\_rv\_cut},
			\texttt{BHB\_SDSS\_SpecDist\_rv\_cut},
			\texttt{DESI\_SpecDist\_cut} Gold Sample,
			\texttt{BOSS\_HALO\_PlxDist\_rv\_cut}) and apply the
			completeness-prioritised $q_v$ threshold ($q_v < 0.5$). The Sagittarius
			stream membership probabilities released with
			Section~\ref{sec:validation} provide spatial coherence filtering that
			effectively compensates for the relaxed kinematic threshold at these
			distances.
		\end{enumerate}
		
		\begin{figure*}
			\centering
			\includegraphics[width=0.49\linewidth]{median_radial_velocity.png}
			\includegraphics[width=0.49\linewidth]{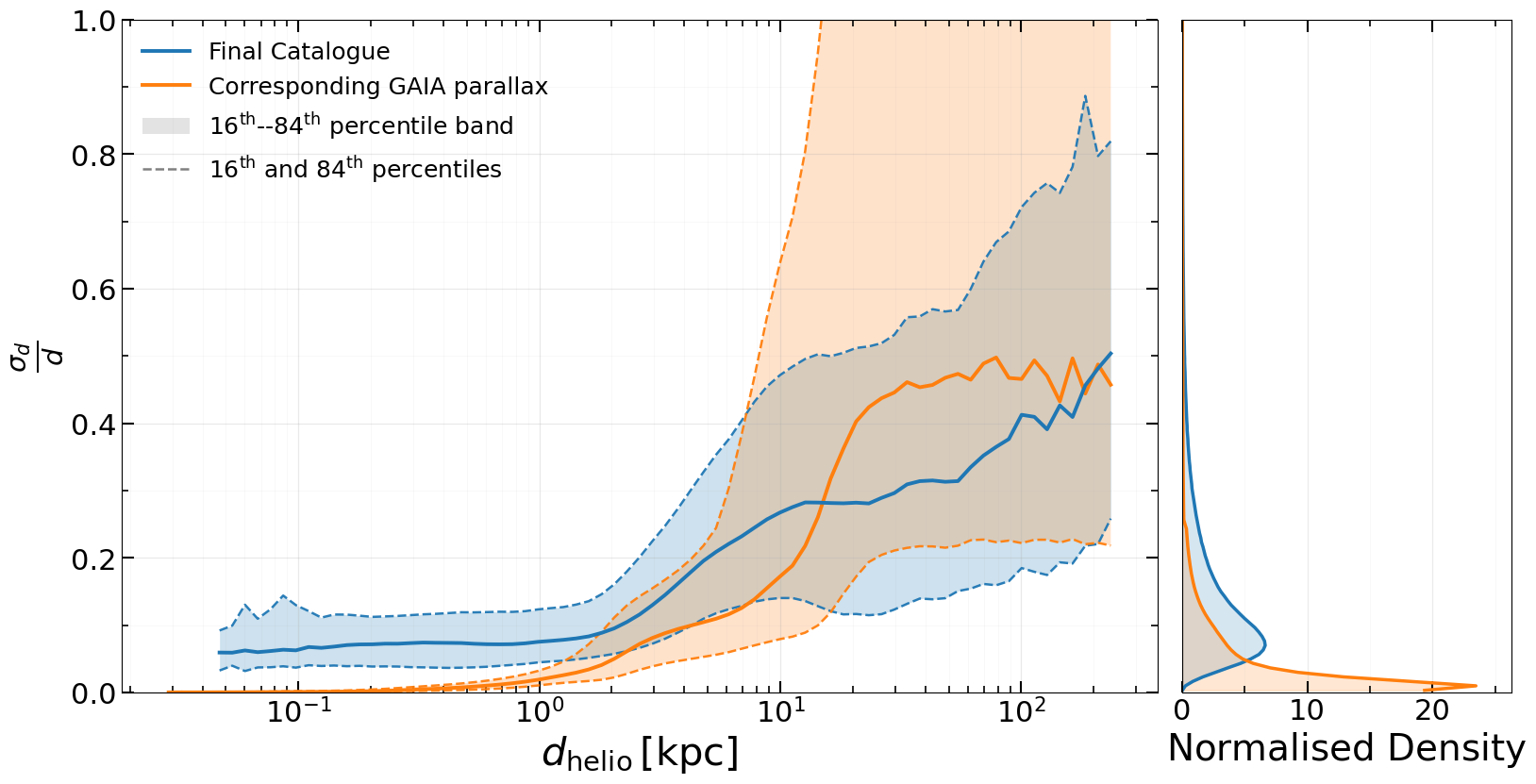}
			
			\caption{ Relative RV (left) and distance (right) uncertainty vs.\ $d_{\rm helio}$: final catalogue (blue) vs.\ Gaia RVS/parallax-only estimates (orange), showing the catalogue's improved precision at larger distances.}
			\label{fig:final_error}
		\end{figure*}
		
		\section{Conclusions}
		
		In this work, we have constructed a homogenised catalogue of
		$32{,}552{,}876$ unique Milky Way stars with full 6D phase-space
		information. The catalogue combines Gaia DR3 astrometry with
		distance and radial-velocity measurements from 14 large surveys and
		several specialised halo-tracer samples.
		
		The main steps in the construction can be summarised as follows:
		\begin{itemize}
			\item The radial-velocity uncertainties of the contributing surveys
			were placed on a common statistical scale using repeated
			measurements within and across surveys.
			
			\item Repeated observations of the same physical star were identified and grouped using a graph-based deduplication procedure.
			
			\item Survey-dependent radial-velocity zero points and their trends with stellar parameters were calibrated relative to Gaia.
			
			\item The calibrated measurements were combined into a single record for each star while retaining the corresponding uncertainties and survey provenance.
		\end{itemize}
		
		The final catalogue has a median fractional radial-velocity uncertainty of $\simeq8.31\%$ (i.e, $50\%$ of the stellar population has relative radial velocity uncertainty less than 0.0831). The median fractional distance uncertainty is $\simeq2.7\%$ for $d_{\rm helio}\leq15$ kpc and $\simeq29\%$ at larger distances (see fig \ref{fig:final_error}). It contains $517{,}123$ halo stars, including K giants, BHB stars, RR Lyrae, and other distant tracers for which Gaia parallaxes alone provides little useful distance information.
		
		We have tested the catalogue using globular clusters, open clusters,
		dwarf galaxies, and the Sagittarius stream. These comparisons show that
		the combined distance, radial-velocity, proper-motion, and stellar
		parameter information can recover coherent stellar populations over a
		wide range of heliocentric distances. The catalogue therefore, provides a common and internally calibrated 6D data set for studies of the Milky Way disc and halo, including Galactic dynamics, substructure, stellar populations, and mass modelling.
		
		\section{Data Availability}
		The catalogues and data products generated in this work will be shared upon reasonable request to the
		corresponding author upon acceptance. The observational data underlying this article are publicly available from the
		respective survey archives cited in the text.
		
		\section{Acknowledgements}
		The authors would like to thank Sergey Koposov and Shadab Alam for multiple constructive discussions that
		helped improve the methodology and rigor of the work, as well as the manuscript. The authors further
		thank Manush Manju for helpful discussions during the initial stages of this project. S. Majumdar would
		like to take this opportunity to thank Pijushpani Bhattacharjee for many discussions on the Milky Way
		dark matter halo and support over the last three decades. Both S. Mukherjee and A. Srivastav would like
		to thank TIFR for academic support. S. Mukherjee further wishes to thank TIFR for financial support, in
		particular, the Department of Theoretical Physics (DTP) for the opportunity to visit TIFR over multiple
		extended periods, during which much of this work was done. The authors also gratefully acknowledge access
		to the Pawna computing cluster of the Department of Theoretical Physics, TIFR, made available through the
		kind support of Shadab Alam. Computations were carried out
		on the computing clusters at the Department of Theoretical Physics, TIFR, Mumbai. This work has strongly benefited from the computational support provided by Ajay Salve and, especially, Kapil Ghadiali. 
		This work is supported by the Department of Atomic Energy, Government of India, under Project Identification Number RTI-4012 and RTI-4013. 
		
		\bibliographystyle{aa}
		\bibliography{apssamp}
		\appendix
		\nolinenumbers
		
		\section{The Expectation--Maximisation (EM) Optimisation}
		
		The EM optimisation alternates between two steps. In the E-step, a posterior membership probability is computed for each star. In the M-step, the model parameters are updated using these probabilities as weights. Thus, instead of assigning each star strictly as either a member or a non-member, each star contributes according to its current probability of membership. At iteration $t$, the member probability is
		\begin{equation}
			P_{\mathrm{mem},i}^{(t)}
			= \frac{\eta^{(t)}\, p_\mathrm{mem}^{(t)}\,
				\mathcal{L}_{\mathrm{mem},i}^{(t)}}
			{\eta^{(t)}\, p_\mathrm{mem}^{(t)}\,
				\mathcal{L}_{\mathrm{mem},i}^{(t)}
				+ (1-\eta^{(t)})\, p_\mathrm{field}^{(t)}\,
				\mathcal{L}_{\mathrm{field},i}^{(t)}},
			\label{eq:e_step}
		\end{equation}
		where $p_\mathrm{mem}^{(t)}$ and $p_\mathrm{field}^{(t)}$ denote the spatial terms and
		$\mathcal{L}_{\mathrm{mem},i}^{(t)}$ and
		$\mathcal{L}_{\mathrm{field},i}^{(t)}$
		denote the products of the remaining active terms. To avoid numerical degeneracy, the posterior probabilities are clipped to stay in the interval
		$[10^{-6},\,1-10^{-6}]$.
		
		In the M-step, the mixture coefficient, member proper-motion centroid, and member covariance are updated
		according to
		\begin{align}
			\eta^{(t+1)} &= \mathrm{clip}\!\left(
			\frac{1}{n_\mathrm{ok}}\sum_{i \in \mathcal{S}_\mathrm{ok}}
			P_{\mathrm{mem},i}^{(t)},\; 0.01,\; 0.99\right),
			\label{eq:m_eta} \\
			\bm{\mu}_\mathrm{mem}^{(t+1)} &=
			\frac{\sum_i w_i^{(t)}\,\bm{\mu}_i}{\sum_i w_i^{(t)}},
			\qquad
			w_i^{(t)} = P_{\mathrm{mem},i}^{(t)},
			\label{eq:m_mu} \\
			\bm{\Sigma}_\mathrm{mem}^{(t+1)} &=
			\frac{\sum_i w_i^{(t)}\,
				(\bm{\mu}_i - \bm{\mu}_\mathrm{mem}^{(t+1)})
				(\bm{\mu}_i - \bm{\mu}_\mathrm{mem}^{(t+1)})^T}
			{\sum_i w_i^{(t)}}
			+ \epsilon_\mathrm{reg}\,\mathbf{I},
			\label{eq:m_sigma}
		\end{align}
		where $\mathcal{S}_\mathrm{ok}$ is the subset of stars with finite proper motions and
		$n_\mathrm{ok}=|\mathcal{S}_\mathrm{ok}|$. To prevent the member centroid from drifting into a field-dominated local solution, the following algorithm is imposed: if the updated
		$\bm{\mu}_\mathrm{mem}^{(t+1)}$
		lies more than $3$~mas\,yr$^{-1}$ from the catalogue prior, it is replaced by
		\[
		\bm{\mu}_\mathrm{mem}^{(t+1)}
		\leftarrow
		0.5\,\bm{\mu}_\mathrm{mem}^{(t+1)}
		+0.5\,\bm{\mu}_\mathrm{prior}.
		\]
		
		The systemic radial velocity is updated through the inverse-variance weighted estimator
		\begin{equation}
			v_\mathrm{sys}^{(t+1)}
			= \frac{\sum_{i \in \mathcal{S}_\mathrm{rv}}
				w_i^{(t)}\,\iota_i\,v_{r,i}}
			{\sum_{i \in \mathcal{S}_\mathrm{rv}}
				w_i^{(t)}\,\iota_i},
			\qquad
			\iota_i = \frac{1}{(\sigma_\mathrm{int}^{(t)})^2
				+ (\sigma_{v,i}^\mathrm{adj})^2},
			\label{eq:vsys_update}
		\end{equation}
		after which $\sigma_\mathrm{int}$ is re-estimated using Eq.~\eqref{eq:mle_sigma}.
		
		Convergence is declared when the fractional change in the marginal log-likelihood satisfies
		\begin{equation}
			\begin{aligned}
				\frac{|\mathcal{L}^{(t+1)} - \mathcal{L}^{(t)}|}
				{|\mathcal{L}^{(t)}| + 10^{-10}}
				&< 10^{-5}, \\
				\mathcal{L}^{(t)}
				&= \sum_{i \in \mathcal{S}_\mathrm{ok}}
				\log\!\Biggl[\eta^{(t)}\,p_\mathrm{mem}^{(t)}\,\mathcal{L}_{\mathrm{mem},i}^{(t)} \\
				&\qquad\qquad + \bigl(1-\eta^{(t)}\bigr)\,p_\mathrm{field}^{(t)}\,\mathcal{L}_{\mathrm{field},i}^{(t)}\Biggr].
			\end{aligned}
			\label{eq:logL}
		\end{equation}
		or when the iteration count reaches 60. The initial conditions are
		$\eta=0.3$,
		$\bm{\mu}_\mathrm{mem}$ equal to the catalogue prior,
		$\bm{\Sigma}_\mathrm{mem}=0.5\,\mathbf{I}$~$(\mathrm{mas\,yr}^{-1})^2$,
		and $\bm{\mu}_\mathrm{field}$ estimated from the outer annulus
		$r>3\,r_h$. The final model quality is summarised using the Bayesian information criterion,
		\begin{equation}
			\mathrm{BIC} = -2\,\mathcal{L}_\mathrm{final} + 11\,\ln(n_\mathrm{ok}),
			\label{eq:bic}
		\end{equation}
		where the 11 free parameters are $\eta$; the two components of
		$\bm{\mu}_\mathrm{mem}$; the three independent elements of
		$\bm{\Sigma}_\mathrm{mem}$; $v_\mathrm{sys}$; $\sigma_\mathrm{int}$; the two components of
		$\bm{\mu}_\mathrm{field}$; and one independent diagonal element of $\bm{\Sigma}_\mathrm{field}$.
		
		\section{\textit{Gaia} astrometric and photometric quality diagnostics}
		\label{app:gaia_quality_diagnostics}
		
		The common \textit{Gaia} quality filters used in this work are designed to
		retain sources with reliable single-source astrometric solutions and internally
		consistent broad-band photometry. The quantity \texttt{RUWE} is the
		renormalised unit-weight error of the astrometric fit; values substantially
		larger than unity indicate that the single-source astrometric model provides a
		poor description of the observations, often because of blending, binarity,
		crowding, or other unmodelled effects. We adopt the commonly used requirement
		\texttt{RUWE}$<1.4$ to remove sources with poor astrometric fits.
		
		The quantity \texttt{visibility\_periods\_used} gives the number of distinct
		groups of observations entering the astrometric solution. Requiring
		\texttt{visibility\_periods\_used}$>8$ ensures that the solution is constrained
		by a sufficiently broad time sampling of the Gaia scanning law, reducing
		the probability of spurious or poorly determined parallaxes and proper motions.
		
		The image-parameter-determination (IPD) diagnostics flag sources whose
		\textit{Gaia} image windows may be affected by crowding, blending, nearby
		neighbours, or scan-angle-dependent structure. We use
		\path{ipd_frac_multi_peak} to identify cases with multiple peaks in the image
		window, and \path{ipd_gof_harmonic_amplitude} to identify scan-angle-dependent
		astrometric residuals. We require:
		\begin{equation}
			\begin{aligned}
				\texttt{ipd\_frac\_multi\_peak} &\le 2,\\
				\texttt{ipd\_gof\_harmonic\_amplitude} &< 0.1,
			\end{aligned}
		\end{equation}
		and remove sources with \path{duplicated_source} set to \path{True}, since these
		entries may correspond to ambiguous or duplicated detections in the
		\textit{Gaia} source processing.
		
		The quantities \texttt{astrometric\_excess\_noise\_sig} and
		\texttt{astrometric\_sigma5d\_max} provide additional checks on the quality of
		the astrometric solution. The former measures the significance of excess noise
		required to explain the residuals of the astrometric fit, while the latter
		summarises the largest semi-major axis of the five-dimensional astrometric
		uncertainty ellipsoid. We require
		\texttt{astrometric\_excess\_noise\_sig}$\le 2$ and
		\texttt{astrometric\_sigma5d\_max}$<1.5$ to remove sources with statistically
		significant excess astrometric residuals or poorly constrained five-parameter
		solutions.
		
		Finally, we filter the BP/RP photometry using the corrected flux-excess
		diagnostic of \citet{Riello2021}. The raw
		\texttt{phot\_bp\_rp\_excess\_factor} is sensitive to colour and crowding, so we
		use the corrected quantity $C^*$ and require
		\begin{equation}
			|C^*| \le 3\sigma_{C^*},
		\end{equation}
		where $\sigma_{C^*}$ is the colour-dependent scatter expected for well-behaved
		sources. This removes objects with inconsistent $G$, $G_{\rm BP}$, and
		$G_{\rm RP}$ photometry while retaining sources whose BP/RP excess is
		consistent with the empirical Gaia photometric locus. These cuts follow the
		quality-control approach recommended for Gaia astrometry and photometry
		\citep{Lindegren2021,GaiaEDR3,GaiaDR3Summary,Riello2021}.
		
		\section{Completeness and Purity of Some Sub-catalogues Constructed from the Final Master Catalogue}
		\label{sec:completeness}
		\label{sec:completeness_purity}
		
		\subsection{Definitions and Mixture Model}
		\label{subsec:cp_defs}
		
		The completeness--purity analysis is applied separately to three quality diagnostics:
		\[
		q_d \equiv \left|\frac{\sigma_d}{d}\right|,
		\qquad
		q_v \equiv \left|\frac{\sigma_{v_r}}{v_r}\right|
		= \frac{\sigma_{v_r}}{|v_r|},
		\qquad
		\texttt{RUWE}.
		\]
		For a diagnostic $x$, let
		\[
		\mathcal{V}_x = \{i:x_i\ {\rm is\ finite}\}
		\]
		be the set of catalogue entries for which that diagnostic is available, and, for an upper threshold $x_{\rm cut}$, let
		\[
		\mathcal{R}_x(x_{\rm cut})
		=
		\{i\in\mathcal{V}_x:x_i<x_{\rm cut}\}
		\]
		be the entries retained by the cut. The quantity called \emph{retention} in the analysis code is
		\begin{equation}
			\mathcal{C}_{\rm ret}(x_{\rm cut})
			=
			\frac{|\mathcal{R}_x(x_{\rm cut})|}
			{|\mathcal{V}_x|}.
			\label{eq:completeness_def}
		\end{equation}
		Thus, the denominator is the number of stars with a finite value of the diagnostic being swept, not necessarily all $N_\star=32{,}552{,}876$ entries in the final master catalogue. This is an internal retained fraction and should not be interpreted as the astrophysical completeness of the Milky Way stellar population.
		
		For each diagnostic, the observed distribution is represented as a two-component mixture,
		\begin{equation}
			p(x)=\eta\,p_{\rm core}(x)+(1-\eta)\,p_{\rm tail}(x),
			\label{eq:cp_mixture}
		\end{equation}
		where the fitted \emph{core} is the component associated with the lower-error, statistically well-behaved part of the diagnostic distribution, and the \emph{tail} represents the higher-error or broader component. In the implementation supplied with this paper, RUWE is fitted with a regularised two-Gaussian mixture. The distance and RV diagnostics are each fitted with both log-normal and gamma mixtures; the gamma-mixture posterior is the primary purity estimate stored as \texttt{purity\_dist} and \texttt{purity\_rv}, while the log-normal fits are retained for comparison.
		
		For an entry with diagnostic value $x_i$, the posterior probability of membership in the fitted core is
		\begin{equation}
			P_{{\rm core},i}
			=
			\frac{\eta\,p_{\rm core}(x_i)}
			{\eta\,p_{\rm core}(x_i)+(1-\eta)\,p_{\rm tail}(x_i)}.
			\label{eq:purity_posterior}
		\end{equation}
		Here, ``good'' means membership in this fitted statistical core; it is not an externally known truth label. The mean posterior purity of the sample retained by a cut is therefore
		\begin{equation}
			\mathcal{P}(x_{\rm cut})
			=
			\frac{1}{|\mathcal{R}_x(x_{\rm cut})|}
			\sum_{i\in\mathcal{R}_x(x_{\rm cut})}P_{{\rm core},i}.
			\label{eq:purity_def}
		\end{equation}
		The analysis also records a model-based core completeness,
		\begin{equation}
			\mathcal{C}_{\rm model}(x_{\rm cut})
			=
			\frac{\sum_{i\in\mathcal{R}_x(x_{\rm cut})}P_{{\rm core},i}}
			{\sum_{i\in\mathcal{V}_x}P_{{\rm core},i}},
			\label{eq:model_completeness_def}
		\end{equation}
		which measures the fraction of the total fitted core probability retained by the threshold.
		
		Finally, for an external reference sample $s$ (GC, OC, or SGR), the code defines an external recovery completeness
		\begin{equation}
			\mathcal{C}_{{\rm ext},s}(x_{\rm cut})
			=
			\frac{N_{{\rm ref},s}^{\rm matched,\ finite,\ pass}(x_{\rm cut})}
			{N_{{\rm ref},s}^{\rm total}}.
			\label{eq:external_completeness_def}
		\end{equation}
		The denominator is the full external reference list, so unmatched reference stars remain in the denominator. Before any $q_d$, $q_v$, or RUWE cut, the one-arcsec positional cross-match recovers $16{,}715/1{,}168{,}746=1.43\%$ of the GC reference entries, $49{,}399/435{,}833=11.33\%$ of the OC entries, and $2{,}687/55{,}192=4.87\%$ of the SGR entries. Consequently, the external-completeness curves in Fig.~\ref{fig:completeness} should not be compared numerically with $\mathcal{C}_{\rm ret}$ as though they used the same denominator.
		
		\subsection{Diagnostic Metrics}
		\label{subsec:cp_metrics}
		
		The threshold grids evaluated by the supplied analysis are
		\begin{align}
			q_d,\ q_v \in \{& 0.005, 0.01, 0.015, 0.02, 0.03, 0.04, 0.05, \notag \\
			& 0.07, 0.10, 0.12, 0.15, 0.20, 0.25, 0.30, \notag \\
			& 0.40, 0.50, 0.60, 0.75, 1.0, 1.5, 2.0\}, \\
			\texttt{RUWE} \in \{& 0.7, 0.8, 0.9, 0.95, 1.0, 1.05, 1.1, 1.15, \notag \\
			& 1.2, 1.3, 1.4, 1.5, 1.6, 1.8, 2.0, 2.5, \notag \\
			& 3.0, 4.0, 5.0\}.
		\end{align}
		All threshold tests use the strict inequality $x<x_{\rm cut}$.
		
\begin{figure*}[htbp]
			\centering
			\includegraphics[width=1\linewidth]{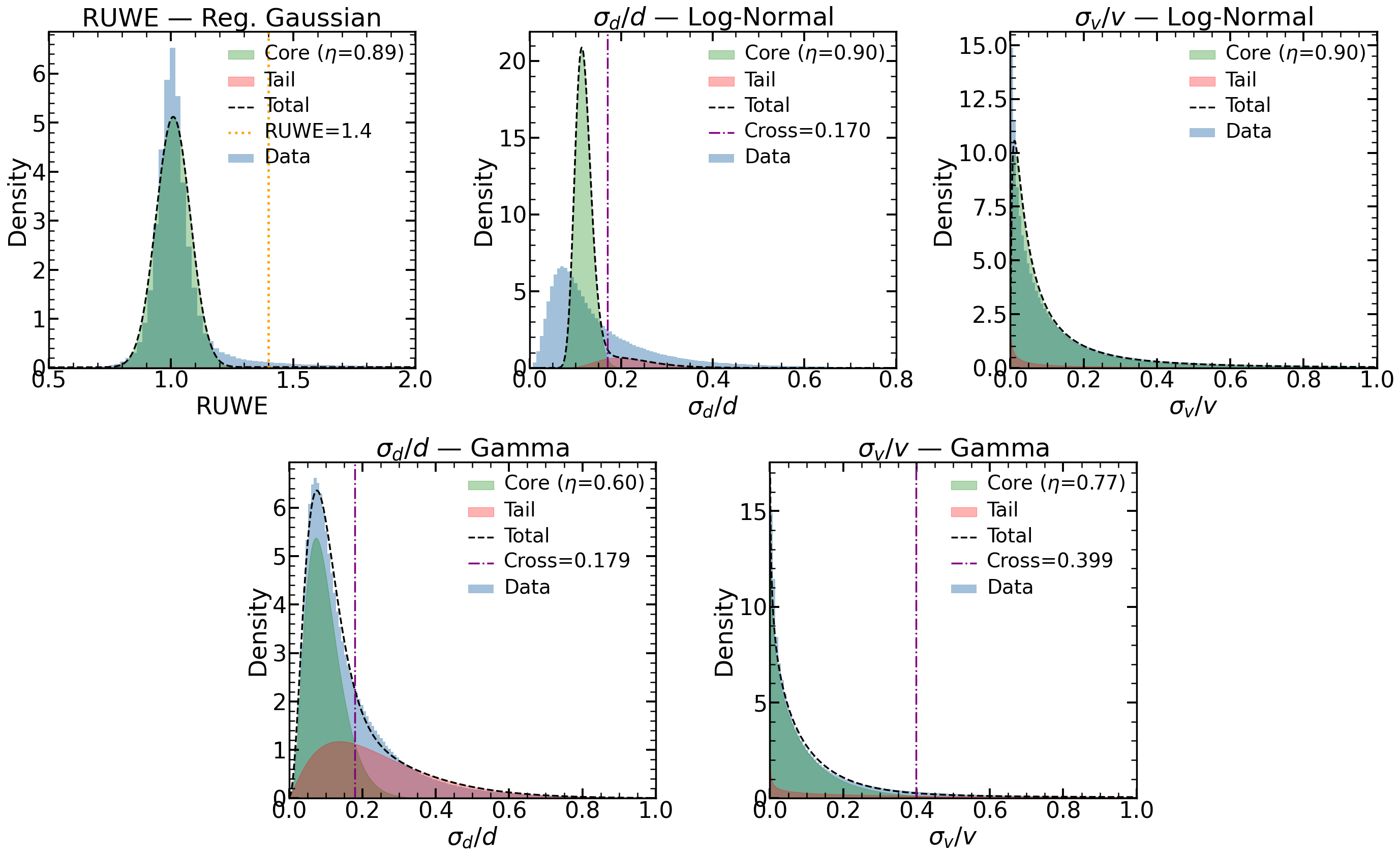}
			\caption{Normalised fitted probability densities for the three quality diagnostics. RUWE is shown with the regularised two-Gaussian mixture in both rows. For $q_d=\sigma_d/d$ and $q_v=\sigma_{v_r}/|v_r|$, the upper row shows the log-normal core--tail fit and the lower row the gamma core--tail fit. Green and red filled regions are the fitted core and tail contributions, respectively, and the black dashed curve is their sum. Orange dotted reference lines mark \texttt{RUWE}$=1.4$ and $q_d=0.20$; purple dash-dotted lines mark the first numerical core--tail crossing when one occurs in the plotted range.}
			\label{fig:purity_metrics}
		\end{figure*}

\begin{figure*}[htbp]
			\centering
			\includegraphics[width=1\linewidth]{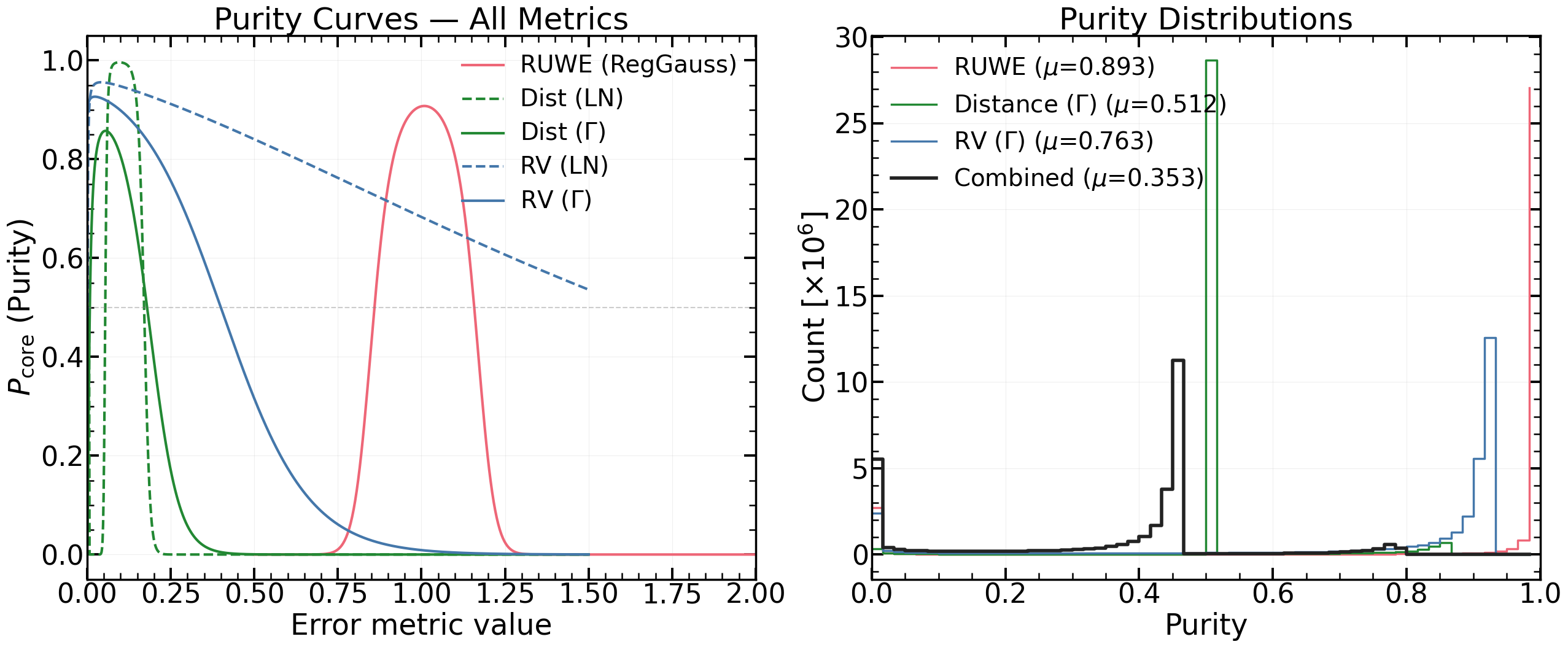}
			\caption{\small Purity diagnostics for the principal quality metrics. Left: posterior core-membership probability $P_{\rm core}(x)$ for RUWE, $q_d$, and $q_v$; solid distance/RV curves are the primary gamma-mixture estimates and dashed curves are the log-normal alternatives. Right: catalogue-wide distributions of the per-entry posterior purities for RUWE, the primary gamma distance and RV models, and their scalar product. The legend reports the corresponding mean of each distribution.}
			\label{fig:purity_curves}
		\end{figure*}

		Figure~\ref{fig:purity_metrics} shows the fitted core--tail decomposition for each diagnostic. RUWE is represented by the regularised-Gaussian mixture in both rows because only one RUWE model is fitted. For $q_d$ and $q_v$, the upper row shows the log-normal alternative and the lower row shows the gamma alternative. The orange reference lines mark \texttt{RUWE}$=1.4$ and $q_d=0.20$, and the purple dash-dotted markers, when present, denote the first numerical crossing of the fitted core and tail contributions. Figure~\ref{fig:purity_curves} shows the corresponding posterior core probabilities and their catalogue-wide distributions; the solid distance and RV curves are the primary gamma results and the dashed curves are the log-normal comparison fits.
		
		The RUWE fit gives a core fraction $\eta=0.893$ and a catalogue-wide mean posterior core probability $\langle P\rangle=0.893$. At the commonly used threshold
		\begin{equation}
			\texttt{RUWE} < 1.4,
			\label{eq:ruwe_cut_cp}
		\end{equation}
		we get $30{,}227{,}662$ stars with finite RUWE that pass the cut, corresponding to $\mathcal{C}_{\rm ret}=92.86\%$, a mean retained-sample purity $\mathcal{P}=0.962$, and $\mathcal{C}_{\rm model}=99.99999\%$.
		
		For $q_d$, the log-normal fit gives $\eta=0.900$ and $\langle P\rangle=0.549$, while the primary gamma decomposition gives $\eta=0.603$ and $\langle P\rangle=0.512$. The primary gamma component means are $0.099$ for the core and $0.253$ for the tail. The distance diagnostic is therefore intrinsically broader and more radius-dependent than RUWE.
		
		For $q_v$, the log-normal and gamma fits give core fractions of $0.900$ and $0.768$, respectively. Their catalogue-wide mean posterior purities are $0.903$ and $0.763$, and the primary gamma component means are $0.091$ for the core and $0.948$ for the tail. The log-normal number $0.903$ is retained as a model-comparison result; all threshold purities quoted below use the primary gamma posterior, in accordance with the supplied code.
		
		The analysis also forms the scalar product $P_{\rm comb}=P_{\rm RUWE}P_{q_d}P_{q_v}$. Its catalogue-wide mean is $0.353$ and its median is $0.442$. Because this product combines three different error mechanisms and is not itself fitted as a calibrated probability model, we do not recommend it as a stand-alone quality selector.
		
\begin{figure}[htbp]
			\centering
			\includegraphics[width=\linewidth]{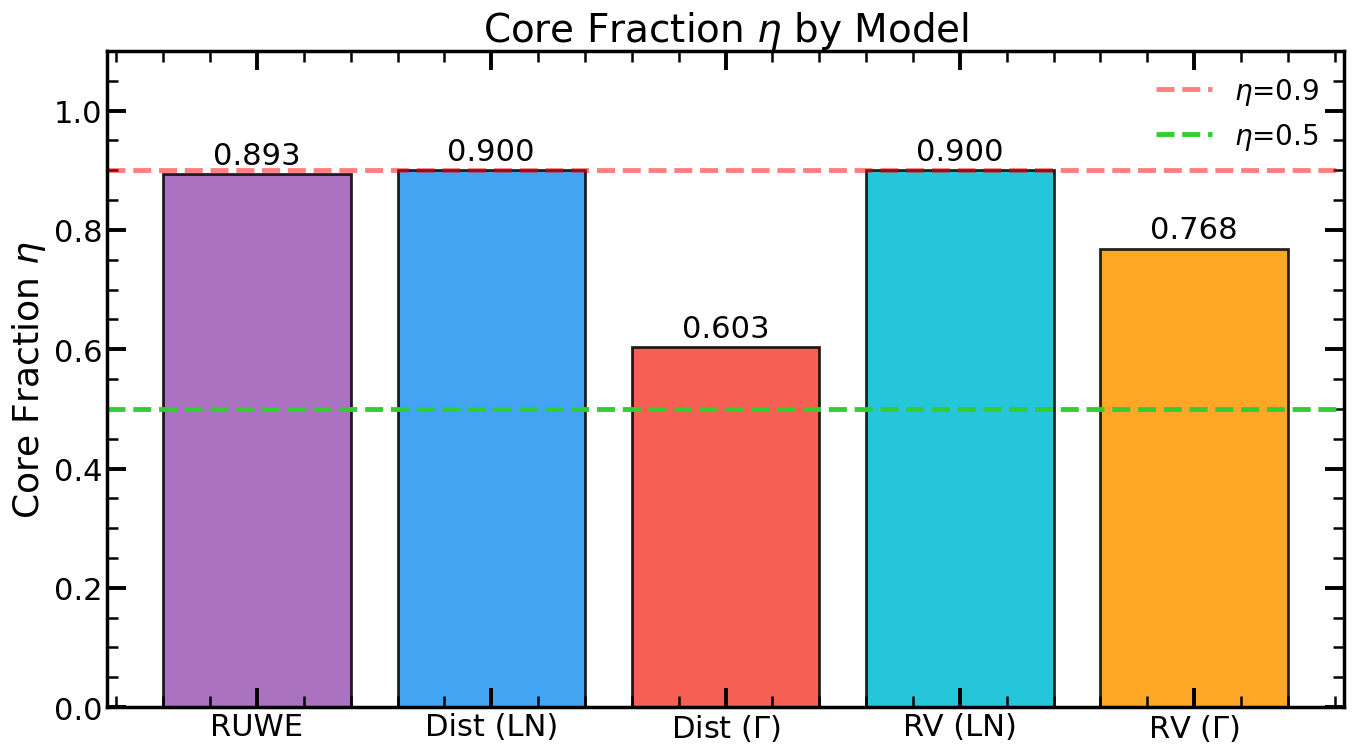}
			\caption{\small Fitted core fraction $\eta$ for RUWE and for the log-normal and gamma models of $q_d$ and $q_v$. The bars report the mixture weights returned by the corresponding fits; the gamma models are the primary models used for the distance and RV threshold-purity calculations.}
			\label{fig:core_fraction}
		\end{figure}

\begin{figure}[htbp]
			\centering
			\includegraphics[width=\linewidth]{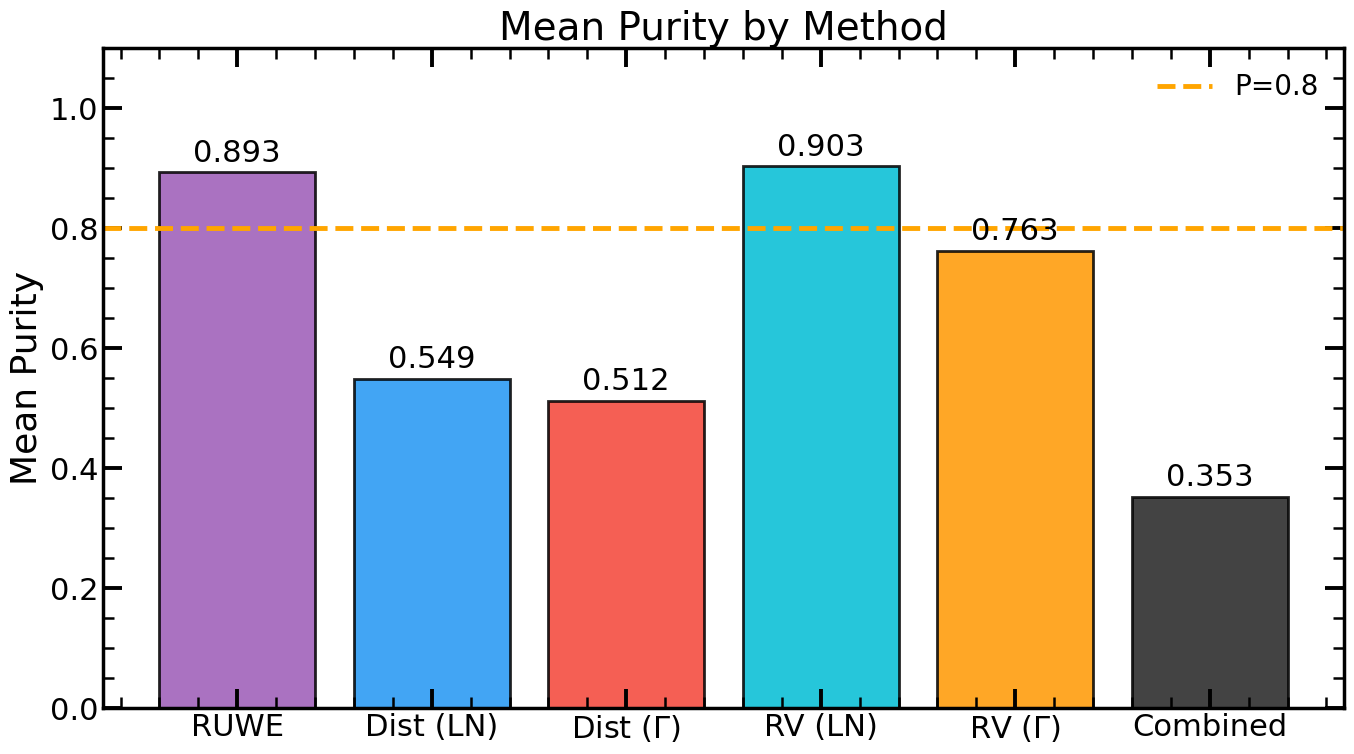}
			\caption{\small Catalogue-wide mean posterior core probability for each fitted diagnostic/model and for the scalar product of the three primary per-entry probabilities. The values are RUWE $0.893$, distance log-normal $0.549$, distance gamma $0.512$, RV log-normal $0.903$, RV gamma $0.763$, and combined product $0.353$.}
			\label{fig:mean_purity}
		\end{figure}

		These fitted fractions and mean probabilities are summarised in Figs.~\ref{fig:core_fraction} and \ref{fig:mean_purity}.
		
		\subsection{Reference Operating Points}
		\label{subsec:cp_operating}
		
		Since, we sweep each diagnostic independently, it does not provide a single measured retention fraction for a simultaneous three-way cut. We therefore report the one-dimensional operating points actually evaluated by the code rather than inferring a joint completeness by multiplying marginal fractions. Table~\ref{tab:cp_operating_points} gives representative thresholds used elsewhere in this paper together with the exact one-dimensional statistics from the sweep.
		
		\begin{table}[htbp]
			\centering
			\caption{Representative one-dimensional quality cuts. $N_{\rm pass}$ is the number of stars that both have a finite value of the indicated diagnostic and satisfy the threshold. $\mathcal{C}_{\rm ret}$ is defined by Eq.~\eqref{eq:completeness_def}, $\langle P\rangle_{\rm pass}$ by Eq.~\eqref{eq:purity_def}, and $\mathcal{C}_{\rm model}$ by Eq.~\eqref{eq:model_completeness_def}. The distance and RV purities use the primary gamma-mixture posteriors. These rows are marginal, one-diagnostic-at-a-time results and should not be read as the retention of a simultaneous multi-cut sample.}
			\label{tab:cp_operating_points}
			\renewcommand{\arraystretch}{1.25}
			\small
			\setlength{\tabcolsep}{3.5pt}
			\resizebox{\columnwidth}{!}{%
				\begin{tabular}{lrrrr}
					\hline\hline
					Cut & $N_{\rm pass}$ & $\mathcal{C}_{\rm ret}$ & $\langle P\rangle_{\rm pass}$ & $\mathcal{C}_{\rm model}$ \\
					\hline
					\texttt{RUWE}$<1.40$ & $30{,}227{,}662$ & $92.86\%$ & $0.962$ & $100.00\%$ \\
					$q_d<0.15$          & $2{,}487{,}519$  & $63.28\%$ & $0.806$ & $84.59\%$ \\
					$q_d<0.25$          & $3{,}268{,}004$  & $83.14\%$ & $0.717$ & $98.83\%$ \\
					$q_d<0.50$          & $3{,}798{,}759$  & $96.64\%$ & $0.624$ & $100.00\%$ \\
					$q_v<0.15$          & $21{,}799{,}915$ & $66.99\%$ & $0.913$ & $80.24\%$ \\
					$q_v<0.30$          & $26{,}182{,}881$ & $80.46\%$ & $0.894$ & $94.35\%$ \\
					$q_v<0.50$          & $28{,}387{,}570$ & $87.24\%$ & $0.866$ & $99.02\%$ \\
					\hline\hline
				\end{tabular}%
			}
		\end{table}
		
		For reference, the broad default thresholds used in the catalogue discussion are
		\begin{equation}
			\texttt{RUWE} < 1.4,
			\qquad
			q_d < 0.5,
			\qquad
			q_v < 0.3,
			\label{eq:default_cuts}
		\end{equation}
		while the more permissive distance/RV example used for high-completeness exploratory work is
		\begin{equation}
			q_d < 0.75,
			\qquad
			q_v < 0.5,
			\label{eq:comp_cuts}
		\end{equation}
		and the more restrictive example used for precision work is
		\begin{equation}
			q_d < 0.25,
			\qquad
			q_v < 0.15.
			\label{eq:pur_cuts}
		\end{equation}
		The value $q_d=0.75$ is used here, because $0.75$ is the corresponding threshold explicitly evaluated in the cut grid. Their joint retention must be measured by applying the cuts simultaneously to the released catalogue; it cannot be reconstructed exactly from the separate one-dimensional sweep arrays.
		
		The $q_d$ and $q_v$ thresholds should therefore be adapted to the science case. In particular, the radial diagnostic plots show that $q_d$ broadens strongly toward the outer halo, so the provenance and uncertainty of the adopted spectrophotometric distance channel become increasingly important there.
		
\begin{figure*}[h]
			\centering
			\includegraphics[width=0.8\linewidth]{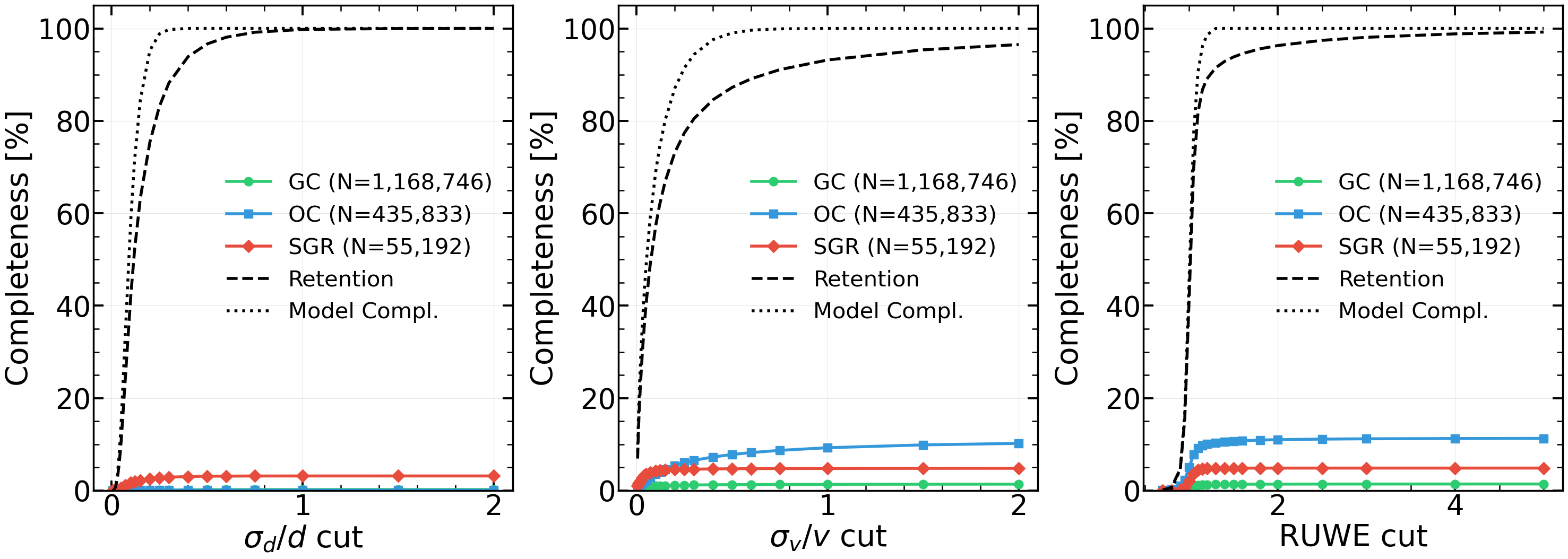}
			\caption{Completeness as a function of the one-dimensional quality threshold for $q_d=\sigma_d/d$, $q_v=\sigma_{v_r}/|v_r|$, and RUWE. The coloured curves show the external recovery completeness of the globular-cluster (GC), open-cluster (OC), and Sagittarius-stream (SGR) reference lists, defined as the number of reference entries that are positionally matched to the master catalogue, have a finite value of the relevant diagnostic, and pass the cut, divided by the full number of entries in that external reference list. The black dashed curve is the internal retained fraction $\mathcal{C}_{\rm ret}$ among master-catalogue entries with a finite value of the diagnostic, and the black dotted curve is the model completeness $\mathcal{C}_{\rm model}$, i.e. the fraction of fitted core probability retained. The external and internal curves therefore use different denominators and should not be interpreted as the same notion of completeness.}
			\label{fig:completeness}
		\end{figure*}

		Three principal conclusions follow from the completeness--purity analysis and are summarised in Fig.~\ref{fig:completeness}.
		
		First, \texttt{RUWE}$<1.4$ is an efficient astrometric-quality cut in the fitted model: it retains $92.86\%$ of finite-RUWE entries while retaining essentially all of the fitted core probability and gives a mean posterior purity of $0.962$ among the passing entries.
		
		Second, $q_v$ provides a high-purity kinematic selection. With the primary gamma posterior, $q_v<0.30$ retains $80.46\%$ of entries with finite $q_v$, has mean posterior purity $0.894$, and retains $94.35\%$ of the fitted RV-core probability. The catalogue-wide RV log-normal mean purity of $0.903$ is a separate model-comparison number and is not the purity at a particular threshold.
		
		Third, $q_d$ is the most radius-dependent of the three diagnostics. For example, $q_d<0.25$ retains $83.14\%$ of entries with finite $q_d$ and $98.83\%$ of the fitted distance-core probability, but its mean retained-sample gamma purity is $0.717$. It should therefore not be used in isolation as a universal outer-halo cleaning criterion.

		\begin{figure*}[htbp]
			\centering
			\includegraphics[angle=90,width=0.68\textheight]{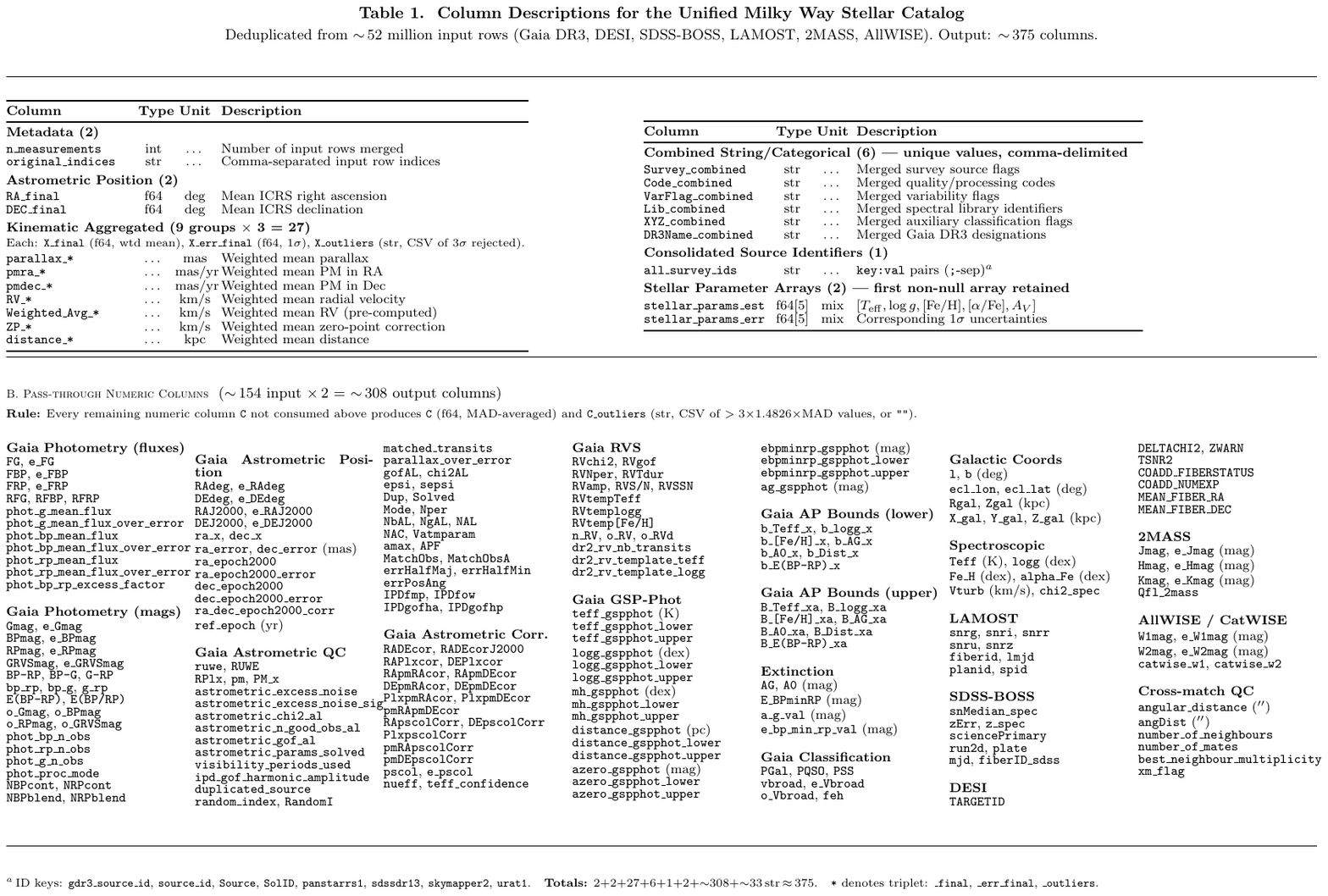}
			\caption{Column descriptions for the final homogenised 6D Milky Way stellar catalogue.}
			\label{fig:column_header_appendix}
		\end{figure*}

		\begin{figure*}[htbp]
			\centering
			
			\begin{minipage}[t]{0.49\textwidth}
				\centering
				\textbf{(a) Relative distance uncertainty, $q_d=\sigma_d/d$}\\[0.3em]
				\includegraphics[width=0.8\linewidth]{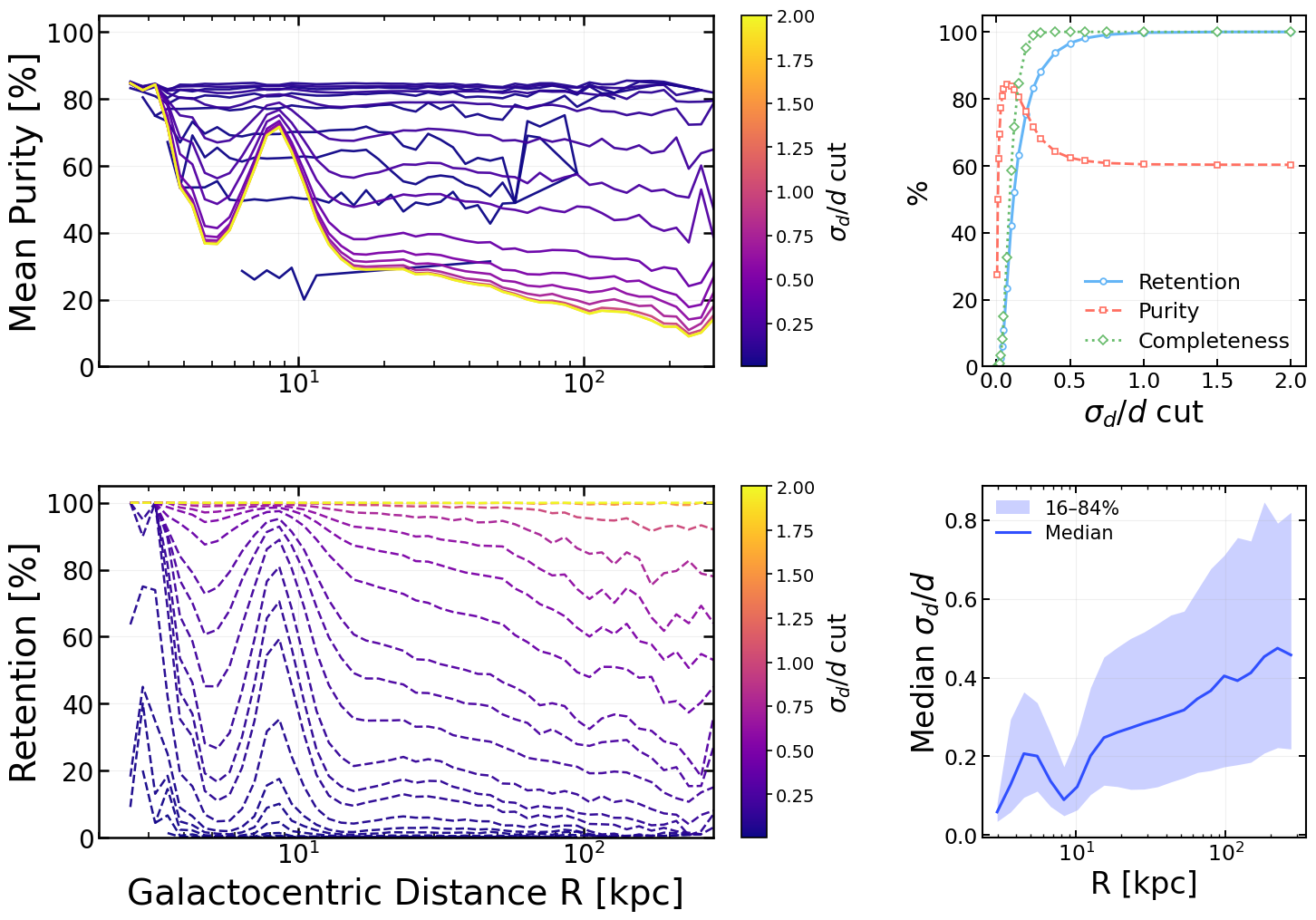}
			\end{minipage}
			\hfill
			\begin{minipage}[t]{0.49\textwidth}
				\centering
				\textbf{(b) Relative radial-velocity uncertainty, $q_v=\sigma_v/|v|$}\\[0.3em]
				\includegraphics[width=0.8\linewidth]{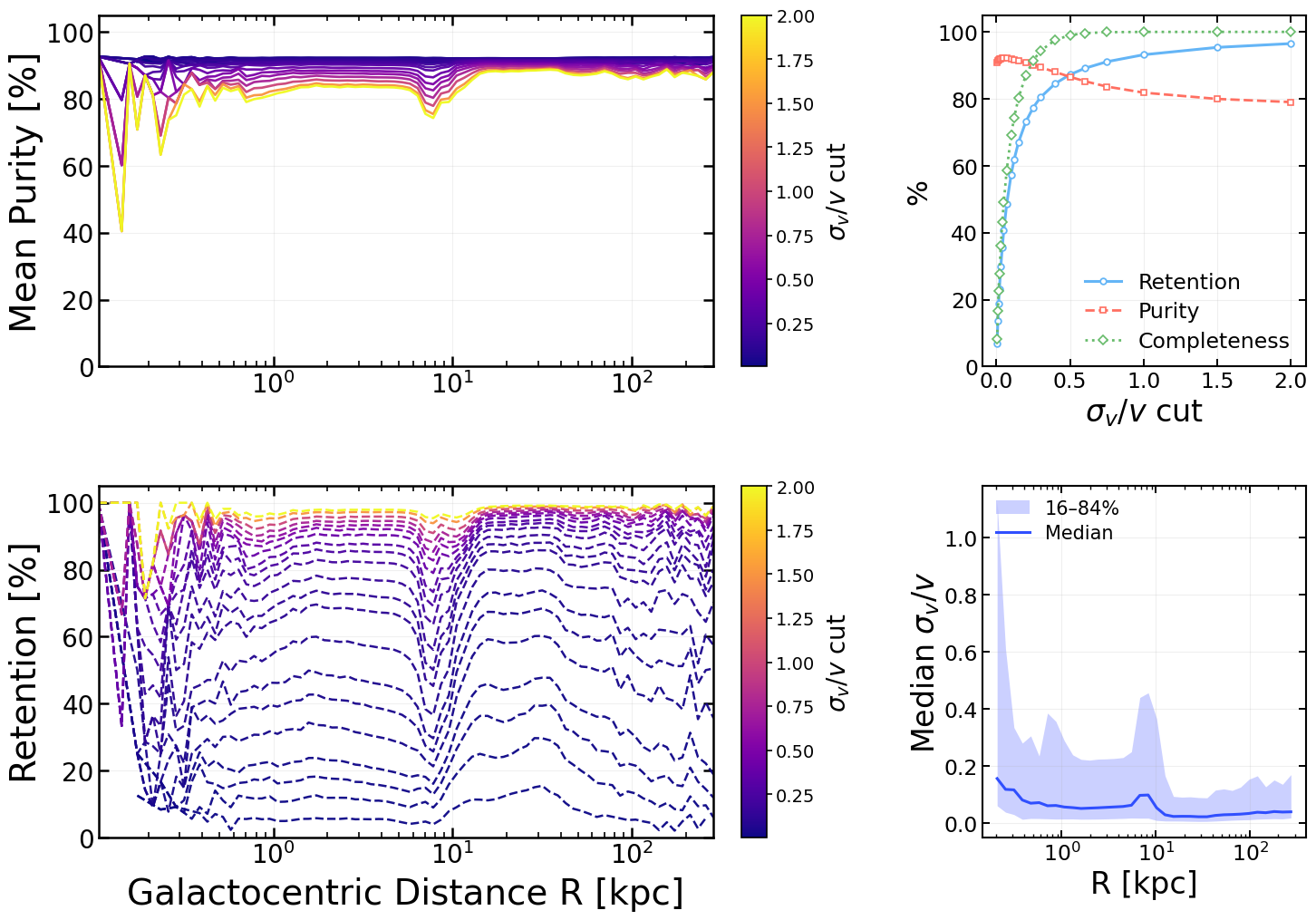}
			\end{minipage}
			
			\begin{minipage}[t]{0.49\textwidth}
				\centering
				\textbf{(c) Astrometric quality, RUWE}\\[0.3em]
				\includegraphics[width=0.8\linewidth]{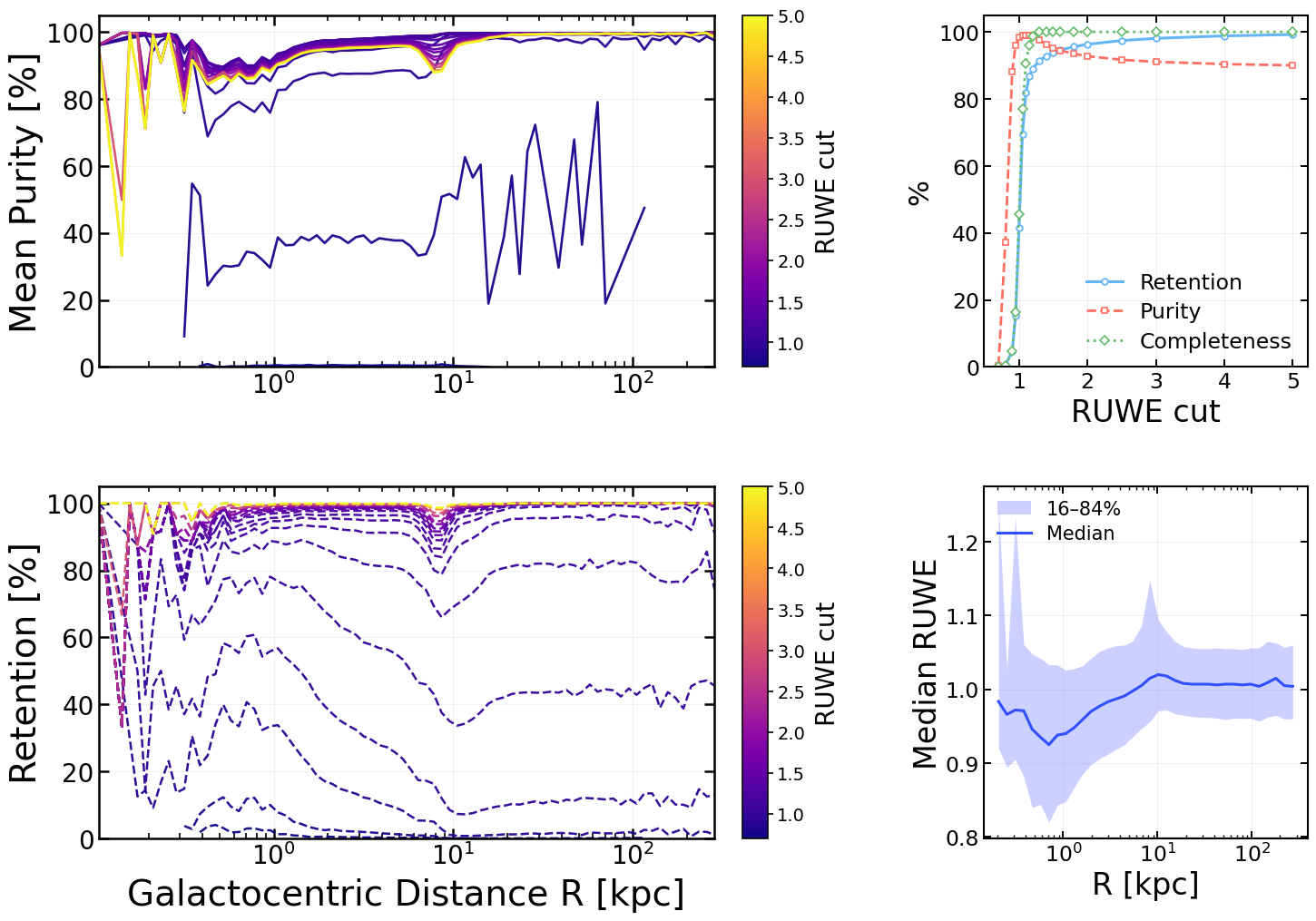}
			\end{minipage}
			\caption{Radial behaviour of the three quality diagnostics: relative distance uncertainty $q_d=\sigma_d/d$, relative radial-velocity uncertainty $q_v=\sigma_{v_r}/|v_r|$, and RUWE. For each diagnostic, the upper panel shows the mean posterior purity of the stars passing each threshold as a function of Galactocentric radius, and the lower panel shows the corresponding retained fraction. The upper inset gives the global one-dimensional trade-off between retention, mean purity, and model completeness as the threshold is varied; the lower inset gives the radial median of the diagnostic with its 16th--84th percentile range. The calculations use only finite values of the diagnostic within each radial bin. RUWE is the least radius-dependent, while $q_d$ broadens most strongly toward the outer halo.}
			\label{fig:appendix_metric_diagnostics}
		\end{figure*}

\begin{figure*}[htbp]
	\centering
	\begin{minipage}[b]{0.48\textwidth}
		\centering
		\includegraphics[width=\textwidth,keepaspectratio]{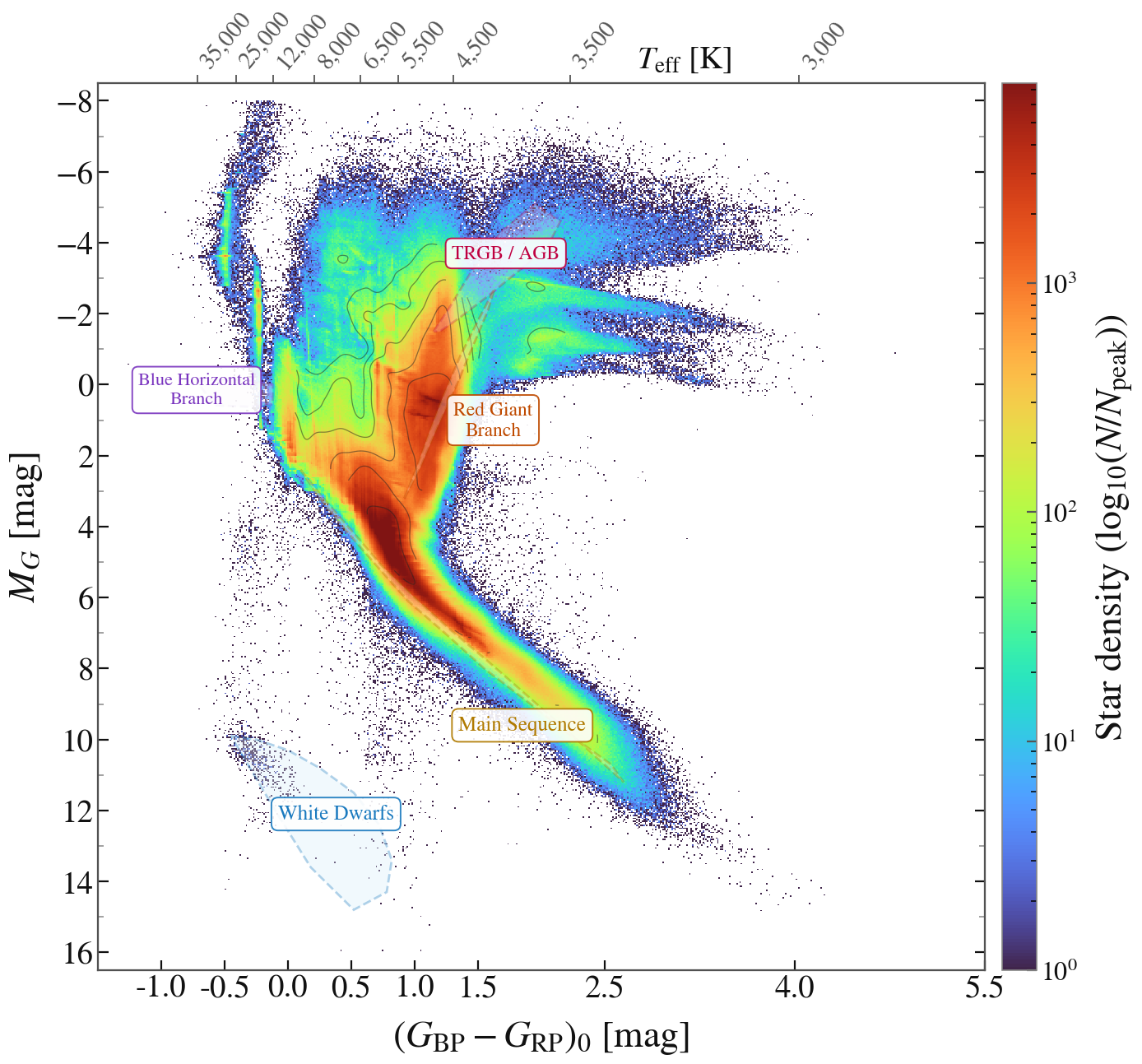}
	\end{minipage}
	\hfill
	\begin{minipage}[b]{0.48\textwidth}
		\centering
		\includegraphics[width=\textwidth,keepaspectratio]{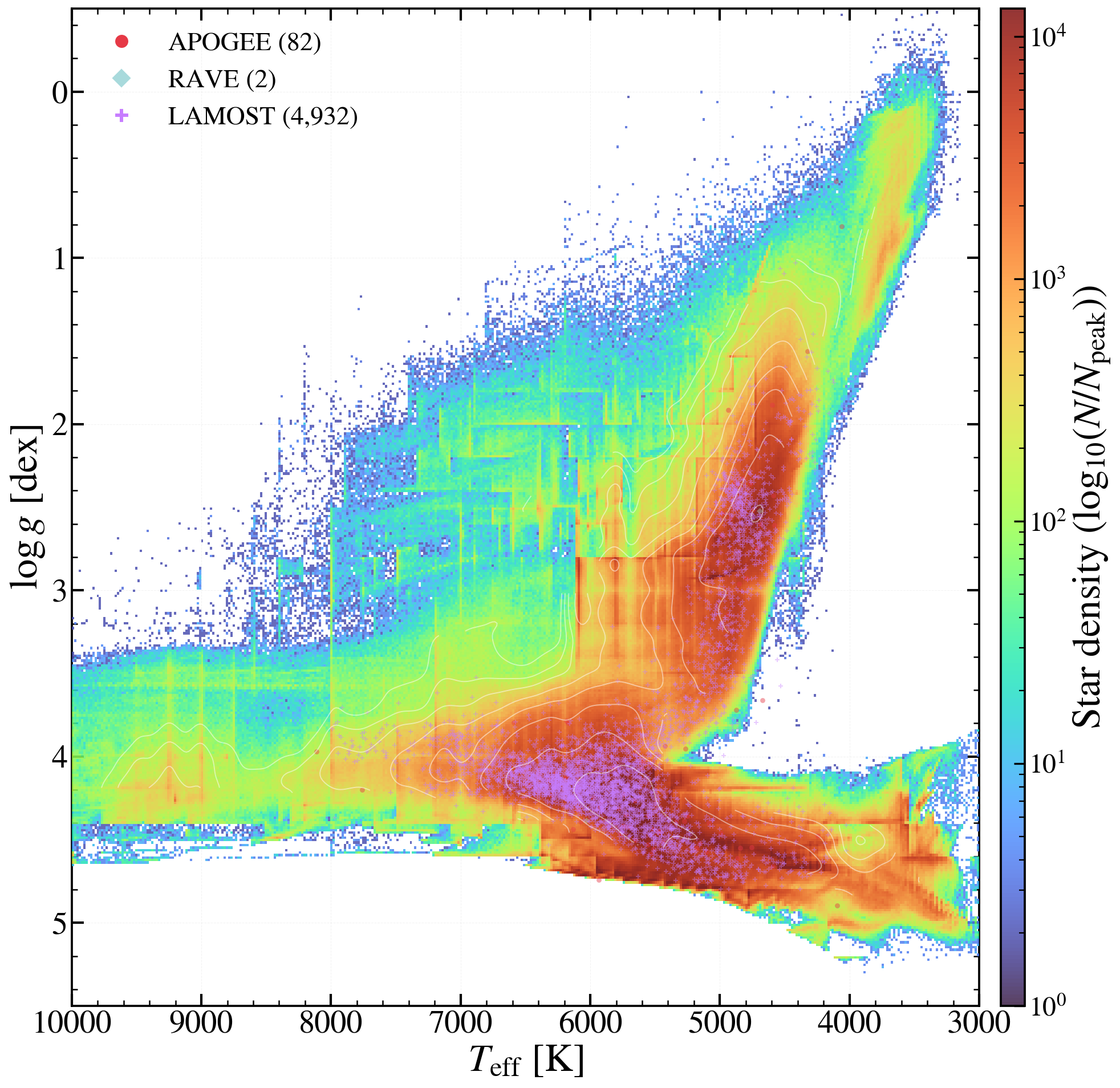}
	\end{minipage}
	\caption{Stellar parameter distributions. Left: Hertzsprung--Russell diagram in extinction-corrected Gaia colour, $(G_{\rm BP}-G_{\rm RP})_0$, and absolute magnitude, $M_G$, colour-coded by $\log_{10}(N/N_{\rm peak})$. Right: Kiel diagram in the $T_{\rm eff}$--$\log g$ plane, with external spectroscopic calibration samples overplotted, colour-coded by $\log_{10}(N/N_{\rm peak})$.}
	\label{fig:hr_and_kiel}
\end{figure*}
		
		\begin{figure*}[htpb]
			\centering
			\begin{subfigure}[b]{0.48\textwidth}
				\centering
				\includegraphics[width=\textwidth]{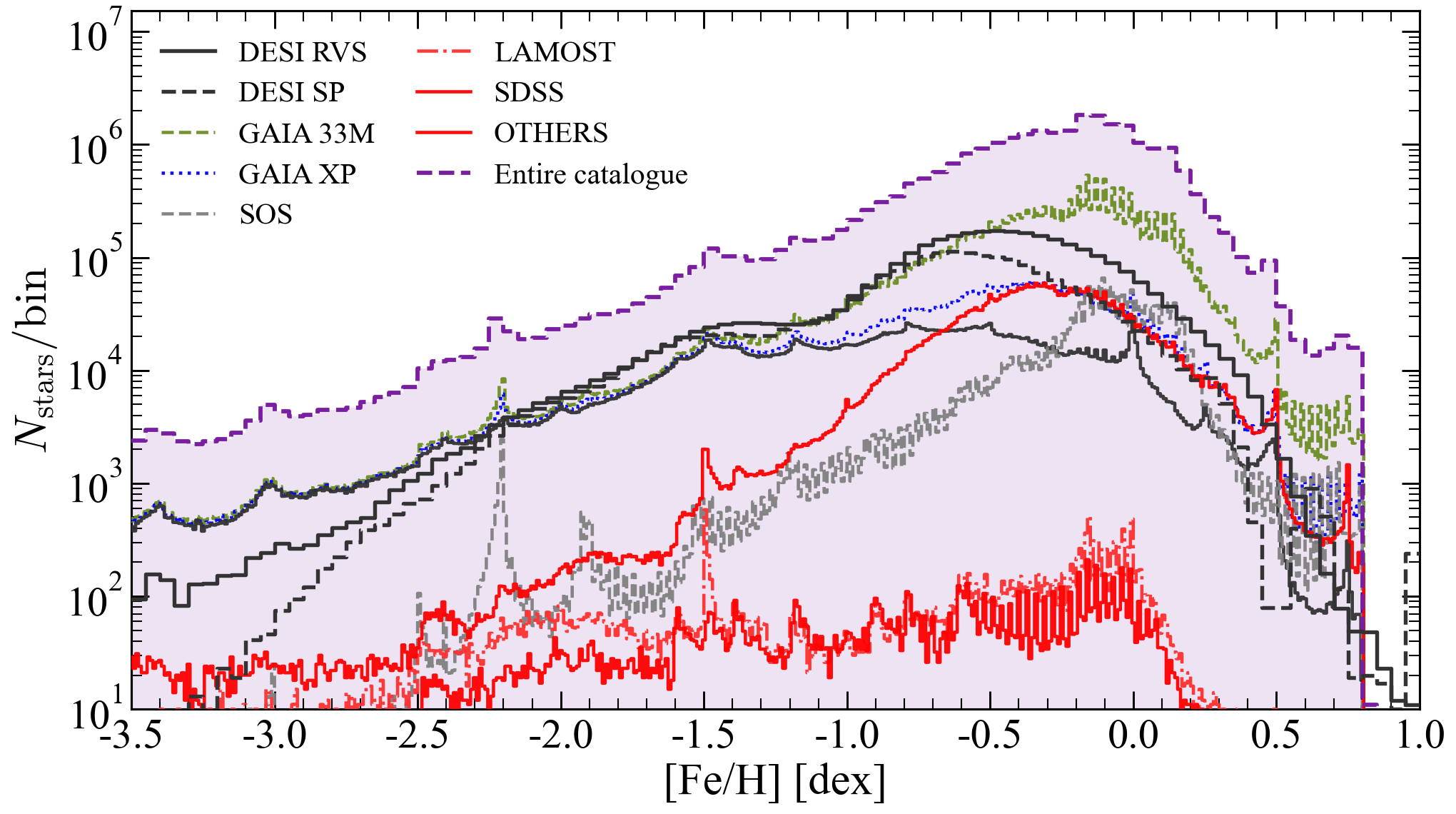}
				\caption{Metallicity distribution}
				\label{fig:feh_plot}
			\end{subfigure}
			\hfill
			\begin{subfigure}[b]{0.48\textwidth}
				\centering
				\includegraphics[width=\textwidth]{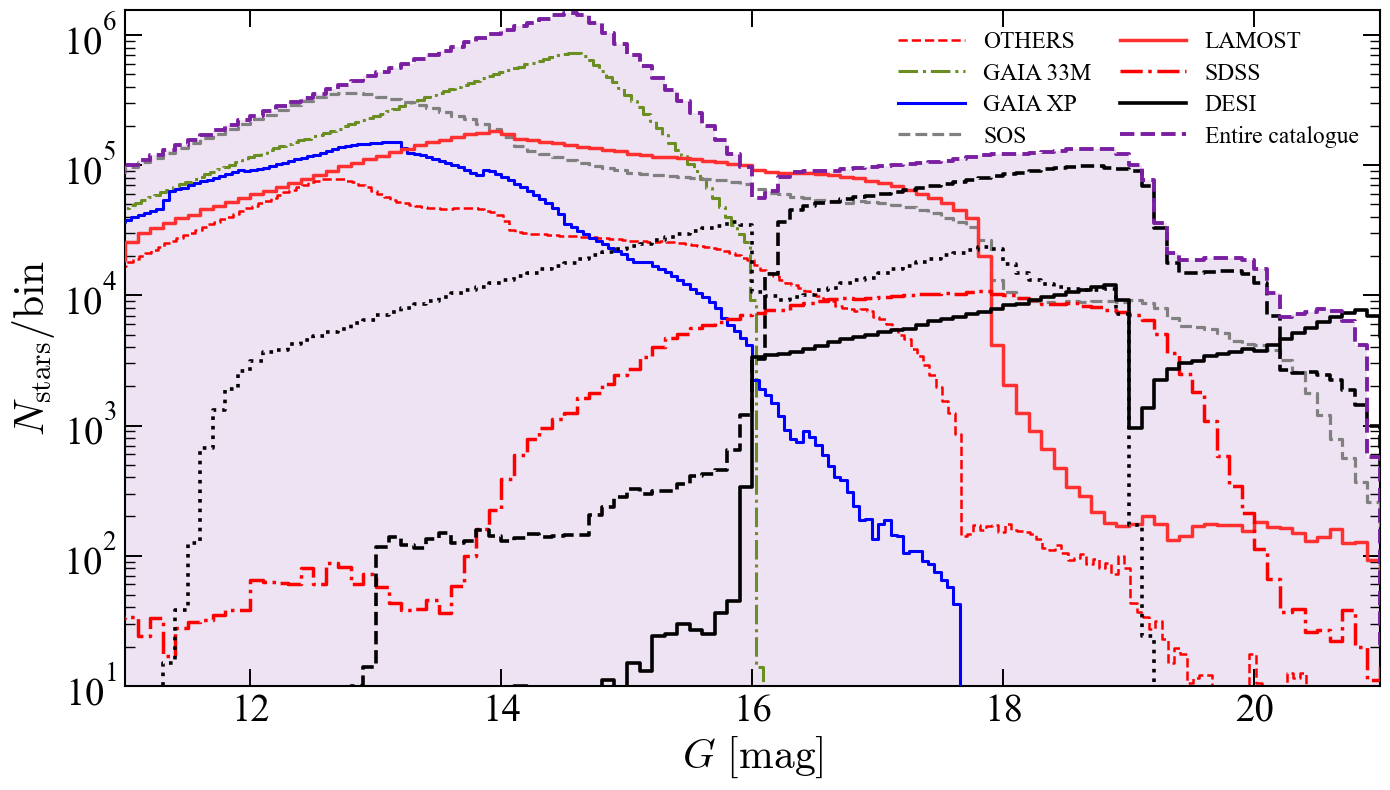}
				\caption{Gaia $G$-band magnitude distribution}
				\label{fig:gmag_plot}
			\end{subfigure}
			\caption{Distributions of stellar metallicity and apparent magnitude for the constituent surveys of the homogenised Milky Way 6D phase-space catalogue. The upper panel displays the number of stars per metallicity bin ([Fe/H]), highlighting the distinct and overlapping contributions of surveys such as DESI, LAMOST, GAIA, and SDSS to the aggregated catalogue. The bottom panel illustrates the stellar counts as a function of the Gaia $G$ broad-band magnitude, showing how the different observational samples complement each other to provide coverage across a wide range of apparent brightnesses.}
			\label{fig:stellar_distributions}
		\end{figure*}
		
		\vspace{-5cm}
		
		\begin{table*}[htbp]]
			\centering
			\caption{Period--luminosity and related calibrations adopted for the distance estimation of variable
				stars in \texttt{CAT\_Variables\_PhotDist\_rv\_cut}. Here $P$ is the pulsation period in days and
				$[\mathrm{Fe/H}]$ is in dex.}
			\label{tab:pl_relations_variables}
			\footnotesize
			\setlength{\tabcolsep}{3.5pt}
			\renewcommand{\arraystretch}{1.08}
			\begin{tabular*}{\textwidth}{@{\extracolsep{\fill}} l p{0.58\textwidth} l}
				\hline\hline
				Class & Relation & Reference \\
				\hline
				Classical Cepheids &
				\begin{tabular}[t]{@{}l@{}}
					$M_V = -2.76\log_{10}P - 1.40$ \\
					$M_I = -2.96\log_{10}P - 1.81$
				\end{tabular}
				& \citet{Madore1991} \\
				
				Type~II Cepheids &
				$M_{K_s} = -2.41\log_{10}P - 1.02$
				& \citet{Matsunaga2006} \\
				
				RR~Lyrae &
				\begin{tabular}[t]{@{}l@{}}
					$M_{K_s} = -2.53\log_{10}P + 0.07\,[\mathrm{Fe/H}] - 0.95$ \\
					$M_G = 0.32\,[\mathrm{Fe/H}] + 1.11$
				\end{tabular}
				& \citet{Muraveva2015,Muraveva2018} \\
				
				Miras &
				\begin{tabular}[t]{@{}l@{}}
					$M_K = -3.51(\log_{10}P - 2.38) - 7.15$ \quad (LMC) \\
					$M_K = -3.50(\log_{10}P - 2.38) - 7.25$ \quad (Galactic)
				\end{tabular}
				& \citet{Whitelock2008} \\
				
				$\delta$~Scuti &
				$M_V = -3.725\log_{10}P - 1.969$
				& \citet{McNamara2011} \\
				
				SX~Phe &
				$M_V = -3.389\log_{10}P - 1.640$
				& \citet{McNamara1997} \\
				
				Overtone correction &
				$\log_{10}P_F = \log_{10}P + 0.128$
				& \citet{Braga2015} \\
				\hline
			\end{tabular*}
		\end{table*}
		
		\begin{table*}[htbp]
			\centering
			\LARGE
			\setlength{\tabcolsep}{2.5pt}
			\caption{Comparison of the iterative-clipped raw-row weighted 1D polynomial fits (Ind.) and survey-wise
				weighted additive multivariate polynomial fits using the same per-parameter orders (Multi.).}
			\label{tab:rv_grand_poly_compare}
			\resizebox{\textwidth}{!}{%
				\begin{tabular}{llrrrrrrrrrrrrrrr}
					\toprule
					Param & $c_k$ & \multicolumn{2}{c}{APOGEE} & \multicolumn{2}{c}{DESI} & \multicolumn{2}{c}{GALAH} & \multicolumn{2}{c}{GES} & \multicolumn{2}{c}{LAMOST} & \multicolumn{2}{c}{RAVE} & \multicolumn{2}{c}{SDSS} & Global \\
					& & Ind. & Multi. & Ind. & Multi. & Ind. & Multi. & Ind. & Multi. & Ind. & Multi. & Ind. & Multi. & Ind. & Multi. & Ind. \\
					\midrule
					\multirow{5}{*}{FeH (2)} & $N$ & 194\,741 & 201\,726 & 3\,356\,610 & 193\,183 & 113\,307 & 118\,550 & 1\,830 & 2\,043 & 1\,210\,148 & 1\,167\,827 & 94\,085 & 68\,092 & 13\,299 & 1\,835 & 4\,984\,020 \\
					& $\chi^2_{\rm red}$ & 0.974 & -- & 0.488 & -- & 0.993 & -- & 0.929 & -- & 1 & -- & 0.863 & -- & 0.675 & -- & 0.65 \\
					& $c2$ & 0.17 & 0.099 & -0.068 & $5.163\!\times\!10^{-3}$ & -0.055 & -0.028 & 0.187 & 0.407 & 0.012 & -0.151 & 0.023 & -0.432 & 0.056 & 0.615 & $1.027\!\times\!10^{-14}$ \\
					& $c1$ & 0.248 & 0.207 & 0.34 & 0.106 & -0.112 & $-4.091\!\times\!10^{-3}$ & 1.821 & 0.67 & -0.101 & -0.266 & 0.175 & -0.047 & 0.486 & 1.087 & $-4.283\!\times\!10^{-14}$ \\
					& $c0$ & -0.33 & -- & -0.238 & -- & -0.405 & -- & -0.61 & -- & -0.161 & -- & 0.603 & -- & -0.411 & -- & $-3.982\!\times\!10^{-14}$ \\
					\midrule
					\multirow{6}{*}{Gmag (3)} & $N$ & 280\,434 & 201\,726 & 3\,896\,979 & 193\,183 & 150\,696 & 118\,550 & 2\,399 & 2\,043 & 1\,441\,145 & 1\,167\,827 & 142\,381 & 68\,092 & 18\,453 & 1\,835 & 5\,932\,487 \\
					& $\chi^2_{\rm red}$ & 0.993 & -- & 0.273 & -- & 0.985 & -- & 0.961 & -- & 0.999 & -- & 0.828 & -- & 0.698 & -- & 0.516 \\
					& $c3$ & $3.506\!\times\!10^{-3}$ & $5.031\!\times\!10^{-3}$ & $3.226\!\times\!10^{-4}$ & $-2.362\!\times\!10^{-3}$ & $-2.971\!\times\!10^{-3}$ & $-8.618\!\times\!10^{-3}$ & -0.048 & 0.041 & -0.013 & 0.054 & 0.041 & 0.154 & 0.03 & 0.075 & $2.454\!\times\!10^{-11}$ \\
					& $c2$ & -0.113 & -0.151 & -0.027 & 0.079 & 0.098 & 0.301 & 1.659 & -1.425 & 0.563 & -1.898 & -1.3 & -5.001 & -1.403 & -3.232 & $-1.024\!\times\!10^{-9}$ \\
					& $c1$ & 1.21 & 1.532 & 0.38 & -0.964 & -1.026 & -3.461 & -18.743 & 15.681 & -7.863 & 21.953 & 13.512 & 54.185 & 21.119 & 46.062 & $1.395\!\times\!10^{-8}$ \\
					& $c0$ & -4.534 & -- & -1.123 & -- & 2.907 & -- & 69.007 & -- & 34.627 & -- & -45.344 & -- & $-1.036\!\times\!10^{2}$ & -- & $-6.212\!\times\!10^{-8}$ \\
					\midrule
					\multirow{6}{*}{RV (3)} & $N$ & 303\,656 & 201\,726 & 3\,897\,147 & 193\,183 & 160\,124 & 118\,550 & 2\,536 & 2\,043 & 1\,510\,407 & 1\,167\,827 & 103\,563 & 68\,092 & 18\,453 & 1\,835 & 5\,995\,886 \\
					& $\chi^2_{\rm red}$ & 0.951 & -- & 0.607 & -- & 0.977 & -- & 0.913 & -- & 0.998 & -- & 0.907 & -- & 0.724 & -- & 0.739 \\
					& $c3$ & $-9.869\!\times\!10^{-8}$ & $-1.373\!\times\!10^{-7}$ & $-1.693\!\times\!10^{-10}$ & $7.621\!\times\!10^{-10}$ & $1.192\!\times\!10^{-8}$ & $1.847\!\times\!10^{-8}$ & $9.369\!\times\!10^{-8}$ & $3.583\!\times\!10^{-8}$ & $2.211\!\times\!10^{-7}$ & $2.523\!\times\!10^{-7}$ & $-6.49\!\times\!10^{-7}$ & $-4.227\!\times\!10^{-7}$ & $3.2\!\times\!10^{-9}$ & $-6.396\!\times\!10^{-10}$ & 0 \\
					& $c2$ & $3.446\!\times\!10^{-7}$ & $2.595\!\times\!10^{-6}$ & $-5.423\!\times\!10^{-6}$ & $1.87\!\times\!10^{-8}$ & $3.092\!\times\!10^{-6}$ & $1.438\!\times\!10^{-6}$ & $-6.857\!\times\!10^{-6}$ & $-3.894\!\times\!10^{-6}$ & $6.414\!\times\!10^{-6}$ & $1.402\!\times\!10^{-5}$ & $2.209\!\times\!10^{-5}$ & $-2.814\!\times\!10^{-5}$ & $7.478\!\times\!10^{-8}$ & $-1.339\!\times\!10^{-7}$ & 0 \\
					& $c1$ & $-5.281\!\times\!10^{-4}$ & $-3.081\!\times\!10^{-4}$ & $3.076\!\times\!10^{-5}$ & $-9.291\!\times\!10^{-5}$ & $-6.608\!\times\!10^{-4}$ & $-9.9\!\times\!10^{-4}$ & $-5.67\!\times\!10^{-3}$ & $-9.927\!\times\!10^{-4}$ & -0.021 & -0.022 & 0.088 & 0.08 & $-1.379\!\times\!10^{-3}$ & $-3.278\!\times\!10^{-4}$ & 0 \\
					& $c0$ & -0.146 & -- & -0.315 & -- & -0.269 & -- & -0.767 & -- & -0.291 & -- & 0.276 & -- & -0.756 & -- & $-4.056\!\times\!10^{-14}$ \\
					\midrule
					\multirow{6}{*}{SNR (3)} & $N$ & 302\,039 & 201\,726 & 226\,831 & 193\,183 & 160\,125 & 118\,550 & 2\,614 & 2\,043 & 1\,383\,252 & 1\,167\,827 & 150\,255 & 68\,092 & 2\,785 & 1\,835 & 2\,227\,901 \\
					& $\chi^2_{\rm red}$ & 0.928 & -- & 0.997 & -- & 0.978 & -- & 0.983 & -- & 0.999 & -- & 0.855 & -- & 0.904 & -- & 0.978 \\
					& $c3$ & $6.252\!\times\!10^{-8}$ & $1.092\!\times\!10^{-7}$ & $2.921\!\times\!10^{-6}$ & $6.152\!\times\!10^{-7}$ & $3.232\!\times\!10^{-8}$ & $7.062\!\times\!10^{-8}$ & $1.828\!\times\!10^{-7}$ & $1.637\!\times\!10^{-6}$ & $-2.544\!\times\!10^{-7}$ & $1.491\!\times\!10^{-7}$ & $1.805\!\times\!10^{-8}$ & $2.047\!\times\!10^{-9}$ & $-8.125\!\times\!10^{-5}$ & $-1.032\!\times\!10^{-4}$ & 0 \\
					& $c2$ & $-6.177\!\times\!10^{-5}$ & $-1.112\!\times\!10^{-4}$ & $-5.18\!\times\!10^{-4}$ & $-7.667\!\times\!10^{-5}$ & $-2.357\!\times\!10^{-5}$ & $-4.569\!\times\!10^{-5}$ & $-8.695\!\times\!10^{-5}$ & $-3.572\!\times\!10^{-4}$ & $1.185\!\times\!10^{-4}$ & $-6.23\!\times\!10^{-5}$ & $-3.329\!\times\!10^{-5}$ & $3.735\!\times\!10^{-6}$ & $6.396\!\times\!10^{-3}$ & 0.011 & 0 \\
					& $c1$ & 0.013 & 0.022 & 0.029 & $5.189\!\times\!10^{-3}$ & $4.977\!\times\!10^{-3}$ & $7.924\!\times\!10^{-3}$ & $8.206\!\times\!10^{-3}$ & $2.092\!\times\!10^{-4}$ & -0.011 & 0.012 & 0.016 & $7.278\!\times\!10^{-3}$ & -0.131 & -0.249 & 0 \\
					& $c0$ & -0.627 & -- & -0.395 & -- & -0.439 & -- & -0.594 & -- & -0.032 & -- & -0.146 & -- & 0.535 & -- & $-4.156\!\times\!10^{-14}$ \\
					\midrule
					\multirow{6}{*}{Teff (3)} & $N$ & 193\,380 & 201\,726 & 3\,356\,610 & 193\,183 & 113\,099 & 118\,550 & 1\,816 & 2\,043 & 1\,210\,548 & 1\,167\,827 & 94\,714 & 68\,092 & 13\,299 & 1\,835 & 4\,983\,466 \\
					& $\chi^2_{\rm red}$ & 0.965 & -- & 0.45 & -- & 0.988 & -- & 0.868 & -- & 1 & -- & 0.871 & -- & 0.667 & -- & 0.625 \\
					& $c3$ & $-3.793\!\times\!10^{-13}$ & $-2.08\!\times\!10^{-14}$ & $3.606\!\times\!10^{-12}$ & $3.638\!\times\!10^{-14}$ & $-7.681\!\times\!10^{-11}$ & $-3.918\!\times\!10^{-12}$ & $-1.754\!\times\!10^{-12}$ & $3.033\!\times\!10^{-11}$ & $5.648\!\times\!10^{-12}$ & $1.142\!\times\!10^{-12}$ & $-3.308\!\times\!10^{-13}$ & $1.983\!\times\!10^{-12}$ & $1.862\!\times\!10^{-12}$ & $4.326\!\times\!10^{-12}$ & 0 \\
					& $c2$ & $2.49\!\times\!10^{-8}$ & $8.193\!\times\!10^{-8}$ & $-7.448\!\times\!10^{-8}$ & $7.672\!\times\!10^{-8}$ & $1.152\!\times\!10^{-6}$ & $-5.025\!\times\!10^{-9}$ & $1.098\!\times\!10^{-7}$ & $-7.616\!\times\!10^{-7}$ & $-1.449\!\times\!10^{-7}$ & $2.491\!\times\!10^{-8}$ & $2.118\!\times\!10^{-8}$ & $-2.757\!\times\!10^{-8}$ & $-2.258\!\times\!10^{-9}$ & $-4.687\!\times\!10^{-8}$ & 0 \\
					& $c1$ & $-4.54\!\times\!10^{-4}$ & $-9.132\!\times\!10^{-4}$ & $7.045\!\times\!10^{-4}$ & $-7.787\!\times\!10^{-4}$ & $-5.689\!\times\!10^{-3}$ & $4.652\!\times\!10^{-4}$ & $-1.711\!\times\!10^{-3}$ & $5.717\!\times\!10^{-3}$ & $8.17\!\times\!10^{-4}$ & $-4.274\!\times\!10^{-4}$ & $-3.812\!\times\!10^{-4}$ & $6.606\!\times\!10^{-4}$ & $8.219\!\times\!10^{-6}$ & $1.79\!\times\!10^{-4}$ & $-1.031\!\times\!10^{-13}$ \\
					& $c0$ & 1.358 & -- & -2.511 & -- & 8.873 & -- & 5.237 & -- & -1.188 & -- & 2.011 & -- & -1.014 & -- & $2.448\!\times\!10^{-10}$ \\
					\midrule
					\multirow{5}{*}{logg (2)} & $N$ & 212\,563 & 201\,726 & 3\,356\,610 & 193\,183 & 120\,380 & 118\,550 & 1\,911 & 2\,043 & 1\,272\,904 & 1\,167\,827 & 98\,350 & 68\,092 & 13\,299 & 1\,835 & 5\,076\,017 \\
					& $\chi^2_{\rm red}$ & 0.93 & -- & 0.476 & -- & 0.981 & -- & 0.863 & -- & 1 & -- & 0.875 & -- & 0.664 & -- & 0.647 \\
					& $c2$ & -0.042 & -0.031 & -0.26 & $-2.546\!\times\!10^{-3}$ & 0.035 & 0.062 & -0.422 & $4.3\!\times\!10^{-3}$ & 0.112 & 0.024 & 0.082 & 0.068 & -0.171 & 0.239 & $2.486\!\times\!10^{-11}$ \\
					& $c1$ & 0.16 & 0.173 & 1.618 & 0.023 & -0.235 & -0.412 & 1.943 & -0.337 & -0.65 & -0.013 & -0.696 & -0.704 & 0.624 & -1.76 & $-1.74\!\times\!10^{-10}$ \\
					& $c0$ & -0.235 & -- & -2.482 & -- & 0.089 & -- & -2.467 & -- & 0.608 & -- & 2.052 & -- & -0.645 & -- & $2.823\!\times\!10^{-10}$ \\
					\midrule
					Shared $\chi^2$ & $\chi^2_{\rm red}$ & -- & 0.87 & -- & 0.995 & -- & 0.935 & -- & 0.948 & -- & 0.997 & -- & 0.905 & -- & 0.875 & -- \\
					\midrule
					Shared Int. & $b_0$ & -- & -4.031 & -- & 6.22 & -- & 11.551 & -- & -66.508 & -- & -83.966 & -- & $-1.975\!\times\!10^{2}$ & -- & $-2.142\!\times\!10^{2}$ & -- \\
					\bottomrule
				\end{tabular}
			}
		\end{table*}

\begin{figure*}[htbp]
	\centering
	\includegraphics[width=0.5\textwidth]{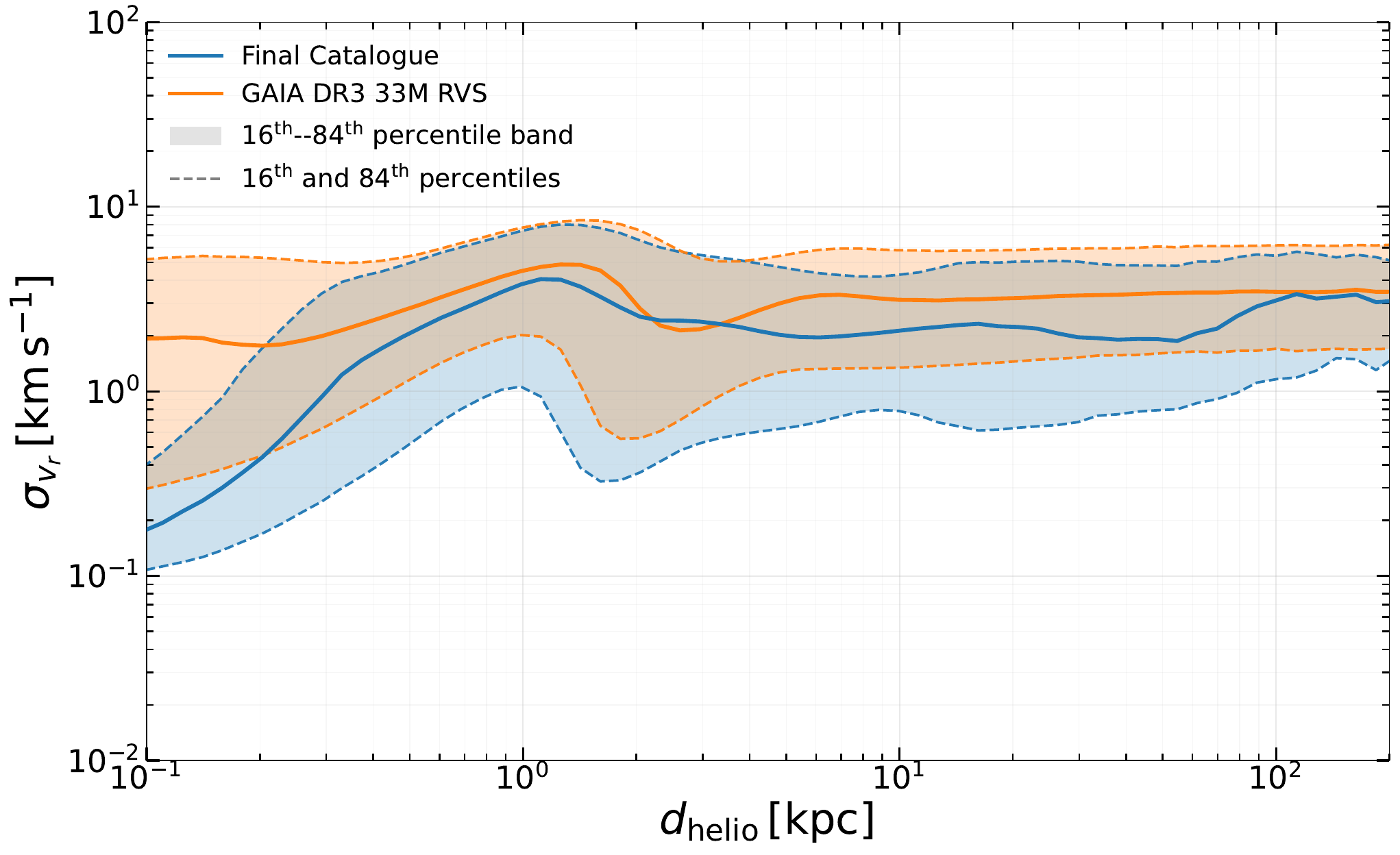}
	\caption{Absolute radial-velocity uncertainty ($\sigma_{v_r}$) as a function of heliocentric distance ($d_{\rm helio}$). The solid blue curve illustrates the median uncertainty for the final merged catalogue, while the solid orange curve represents the estimates from the GAIA DR3 33M RVS sample. The shaded regions and dashed lines correspond to the $16^{\rm th}$--$84^{\rm th}$ percentile bands for each respective dataset}
	\label{fig:rv_uncertainty_dist}
\end{figure*}
	\end{document}